\documentclass[%
reprint,
twocolumn,
amsmath,amssymb,
aps, 
pra,
showkeys
]{revtex4-2}
\usepackage[shortlabels]{enumitem}
\usepackage{tikz}
\usetikzlibrary{arrows.meta,positioning}
\usepackage{graphicx}
\graphicspath{{./}{figures/}}
\usepackage{xcolor}
\usepackage{comment}
\begin{document}


\title{Directional transfer from coherence to longitudinal polarization and algebraic storage in a transmon with
Markovian dephasing and a glassy environment: A non-Markovian fractional renewal model}


\author{Ruha Uğraş Erdoğan}
\email{ugras.erdogan@deu.edu.tr}  
\affiliation{Department of Biophysics, Faculty of Medicine, Dokuz Eylul University, Izmir, Türkiye}

\author{Ferit Acar Savacı}
\email{acarsavaci@iyte.edu.tr}
\thanks{Alternate email: savaciacar@gmail.com}
\affiliation{Department of Electrical and Electronics Engineering, Izmir Institute of Technology, Izmir, Türkiye}
\begin{abstract}
We develop a completely positive and trace-preserving non-Markovian model of a transmon qubit under longitudinal Markovian dephasing and transverse renewal events motivated by a glassy two-level-system environment. A microscopic Hamiltonian identifies the coupling subspaces; slow defect reconfigurations are coarse grained phenomenologically through a normalized Prabhakar waiting-time law. Forward chronological renewal histories yield an exact map-level Volterra equation, a noncommuting kernel--resolvent master equation, and an equivalent drift-conjugated Caputo-type representation with its required initial-value term. Without energy relaxation, Pauli-space dynamics separates into two one-dimensional subspaces and a coupled $(\sigma_y,\sigma_z)$ subspace supporting directional transfer between coherence and longitudinal qubit polarization; $\sigma_x$ polarization is renewal-insensitive. For heavy-tailed waiting times with $0<\alpha<1$, branch-cut analysis gives a generic $t^{-\alpha}$ storage tail in the longitudinal polarization generated from initial coherence and $O(t^{-1-\alpha})$ reverse transfer; constant-rate Poisson renewal yields pole-generated, exponentially decaying transients. Independent zero-temperature energy relaxation at finite $T_1$ exponentially cuts off these tails when its rate is below twice the pure-dephasing rate. To assess numerical accuracy, we compare independent inverse-Laplace reconstructions of the waiting-time density, memory kernel, and selected qubit-response components. For the studied Prabhakar parameters, Prabhakar renewal gives larger forward-transfer magnitudes at the prescribed readout than both a scale-matched constant-rate Poisson reference and an inhomogeneous Poisson reference matching the entire expected-count profile, including in finite-$T_1$ tests. Both Poisson references yield larger evaluated finite-window peaks. Thus disorder-motivated renewal memory reshapes transmon decoherence selectively by state and direction.
\end{abstract}

\maketitle

\section{\label{sec:INTRO}Introduction}
Decoherence remains one of the central obstacles to the realization of scalable quantum systems in information processing~\cite{BreuerPetruccione2007,NielsenChuang2010c,Watrous2018}, quantum communication~\cite{Wilde2017} and quantum control~\cite{WisemanMilburn2010,Jacobs2014}. The microscopic origins of quantum noise and its measurement back-action in mesoscopic systems are comprehensively reviewed in Ref.~\cite{ClerkDevoretGirvinMarquardtSchoelkopf2010}.

In superconducting circuits, transmon qubits embedded in the circuit quantum electrodynamics (cQED) architecture~\cite{SchoelkopfClerkGirvinLehnertDevoret2003,KrantzKjaergaardYanOrlandoGustavssonOliver2019} provide a scalable platform for quantum information processing~\cite{KochYuGambettaHouckSchusterMajerBlaisDevoretGirvinSchoelkopf2007,
SchreierHouckKochSchusterJohnsonChowGambettaMajerFrunzioDevoretGirvinSchoelkopf2008,
BlaisHuangWallraffGirvinSchoelkopf2004,
BlaisGrimsmoGirvinWallraff2021}. However, decoherence limits gate fidelities, error-correction thresholds~\cite{LidarBrun2013} and device scalability. 

Accordingly, we investigate how decoherence can be reshaped within a quantum fractional renewal framework based on the extensive literature on the dynamics of glassy two-level systems (TLSs) and transmon qubits. First, we delineate our main contributions and then briefly review the relevant literature to motivate our approach in the following subsections.

\subsection{\label{sec:Contrib}Contributions of this work}

We incorporate the established gamma-mixture representation of the Prabhakar waiting-time law, obtained by evaluating a one-sided $\alpha$-stable L\'evy subordinator at an independent gamma-distributed argument~\cite{CahoyPolito2013}, within an explicit quantum-renewal model of a transmon subject to longitudinal Markovian dephasing and transverse rotation kicks. This representation specifies the classical renewal-event statistics, while chronologically ordered, generally noncommuting drift and kick maps determine the reduced quantum evolution. For this explicit quantum specialization, we derive the
directional coherence--polarization response, its asymmetric long-time algebraic tails, and their exponential cutoff under independent zero-temperature energy relaxation, subject to the stated asymptotic and rate conditions. These results quantify both the directional persistence supported by Prabhakar renewal
and its limitation at finite $T_1$. The microscopic Hamiltonian identifies the longitudinal and transverse coupling sectors, while slow TLS-ensemble reconfigurations are modeled phenomenologically as a classical renewal process.

In the reference model illustrated in Fig.~\ref{fig:conceptual_framework}, $\Phi_X$ denotes the qubit
map describing the transverse rotation applied at each renewal event. Without energy relaxation, storage denotes the $y\to z$ transfer from coherence to longitudinal polarization; return denotes the reverse $z\to y$ transfer. Here $\alpha$ is the fractional exponent of the Prabhakar waiting-time law, $\Gamma_Z$ is the interevent pure-dephasing rate, and $\theta$ is the transverse rotation angle per event. For $0<\alpha<1$, $\Gamma_Z>0$, and $\sin\theta\neq0$, under the branch and pole conditions in
Sec.~\ref{sec:Prabhakar_asymptotics} and Appendix~\ref{app:Wyz_singularity_analysis}, the forward transfer has a leading $t^{-\alpha}$ tail, whereas the reverse transfer is generically
$O(t^{-1-\alpha})$. The constant-rate Poisson-renewal counterpart yields pole-generated, exponentially decaying transients (see Appendix~\ref{app:Poisson_resolvent_pole}).

We derive an exact factorization of the homogeneous transfer block associated with the Pauli operators
$\sigma_y$ and $\sigma_z$ when independent zero-temperature energy relaxation, with relaxation time $T_1$ and rate
$\Gamma_1=T_1^{-1}$, is added to the reference model. Differences between evolutions from opposite initial
Bloch vectors cancel the preparation-independent polarization offset of the resulting affine Bloch map
and isolate the directional response, which remains attenuated by relaxation. For $0<\Gamma_1<2\Gamma_Z$,
this factorization yields an exponential cutoff of the directional algebraic tails under the stated asymptotic
hypotheses. Appendix~\ref{app:finite_t1_directional} provides the derivation, and
Fig.~\ref{fig:sim108_finite_t1} quantifies the resulting relaxation-limited scope of directional storage.

Building on established quantum-renewal theory~\cite{Budini2004,Vacchini2020}, we construct the completely positive and trace-preserving (CPTP) map for the present model from chronological
histories and obtain the first-event Volterra equation, Eq.~\eqref{eq:23} (see Appendix~\ref{app:renewal_structure}). Its exact shifted kernel--resolvent form,
Eq.~\eqref{eq:exact_kernel_resolvent}, serves as our primary analytical formulation. This operator-valued Montroll--Weiss representation retains the noncommuting drift--kick order in the inverse operator pencil $\mathcal M_{\mathrm{fwd}}(s)$ of Eq.~\eqref{eq:M_def} (see Appendix~\ref{app:laplace_homogeneous_closure}). We derive the minimal common invariant decomposition, isolating the renewal-insensitive control subspace $W_x$ and reducing the active dynamics to the two-dimensional $W_{yz}$ block. The explicit reduced response is given by Eq.~\eqref{eq:cyz_block_eq}. For $0<\alpha<1$, the nonanalytic Prabhakar kernels require branch-cut
analysis beyond the pole equation. In Appendix~\ref{app:Wyz_singularity_analysis}, we derive the pole--branch-cut decomposition and the explicit coefficients supporting the directional tail laws stated above.

As an equivalent auxiliary time-domain form, we derive the drift-conjugated Caputo-type equation with its required initial-value term in Eq.~\eqref{eq:rho_fractional}. In Appendix~\ref{app:prabhakar_forward_series}, we prove local $L^1$ convergence of the renewal-density series for $m(t)$ in Eq.~\eqref{eq:D_m_series} and of the drift-dressed Prabhakar-convolution expansion. We establish absolute and uniform convergence of the differentiated renewal-density series on every compact interval in $(0,\infty)$, justifying the positive-time kernel series in Eq.~\eqref{eq:memKt}, and show that the drift-dressed expansion resums to the same forward-ordered resolvent.

Numerical comparisons use both a scale-matched constant-rate Poisson reference and an inhomogeneous Poisson reference
that matches the Prabhakar expected event count at every time. For the parameters tested in Fig.~\ref{fig:sim108_finite_t1}, including the finite-$T_1$
cases, Prabhakar renewal gives a larger forward-transfer magnitude at the prescribed readout, whereas both Poisson
references give larger evaluated finite-window peaks. The readout difference therefore cannot be attributed solely to unequal mean event activity.
Independent inverse-Laplace reconstructions test selected entries of the reduced resolvent; Appendix~\ref{app:numerical_inverse_laplace} specifies
their scope.

These results build on the first author's Ph.D. thesis~\cite{erdogan2025}. Figure~\ref{fig:conceptual_framework} summarizes the coupling architecture, coarse-grained renewal clock, and resulting $W_{yz}$ storage--return subspace.

\subsection{\label{sec:DGTLSSQ}Disorder engineering of glassy two-level systems as non-Markovian environments in superconducting qubits}
Amorphous surface and interfacial oxides host tunneling two-level systems (TLSs) that contribute to dielectric loss and noise in superconducting circuits~\cite{MullerColeLisenfeld2019,Siddiqi2021}. In amorphous $AlO_{x}$ layers and interfaces, these TLSs behave as disordered dipolar relaxors with switching-rate distributions spanning hierarchical timescales~\cite{FaoroIoffe2006,FaoroIoffe2015,MullerShnirmanMakhlin2009,
ChurkinMatityahuMaksymovBurinSchechter2021}. Structural disorder produces heterogeneous defect
parameters, while electric and elastic TLS interactions provide mechanisms for slow, heterogeneous dynamics,
spectral diffusion, and correlated fluctuations~\cite{FaoroIoffe2006,FaoroIoffe2015,ChurkinMatityahuMaksymovBurinSchechter2021}.
Broad switching-rate distributions provide a microscopic origin of $1/f^\eta$ noise~\cite{Weissman1988}, and
TLS-induced noise is a dominant source of qubit dephasing~\cite{PaladinoGalperinFalciAltshuler2014}.
Broad relaxation scales and defect switching can also generate nonexponential coherence dynamics, limiting
descriptions based on a single Markovian dephasing rate~\cite{PaladinoGalperinFalciAltshuler2014,
BergliGalperinAltshuler2009}.

Alongside materials and interface optimization~\cite{Murray2021}, modifying defect environments can alter their dynamics. Immersion in liquid $^3\mathrm{He}$ enhances TLS energy relaxation and enables suppression of low-temperature frequency noise in superconducting resonators~\cite{LucasDanilovLevitinJayaramanCaseyFaoroTzalenchukKubatkinSaundersdeGraaf2023}.
Phononic-bandgap engineering provides a complementary way to modify TLS relaxation; its mechanism and experimental signatures are discussed in Sec.~\ref{sec:GTSL}.

Measurements linking material disorder to flux-noise amplitudes in fluxonium devices constrain superconducting-material design~\cite{GaoWuSunChenDengMaMiaoSongWanWangXiaYingZhangShiZhaoDeng2025}.
At the circuit level, theoretical proposals for dynamical sweet spots offer complementary reduction of sensitivity to low-frequency noise through periodic driving~\cite{HuangMundadaGyenisSchusterHouckKoch2021}.
Theoretical studies also show that introducing spectral gaps and varying their edge sharpness can control trace-distance revivals and polarization relaxation in bosonic and fermionic reservoir models~\cite{Luo2026}.

The possibility of tailoring transport statistics through engineered structural disorder is illustrated by optical L\'evy glasses. A prescribed diameter distribution of
refractive-index-matched glass microspheres creates spatial variations in the density of scattering particles, producing a heavy-tailed distribution of propagation step lengths over a finite range~\cite{BarthelemyBertolottiWiersma2008}. In our model, heavy-tailed statistics are instead assigned to the time intervals between coarse-grained TLS ensemble reconfigurations.
\subsection{\label{sec:NDRHN}Non-Debye dielectric relaxation and the Prabhakar representation of the Havriliak–Negami response}
The broad, heterogeneous dynamics of TLS-bearing amorphous dielectrics are consistent with non-Debye relaxation in structurally disordered materials~\cite{BouchaudGeorges1990,Weissman1988, MullerColeLisenfeld2019}. The Havriliak--Negami (HN) response empirically describes such broad relaxation spectra~\cite{HavrilikNegami1967,GorskaHorzelaBratekDattoliPenson2018,jurlewicz2003stochastic}. The inverse Laplace transform of its normalized susceptibility is a Prabhakar (three-parameter Mittag--Leffler function) response density~\cite{GarraGarrappa2018}, whose normalized survival integral gives the associated relaxation
function. Within its completely monotone (CM) parameter domain, the Prabhakar family provides a physically admissible generalization of Debye and standard Mittag--Leffler relaxation~\cite{MainardiGarrappa2015,GorskaHorzelaBratekDattoliPenson2018}. CM of the normalized
relaxation function guarantees a positive Bernstein representation in exponential modes, permitting an effective relaxation-rate spectrum without requiring a one-to-one correspondence with microscopic TLS rates.

For the Cole--Davidson (CD) specialization, $\alpha=1$, and the shape parameter $\gamma$ satisfies $0<\gamma\leq1$. In this range, the normalized relaxation function is CM and admits the Bernstein representation
\begin{equation}
\begin{aligned}
&\Phi_{\mathrm{CD}}(t)
=\int_{[0,\infty)} e^{-rt}\,\mu_{\mathrm{CD}}(dr),\\
&\int_{[0,\infty)}\mu_{\mathrm{CD}}(dr)=1,
\end{aligned}
\label{eq:CD_Bernstein_spectrum}
\end{equation}
where the positive Borel probability measure $\mu_{\mathrm{CD}}$ weights exponential modes with effective Debye rates $r\geq0$. Equation~\eqref{eq:CD_Bernstein_spectrum} thus describes relaxation-rate heterogeneity. The construction in Sec.~\ref{subsec:gamma_stable_subord} generates renewal waiting times through a gamma mixture of
one-sided $\alpha$-stable L\'evy laws~\cite{GorskaHorzelaBratekDattoliPenson2018,weron2005havriliak}. Its gamma-distributed variable $U$ specifies the argument of the one-sided $\alpha$-stable Lévy subordinator and must be distinguished from the relaxation rate $r$.

The dielectric relaxation function $\Phi_{\mathrm{HN/CD}}(t)$ and the renewal survival probability $\Psi(t)$ introduced in Sec.~\ref{sec:PMRD} are therefore not identified as the same microscopic observable. The HN/Prabhakar family, its CM properties and its positive-mixture representations instead provide phenomenological motivation for the normalized coarse-grained waiting-time law assigned to TLS ensemble reconfiguration events in Sec.~\ref{subsec:ASLS}. This coarse-grained description is motivated by the broad structural and interaction disorder of amorphous TLS ensembles~\cite{FaoroIoffe2006,FaoroIoffe2015,
MullerColeLisenfeld2019}.

\subsection{\label{sec:GTSL}Experimental signatures of interacting glassy TLS ensembles as non-Markovian environments}

Experiments probe the TLS dynamics discussed in Sec.~\ref{sec:DGTLSSQ}. A single slow TLS can dominate qubit coherence over hours~\cite{SchlorLisenfeldMullerBilmesSchneiderPappasUstinovWeides2019}.
Waiting-time distributions and tunneling rates have been extracted for a charge fluctuator coupled to a
superconducting charge detector~\cite{jenei2019waiting}, while relaxation-time measurements combined with
simulations link stochastic qubit fluctuations to TLS interactions~\cite{BejaninEarnestSharafeldinMariantoni2021}. Adaptive Bayesian measurements estimate transmon relaxation times within a few milliseconds; statistical analysis indicates TLS switching rates up to approximately $10\,\mathrm{Hz}$~\cite{Berritta2026}. By modifying the available phonon
modes~\cite{ChenOwensPuttermanSchaferPainter2024}, phononic-bandgap engineering suppresses TLS--phonon emission, extends TLS lifetimes into the millisecond regime, and induces explicitly non-Markovian
behavior~\cite{OdehGodeneliLiTangiralaZhouZhangZhangSipahigil2025}. Non-Poissonian quantum-jump statistics have been reported and interpreted in terms of long-lived TLSs~\cite{Gosling2026}.

Electric-field tuning reveals hysteretic memory lasting seconds in the defect bath of a superconducting
qubit~\cite{Agarwal2026}. Together, these studies indicate that, in relevant parameter regimes, the long-lived temporal structure of amorphous TLS environments is inadequately described by a single-rate exponential reservoir.

Despite their structural disorder, TLSs are often coherently addressable and tunable~\cite{RosenKhalilBurinOsborn2016}. Electric-field and strain spectroscopy directly map TLS
parameters and reveal coherent qubit--TLS interactions~\cite{LisenfeldGrabovskijMullerColeWeissUstinov2015,
LisenfeldBilmesMegrantBarendsKellyKlimovWeissMartinisUstinov2019,MullerColeLisenfeld2019,LisenfeldBilmesUstinov2023}.
DC electric-field tuning optimizes qubit relaxation times by shifting individual TLSs into or out of resonance with the qubit transition~\cite{LisenfeldBilmesUstinov2023}. Theoretical studies propose microwave dressing to mitigate TLS-induced relaxation fluctuations~\cite{ZhaoMaJinYu2022} and analyze how external driving redistributes TLS noise spectra~\cite{BroxBergliGalperin2011}.

These developments motivate treating TLS environments as potentially controllable sources of noise and memory whose modification may suppress particular contributions to qubit decoherence. To connect this phenomenology with the reduced qubit dynamics, we represent coarse-grained TLS reconfigurations as discrete events separated by random waiting times. Independent successive event intervals and the Prabhakar waiting-time law are modeling assumptions specified in Sec.~\ref{subsec:Physmot_fract}; they are not established by the cited experiments.

\subsection{\label{sec:RTNM}Quantum fractional renewal theory and non-Markovian dynamics of decoherence}

Non-Markovian open-system dynamics are studied through time-local master equations, divisibility criteria, and information-theoretic measures of memory~\cite{BreuerPetruccione2007, RivasVargasHuelga2012,WolfEisertCubittCirac2008,BylickaChruscinskiManiscalco2014,DeVegaAlonso2017}. Non-Markovian transmon dynamics in an open multimode resonator has also been studied theoretically~\cite{MalekakhlaghPetrescuTureci2016}.

Budini constructed completely positive non-Markovian master equations by subordinating a Markovian Lindblad semigroup with a classical waiting-time distribution~\cite{Budini2004,Budini2005}. Breuer and Vacchini formulated the quantum semi-Markov framework and general conditions for complete positivity (CP)~\cite{BreuerVacchini2008}, while Vacchini systematized more general quantum-renewal constructions~\cite{Vacchini2020}. Peng and Zhang~\cite{PengZhang2026} relate Caputo fractional master equations to memory-kernel dynamics and, for a time-independent Lindblad generator and $0<\alpha<1$, express the fractional evolution as an average of the corresponding semigroup over an inverse-stable random operational time. Short-memory kernels can also modify decoherence while preserving CP~\cite{DafferWodkiewiczCresserMcIver2004,MarshallCamposVenutiZanardi2017a}, and classical control protocols have experimentally emulated non-Markovian master equations in superconducting qubits~\cite{VlachosZhangMauryaMarshallAlbashLevenson-Falk2022a}.

Prabhakar renewal statistics have also been applied to photon detection. Tey et al.~\cite{TeyChongMuniandy2026} fitted a dead-time-modified generalized fractional Poisson model to
photon-detection interarrival data from a spontaneous parametric down-conversion source and compared Monte Carlo photon-count distributions with measurements.

We specialize quantum-renewal theory to Prabhakar-distributed waiting times, longitudinal Markovian dephasing, and transverse kicks, retaining the chronological order of the generally noncommuting drift and kick maps~\cite{Vacchini2020}. Renewal preparation and aging are discussed in Sec.~\ref{subsec:Physmot_fract}~\cite{GodrecheLuck2001,BarkaiCheng2003}. Here non-Markovianity denotes memory associated with nonexponential renewal waiting times and their nonlocal contribution to the reduced evolution~\cite{Megier2021}. This usage does not by itself establish trace-distance backflow or CP-indivisibility of the reduced dynamical map; these criteria are not evaluated here.

\subsection{Outline}
Section~\ref{sec:PMRD} introduces the two-environment transmon model and coarse-graining convention, derives the normalized Prabhakar waiting-time law with its admissible domain and asymptotics, and presents its gamma-mixture representation by evaluating a one-sided $\alpha$-stable L\'evy subordinator at an independent gamma-distributed argument. Section~\ref{sec:gen_master_qrenewal} constructs the forward quantum-renewal map and its Volterra, kernel--resolvent, operator-space, and drift-conjugated Caputo-type representations. Section~\ref{sec:dynamics_decoherence} specializes this construction to the transmon, deriving the exact invariant-subspace reduction and analyzing storage--return dynamics and branch-cut asymptotics.
Comparisons with Poisson renewal include surviving directional transfer and finite-window peaks under independent energy relaxation at finite $T_1$. Section~\ref{sec:CONCL} concludes the paper.

Appendix~\ref{app:prabhakar_admissibility} distinguishes scalar CM conditions from CP of the quantum map. In Appendix~\ref{app:renewal_structure}, we derive the first-event factorization and homogeneous memory closure for the present model. Appendices~\ref{app:Wyz_singularity_analysis} and~\ref{app:prabhakar_forward_series} present our supporting
derivations for the reduced spectral analysis and the convergence, differentiation, and resummation of the Prabhakar series, respectively. Appendix~\ref{app:numerical_inverse_laplace} documents the numerical inverse-Laplace cross-checks. In Appendix~\ref{app:finite_t1_directional}, we derive the finite-$T_1$ extension and specify and numerically validate the count-matched inhomogeneous Poisson benchmark.

\begin{figure*}[t!]
  \centering
  \includegraphics[width=\textwidth]{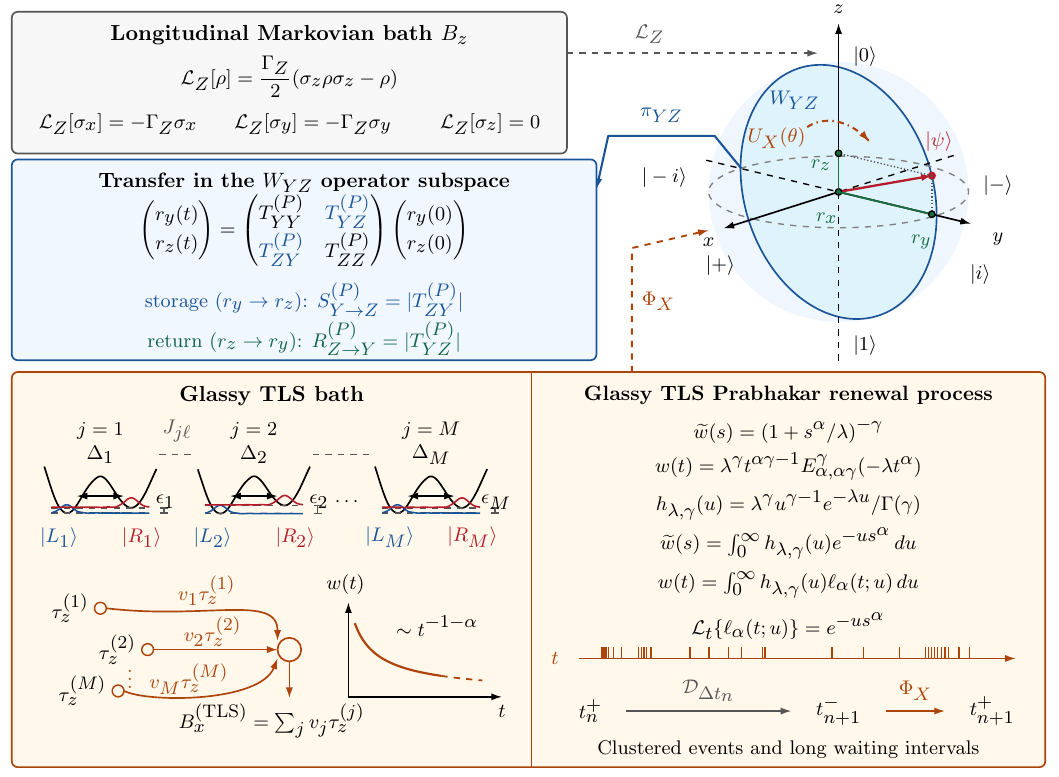}
  \caption{Conceptual structure of the fractional-renewal transmon model. Gray and orange connectors indicate longitudinal Markovian dephasing $\mathcal L_Z$ and transverse TLS kicks $\Phi_X$, respectively; TLS notation follows Eqs.~\eqref{eq:3a}--\eqref{eq:3c}. Independent Prabhakar waiting times are drawn from the law obtained by evaluating a one-sided $\alpha$-stable L\'evy subordinator at an independent gamma-distributed argument. This scalar construction does not subordinate a quantum semigroup. The waiting-time curve and event timeline are schematic; the illustrated algebraic tail applies for $0<\alpha<1$. The strip shows conditioned-state drift $\mathcal D_{\Delta t_n}=e^{\Delta t_n\mathcal L_Z}$ over $\Delta t_n=t_{n+1}-t_n$, followed by an instantaneous $x$-axis rotation kick. Superscripts $-$ and $+$ denote pre- and post-event states. The blue connector represents the coordinate projection $\pi_{YZ}:\rho\mapsto(r_y,r_z)^{\mathsf T}$ onto $W_{YZ}=\mathrm{span}\{\sigma_y,\sigma_z\}$, not an additional dynamical map. The reduced matrix distinguishes storage $y\to z$ through $T_{ZY}^{(P)}$ from return $z\to y$ through $T_{YZ}^{(P)}$. R.U.Erdogan designed the schematic and used ChatGPT (OpenAI; GPT-5 Sol Ultra) to generate the MATLAB and \LaTeX{}/TikZ code used to draw it. He reviewed the code for consistency with the mathematical and physical model.}
\label{fig:conceptual_framework}
\end{figure*}

\section{\label{sec:PMRD}Fractional renewal description of the transmon qubit coupled to an amorphous TLS environment}
We consider a superconducting transmon coupled to two physically distinct environments: a longitudinal Markovian bath and a transversely coupled amorphous environment dominated by TLS defects. The latter represents oxide and interface defects in superconducting circuits whose broad, intermittent switching produces low-frequency noise and motivates the coarse-grained heavy-tailed renewal model~\cite{PaladinoGalperinFalciAltshuler2014,SchlorLisenfeldMullerBilmesSchneiderPappasUstinovWeides2019,
FaoroIoffe2006}. The total and qubit Hamiltonians are
\begin{subequations}
\begin{alignat}{2}
&H&\;=\; &H_S+H_{B_z}+H_{B_x}^{(\mathrm{TLS})}
+H_{SB_z}+H_{SB_x},\label{eq:1a}\\
&H_S&\;=\; &-\frac{\hbar\omega_q}{2}\,\sigma_z.
\label{eq:1b}
\end{alignat}
\end{subequations}
Here $\omega_q$ is the qubit transition frequency, and $H_S$ is written in the energy eigenbasis of the effective two-level transmon. We set $\hbar=1$ throughout.

The microscopic Hamiltonian $H$ is autonomous in the Schr\"odinger picture. Although $B_x^{(\mathrm{TLS})}$ has no explicit time dependence in this picture, its interaction-picture representative $B_x^{(\mathrm{TLS})}(t)$ evolves under $H_{B_x}^{(\mathrm{TLS})}$ in Eq.~\eqref{eq:3a}.

\subsection{Microscopic bath models and modeling convention}
We model the longitudinal environment as a bosonic quantum reservoir of fast electromagnetic or phononic fluctuations,
\begin{subequations}
\begin{alignat}{2}
&H_{B_z}&\;=\; &\sum_k \omega_k a_k^\dagger a_k,
\label{eq:2a}\\
&H_{SB_z}&\;=\; &\sigma_z \otimes B_z,
\label{eq:2b}\\
&B_z&\;=\; &\sum_k
\left(g_k a_k+g_k^\ast a_k^\dagger\right),
\label{eq:2c}
\end{alignat}
\end{subequations}
\noindent
where $a_k$ and $a_k^\dagger$ are bosonic annihilation and creation operators for the independent modes of angular frequency $\omega_k$ in Eq.~\eqref{eq:2a}, satisfying
\begin{equation}
[a_k,a_{k'}^\dagger]=\delta_{kk'}.
\label{eq:CRa_adag}
\end{equation}
The complex coupling amplitudes $g_k$ define the Hermitian bath operator $B_z$ in Eq.~\eqref{eq:2c}, which couples to $\sigma_z$ through Eq.~\eqref{eq:2b}. We assume rapidly decaying reservoir correlations and apply the Born--Markov and secular approximations~\cite{BreuerPetruccione2007,
GoriniKossakowskiSudarshan1976,Lindblad1976}. Tracing out the longitudinal reservoir gives the Markovian pure-dephasing generator
\begin{equation}
\mathcal L_Z[\rho]
=
\frac{\Gamma_Z}{2}
\left(\sigma_z\rho\sigma_z-\rho\right),
\label{eq:LZ_generator}
\end{equation}
acting on the reduced qubit density operator in the transmon energy eigenbasis.

Before this reduction, the longitudinal interaction picture is defined with respect to
\begin{subequations}
\begin{equation}
H_0=H_S+H_{B_z},
\label{eq:H0}
\end{equation}
giving
\begin{equation}
V_z^I(t)
=e^{iH_0t}H_{SB_z}e^{-iH_0t}
=\sigma_z\otimes B_z^I(t).
\label{eq:long_interact}
\end{equation}
\end{subequations}

The TLS Hamiltonian combines standard tunneling-model terms for asymmetric double-well excitations in amorphous
solids~\cite{AndersonHalperinVarma1972,Phillips1972} with effective interactions between
defects~\cite{FaoroIoffe2006,FaoroIoffe2015, MullerShnirmanMakhlin2009},
\begin{subequations}
\begin{alignat}{2}
&H_{B_x}^{(\mathrm{TLS})}&\;=\; &
\sum_j\left(
\frac{\epsilon_j}{2}\,\tau_z^{(j)}
+\frac{\Delta_j}{2}\,\tau_x^{(j)}
\right)
+\sum_{j<\ell}J_{j\ell}\,\tau_z^{(j)}\tau_z^{(\ell)},
\label{eq:3a}\\
&H_{SB_x}&\;=\; &\sigma_x\otimes B_x^{(\mathrm{TLS})},
\label{eq:3b}\\
&B_x^{(\mathrm{TLS})}&\;=\; &\sum_j v_j\,\tau_z^{(j)}.
\label{eq:3c}
\end{alignat}
\end{subequations}
\noindent
In Eqs.~\eqref{eq:3a}--\eqref{eq:3c}, $\tau_{x,z}^{(j)}$ are Pauli operators on the $j$th TLS
Hilbert space, $\epsilon_j$ and $\Delta_j$ are its asymmetry energy and tunneling amplitude, $J_{j\ell}$ parametrizes the effective interaction between the $j$th and $\ell$th TLSs and the real $v_j$ are transverse qubit--TLS coupling coefficients. The uncoupled $j$th TLS has the bare splitting $E_j=(\epsilon_j^2+\Delta_j^2)^{1/2}$. Equation~\eqref{eq:3c} defines the collective TLS bath operator $B_x^{(\mathrm{TLS})}$ coupled to $\sigma_x$ in Eq.~\eqref{eq:3b}. The operator $\tau_z^{(j)}$ is defined in the localized (configurational) basis, so its fluctuations describe configurational switching. After coarse-graining, these events induce transverse qubit noise through the effective kick map $\Phi_X$, consistent with microscopic models in which TLS switching modulates transverse tunneling amplitudes or effective fields in superconducting circuits~\cite{FaoroIoffe2006,MullerColeLisenfeld2019}.

Although $H_{B_x}^{(\mathrm{TLS})}$ is a quantum Hamiltonian, the TLS ensemble is not treated as a coherent quantum reservoir in the reduced dynamics. Its slow, incoherent, strongly disordered configurational dynamics are coarse-grained into a classical stochastic process of ensemble reconfigurations~\cite{PaladinoGalperinFalciAltshuler2014,FaoroIoffe2015,BouchaudGeorges1990}.
This replaces $H_{B_x}^{(\mathrm{TLS})}+H_{SB_x}$ by a classical renewal history: the qubit undergoes Markovian drift generated by $\mathcal L_Z$ between events and a CPTP kick $\Phi_X$ at each event.

This classical description is appropriate when the fluctuators' internal decoherence dominates their coherent
coupling to the qubit~\cite{WoldBroxGalperinBergli2012,SairaBergholmOjanenMottonen2007}, consistent with
classical--quantum correspondence in open quantum systems~\cite{GuFranco2019}. Motivated by the glassy nature of amorphous TLS ensembles, it imposes no equilibrium or Kubo--Martin--Schwinger (KMS) conditions~\cite{BreuerPetruccione2007} on the transverse non-Markovian environment.

\subsection{Open-system dynamics}
At the microscopic level, the joint density operator $\rho_{\mathrm{joint}}(t)$ of the qubit, longitudinal bosonic
bath, and TLS ensemble obeys 
\begin{equation}
\frac{d}{dt}\rho_{\mathrm{joint}}(t)
=-i\left[H,\rho_{\mathrm{joint}}(t)\right],
\label{eq:microscopic_LvN}
\end{equation}
with $\hbar=1$. This autonomous quantum description motivates the longitudinal coupling to the bosonic bath and the transverse coupling to the TLS ensemble.

The effective model omits the explicit TLS Hilbert space and represents its slow configurational dynamics by a classical renewal history $\omega=\{t_1,t_2,\ldots\}$ of coarse-grained ensemble reconfigurations. For fixed $\omega$, let $\varrho_\omega(t)$ denote the density operator of the qubit and longitudinal bath. It evolves under the longitudinal system--bath Hamiltonian between events and at each renewal time $t_n$, obeys
\begin{equation}
\varrho_\omega(t_n^+)
=
\left(\Phi_X\otimes\mathrm{id}_{B_z}\right)
\!\left[\varrho_\omega(t_n^-)\right],
\label{eq:conditioned_kick}
\end{equation}
where $\mathrm{id}_{B_z}$ is the identity superoperator on the longitudinal bath's operator space. The kick acts only on the qubit, leaving the longitudinal bath unchanged at that instant.

The conditioned qubit state is
\begin{equation}
\rho_\omega(t)
=
\operatorname{Tr}_{B_z}
\!\left[\varrho_\omega(t)\right],
\label{eq:conditioned_reduced_state}
\end{equation}
and averaging over renewal histories gives the physical reduced state,
\begin{equation}
\rho(t)
=
\mathbb E_\omega\!\left[\rho_\omega(t)\right]
=
\mathbb E_\omega
\!\left[\operatorname{Tr}_{B_z}\varrho_\omega(t)\right].
\label{eq:renewal_averaged_reduced_state}
\end{equation}
Equation~\eqref{eq:renewal_averaged_reduced_state} combines the quantum trace over the longitudinal bath with a classical average over TLS renewal histories. The latter, together with the transverse CPTP map $\Phi_X$, represents the TLS ensemble as an effective nonequilibrium, nonthermal environment. This separation is consistent with quantum-renewal constructions~\cite{Budini2004,Budini2005,Vacchini2020}.

Within the Born--Markov treatment, we assume initially uncorrelated qubit and longitudinal-bath
states~\cite{BreuerPetruccione2007}, together with statistical independence of the classical TLS renewal history from both initial states. Thus, for every $\omega$,
\begin{equation}
\varrho_\omega(0)
=
\rho(0)\otimes\rho_{B_z}.
\label{eq:initial_factorized_state}
\end{equation}
Here $\rho_{B_z}$ is the initial longitudinal-bath state, and $\rho(0)$, written in Eq.~\eqref{eq:initial_bloch_state}, is an arbitrary initial qubit density operator. The renewal-averaged evolution consequently defines a dynamical map $\Lambda(t)$ on the full qubit state space, without restricting the preparation to a particular pure state.

We write the initial qubit state in Bloch form,
\begin{equation}
\rho(0)=\frac{1}{2}
\left[
I+r_x(0)\sigma_x+r_y(0)\sigma_y+r_z(0)\sigma_z
\right].
\label{eq:initial_bloch_state}
\end{equation}
The preparation
$|+x\rangle\langle+x|=\frac12(I+\sigma_x)$
serves only as a control trajectory: its identity component lies in $W_0$ and its traceless component in the jump-invariant subspace $W_x=\mathrm{span}\{\sigma_x\}$, with no component in $W_{yz}$. Initial states with nonzero $r_y(0)$ or $r_z(0)$ overlap with $W_{yz}$ and can exhibit the storage--return mechanism analyzed in Sec.~\ref{sec:dynamics_decoherence}.

\subsection{Physical motivation for fractional renewal statistics}
\label{subsec:Physmot_fract}
We model coarse-grained TLS ensemble reconfigurations as a renewal process with statistically independent and identically distributed inter-event intervals and common waiting-time density $w(t)$~\cite{feller1991introduction,Cox1962}. The form and parameters of $w(t)$ encode heterogeneity and interaction effects phenomenologically; correlations between successive intervals are excluded. Section~\ref{sec:GTSL} summarizes the experimental motivation.

Unless stated otherwise, $t=0$ is a renewal epoch, with the same density $w$ for the first and subsequent intervals. This specifies the effective clock's preparation; preparing the qubit alone need not reset the physical TLS environment. For $0<\alpha<1$, the normalized Prabhakar law has an infinite mean waiting time. Observing a previously prepared clock requires its forward-recurrence distribution, which depends on the elapsed time since clock preparation. Such renewal aging is compatible with independent successive
intervals~\cite{GodrecheLuck2001,BarkaiCheng2003}.

We denote a generic low-frequency noise spectrum by $1/f^\eta$~\cite{Weissman1988,PaladinoGalperinFalciAltshuler2014} and reserve $\alpha$
for the Prabhakar waiting-time exponent, without identifying the two exponents.

The HN/Prabhakar phenomenology in Sec.~\ref{sec:NDRHN} motivates the common inter-event density ansatz
\begin{equation}
w(t)=C\,t^{\beta-1}
E_{\alpha,\beta}^{\gamma}\!\left(-\lambda t^\alpha\right),
\qquad
C>0,\quad \lambda>0,
\label{eq:11a}
\end{equation}
where
\begin{equation}
E_{\alpha,\beta}^{\gamma}(z)
=
\sum_{n=0}^{\infty}
\frac{(\gamma)_n z^n}
{n!\,\Gamma(\alpha n+\beta)}
\end{equation}
is the three-parameter Mittag--Leffler, or Prabhakar, function and $(\gamma)_n$ is the rising Pochhammer
symbol~\cite{prabhakar1971singular,GarraGarrappa2018,MainardiGarrappa2015}.

For the waiting-time ansatz, the parameters $\alpha$, $\beta$ and $\gamma$ are positive. Whenever differentiation produces a nonpositive second parameter, we interpret the same series using the entire reciprocal-gamma function $1/\Gamma(\alpha n+\beta)$; terms at gamma-function poles have zero reciprocal-gamma coefficient. This convention defines the functions in Eq.~\eqref{eq:memKt} without extending the waiting-time density's admissible parameters. Section~\ref{subsec:ASLS} derives the normalization conditions
for Eq.~\eqref{eq:11a} and distinguishes probabilistic admissibility from the stronger requirement that $w(t)$ be CM.

Let $N(t)$ count renewals up to time $t$ and $M(t)=\mathbb E[N(t)]$ be the mean renewal function. When $M$ is absolutely continuous,
\begin{subequations}
\label{eq:renewal_statistics}
\begin{equation}
\begin{aligned}
m(t)
&:=\frac{dM(t)}{dt}
=\sum_{n\geq1}w^{\ast n}(t),
\end{aligned}
\label{eq:m_t}
\end{equation}
is the renewal density or ensemble-averaged event-rate density, not a normalized probability density.
Here $n$ counts renewal events and $w^{\ast n}(t)$ is the probability density of the $n$th renewal
time where $\ast$ denotes the convolution operator~\cite{feller1991introduction,Cox1962,FerraroManziniMasoeroScalas2009,GodrecheLuck2001}.
Laplace transformation and the convolution theorem give
\begin{equation}
\begin{aligned}
\widetilde m(s)
&=\frac{\widetilde w(s)}{1-\widetilde w(s)},
\qquad
\lvert\tilde w(s)\rvert<1,\quad \Re s>0.
\end{aligned}
\label{eq:m_s}
\end{equation}
\end{subequations}

The causal renewal-resolvent kernel is defined by
\begin{equation}
\widetilde K(s)
=s\widetilde m(s)
=\frac{s\widetilde w(s)}{1-\widetilde w(s)}
=s\sum_{n=1}^{\infty}\widetilde w(s)^n,
\qquad \Re s>0.
\label{eq:30a}
\end{equation}
It is determined by the one-interval renewal law and must be distinguished from a two-time bath correlation function which generally depends on both observation times in an aging, nonstationary environment. For the general time-domain interpretation, let $H$ be the Heaviside function, equal to one for $t>0$ and zero
for $t\leq0$ (see Eq.~(1.16.13) of Ref.~\cite{OlverEtAl2010}). The notation $Hm$ denotes the renewal density in Eq.~\eqref{eq:m_t} extended by zero to nonpositive times; this extension does not require evaluating $m(0^+)$. We define the causal kernel by $K:=D[Hm]$, where $D$ is the distributional derivative defined in Eq.~(1.16.8) of Ref.~\cite{OlverEtAl2010}. The local integrability established in Appendix~\ref{app:prabhakar_forward_series} makes $Hm$ a regular distribution in the sense of Eq.~(1.16.3) of the same reference, so this definition remains
meaningful when $m(0^+)$ diverges. By Theorem~14.3 of Ref.~\cite{Doetsch1974}, distributional differentiation corresponds to multiplication by $s$ in the Laplace domain. Applied to $Hm$, whose transform is given in Eq.~\eqref{eq:m_s}, this rule yields precisely Eq.~\eqref{eq:30a}. If $m$ extends to an absolutely continuous function on every finite interval $[0,T]$, with finite $m(0^+)$, integration by parts gives the first-order causal derivative formula (see Sec.~14, Eq.~(6), of Ref.~\cite{Doetsch1974}),
\begin{equation}
K(t)=\dot m(t)+m(0^+)\delta(t).
\label{eq:30b}
\end{equation}
Here $\dot m$ denotes the almost-everywhere ordinary derivative on $t>0$, extended by zero to nonpositive times and $\delta$ is the Dirac distribution (see Eq.~(1.16.10) of Ref.~\cite{OlverEtAl2010}).

The ordinary positive-time part of $K$, when it exists, need not be nonnegative. Nonexponential waiting times
produce a temporally nonlocal renewal contribution, whereas exponential waiting times recover the local
Poisson generator in Eq.~\eqref{eq:D_poisson_master_check}. The Prabhakar kernel series is introduced in
Eq.~\eqref{eq:memKt}.
\subsection{\label{subsec:ASLS}{Prabhakar waiting--time law, normalization, and admissibility}}
The non-Debye phenomenology in Sec.~\ref{sec:NDRHN} motivates the renewal ansatz. Its implications for fractional evolution and transmon dynamics follow from the quantum-renewal construction in Sec.~\ref{sec:gen_master_qrenewal} and the analysis in Sec.~\ref{sec:dynamics_decoherence}. Here we normalize the ansatz, distinguish probabilistic admissibility from CM, and derive its asymptotics.

In Eq.~\eqref{eq:11a}, take $0<\alpha\leq1$ and $\beta,\gamma,\lambda>0$, with $\beta$ an independent shape parameter before normalization. For a nonnegative normalized waiting-time density, the no-renewal probability is
\begin{equation}
\Psi(t)=1-\int_0^t w(\tau)\,d\tau.
\label{eq:11psi}
\end{equation}
The defining Prabhakar series gives
\begin{equation}
w(t)\sim\frac{C}{\Gamma(\beta)}\,t^{\beta-1},
\qquad t\to0^+.
\end{equation}
After normalization, $\beta$ controls the weight of short intervals between successive coarse-grained TLS ensemble reconfigurations. For $0<\beta<1$, the divergence is integrable: the probability of an interval shorter than $\varepsilon$ vanishes as $\varepsilon\to0^+$, so there is no probability mass at zero waiting time.

The standard Prabhakar transform formula~\cite{prabhakar1971singular,GarraGarrappa2018} yields
\begin{equation}
\begin{aligned}
\widetilde w(s)
&=\mathcal L\!\left\{
C\,t^{\beta-1}E_{\alpha,\beta}^{\gamma}
\!\left(-\lambda t^\alpha\right)\right\}\\
&=\frac{C\,s^{\alpha\gamma-\beta}}
{(s^\alpha+\lambda)^\gamma},
\qquad \Re s>0.
\label{eq:11b}
\end{aligned}
\end{equation}
For a nonnegative density, normalization requires
$\lim_{s\to0^+}\widetilde w(s)=1$.
A finite nonzero limit in Eq.~\eqref{eq:11b} fixes
\begin{equation}
\beta=\alpha\gamma,
\label{eq:11c}
\end{equation}
and unit normalization gives
\begin{equation}
C=\lambda^\gamma.
\label{eq:11d}
\end{equation}
Normalization leaves two independent shape parameters, $\alpha$ and $\gamma$, which jointly determine the short-time exponent and one scale parameter $\lambda$. With units $T^{-\alpha}$, $\lambda$ sets the characteristic time $\tau_\star=\lambda^{-1/\alpha}$ but does not by itself represent a microscopic TLS switching rate.

\begin{figure*}[t]
\centering
\includegraphics[width=\textwidth]{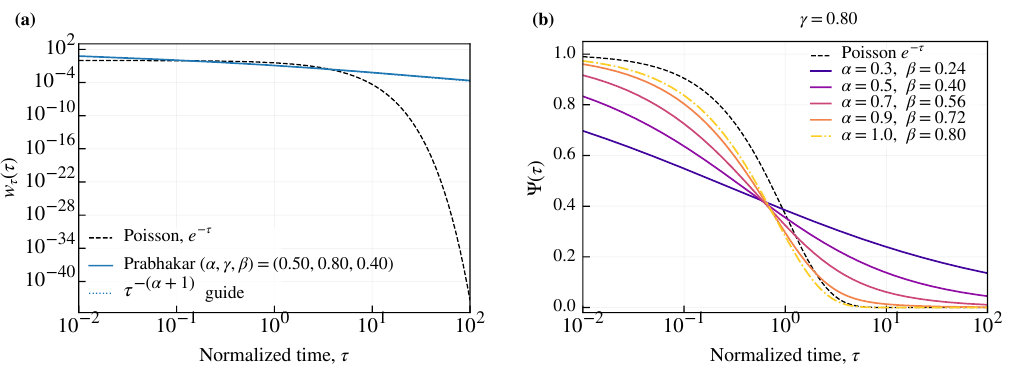}
\caption{Normalized Prabhakar renewal statistics for $\gamma=0.8$, $\beta=\alpha\gamma$ and dimensionless time $\tau=\lambda^{1/\alpha}t$. (a) Dimensionless waiting-time density $w_\tau(\tau)=\lambda^{-1/\alpha} w(\tau/\lambda^{1/\alpha})$ at $\alpha=0.5$, compared with the Poisson density $e^{-\tau}$ and the algebraic guide $\tau^{-1-\alpha}$. (b) Survival probability for $\alpha\in\{0.3,0.5,0.7,0.9,1.0\}$, compared with the Poisson survival $e^{-\tau}$. For $0<\alpha<1$, the algebraic survival tail in Eq.~\eqref{eq:asymp_Psi} decays more slowly than the Poisson survival. The $\alpha=1$ curve corresponds to gamma-distributed waiting times. ChatGPT (OpenAI; GPT-5 Sol Ultra) generated the Python evaluation and plotting code from R.U.Erdogan's theoretical and numerical specifications; R.U.Erdogan reviewed it against the analytical expressions and numerical methods.}
\label{fig:sim9-prabhakar-renewal-clock}
\end{figure*}

The construction in Sec.~\ref{subsec:gamma_stable_subord}, based on evaluating a one-sided $\alpha$-stable L\'evy subordinator at an independent gamma-distributed argument, yields a nonnegative, normalized waiting-time
law for all $0<\alpha\leq1$, $\gamma>0$, and $\lambda>0$. We select the more restricted class of CM densities. For the general ansatz in Eq.~\eqref{eq:11a}, the CM conditions are~\cite{MainardiGarrappa2015}
\begin{equation}
0<\alpha\leq1,\qquad 0<\beta\leq1,\qquad
0<\alpha\gamma\leq\beta.
\label{eq:11f}
\end{equation}
On the normalized slice of the domain in Eq.~\eqref{eq:11f}, $w(t)$ and $\Psi(t)$ are CM and their Laplace transforms $\widetilde w(s)$ and $\widetilde\Psi(s)$ are Stieltjes functions~\cite{Schilling2010}. Appendix~\ref{app:prabhakar_admissibility} discriminates scalar admissibility from CP of the reduced quantum dynamics.

Under Eqs.~\eqref{eq:11c} and~\eqref{eq:11d}, the causal kernel in Eq.~\eqref{eq:30a} has the positive-time representative
\begin{equation}
K(t)=\sum_{n=1}^{\infty}\lambda^{\gamma n}
\Bigl[t^{n\beta-2}
E_{\alpha,n\beta-1}^{n\gamma}(-\lambda t^\alpha)\Bigr],
\qquad t>0.
\label{eq:memKt}
\end{equation}
By Eqs.~\eqref{eq:D_differentiated_series_tail} and~\eqref{eq:D_ordinary_differentiated_series} of
Appendix~\ref{app:prabhakar_forward_series}, the series converges absolutely and uniformly on every compact
interval $[\varepsilon,T]\subset(0,\infty)$ and equals $\dot m(t)$ there. It represents the restriction to $t>0$
of the causal kernel defined in Sec.~\ref{subsec:Physmot_fract}; its behavior at the origin is understood distributionally through Eq.~\eqref{eq:30a}.

\begin{figure}
\centering
\includegraphics[width=\columnwidth]{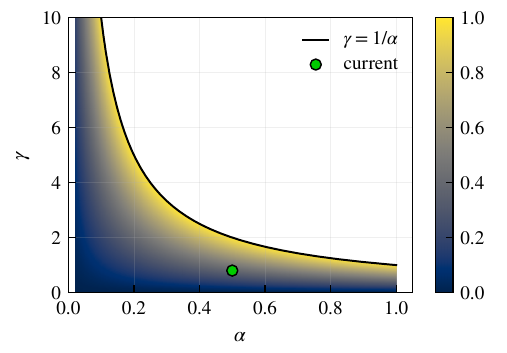}
\caption{Complete-monotonicity (CM) domain of the normalized Prabhakar waiting-time density for $\lambda>0$ [Eq.~\eqref{eq:11ee}].
The color field shows $\beta=\alpha\gamma$ within $0<\alpha\leq1$, $0<\gamma\leq1/\alpha$. The black boundary $\gamma=1/\alpha$ ($\beta=1$) is included.
The uncolored region above it represents nonnegative normalized densities that are not CM. The green marker denotes the default simulation parameters
$(\alpha,\gamma,\beta)=(0.5,0.8,0.4)$. ChatGPT (OpenAI; GPT-5 Sol Ultra) generated the Python plotting code from R.U.Erdogan's theoretical and numerical specifications; R.U.Erdogan reviewed it against the analytical CM conditions.}
\label{fig:sim13-cm-normalized-domain}
\end{figure}

The normalized transform and selected CM range are, respectively,
\begin{subequations}
\begin{align}
&\widetilde w(s)=\left(1+\frac{s^\alpha}{\lambda}\right)^{-\gamma},
\label{eq:11e}\\
&0<\alpha\leq1,\qquad 0<\gamma\leq\frac1\alpha.
\label{eq:11ee}
\end{align}
\end{subequations}
Equation~\eqref{eq:11e} is the generalized Mittag--Leffler waiting-time transform studied by Cahoy and
Polito~\cite{CahoyPolito2013}, with $\nu=\alpha$ and $\delta=\gamma$ in their notation; Eq.~\eqref{eq:11ee} specifies its CM subclass.
At $\alpha=1$, the law is gamma distributed: it remains a valid waiting-time density for $\gamma>1$ but is then not CM.
Figures~\ref{fig:sim9-prabhakar-renewal-clock} and~\ref{fig:sim13-cm-normalized-domain} illustrate the renewal statistics and selected CM domain, respectively.

Within that domain, Bernstein's theorem~\cite{Widder2010,Schilling2010} gives a unique positive probability measure $\mu_{\mathrm{eff}}$ on
$[0,\infty)$ such that
\begin{equation}
\Psi(t)=\int_{[0,\infty)}e^{-rt}\,\mu_{\mathrm{eff}}(dr),
\label{eq:survival_Bernstein_measure}
\end{equation}
with $\mu_{\mathrm{eff}}([0,\infty))=1$. In statistical-mechanical terms, this represents the coarse-grained no-renewal probability as a positive mixture
of exponential survival modes with heterogeneous effective rates, without specifying a microscopic TLS realization.

The measure in Eq.~\eqref{eq:survival_Bernstein_measure} defines an effective rate variable $R\sim\mu_{\mathrm{eff}}$, with units of inverse time;
Sec.~\ref{subsec:gamma_stable_subord} distinguishes it from the gamma mixing variable $U$.

For $0<\alpha<1$, the normalized transform has the expansion 
\begin{equation}
\widetilde w(s)=1-\frac\gamma\lambda s^\alpha
+O(s^{2\alpha}),\qquad s\to0^+.
\label{eq:prabhakar_small_s}
\end{equation}
Together with
\begin{equation}
\widetilde\Psi(s)=\frac{1-\widetilde w(s)}s,
\label{eq:Psi_laplace_s}
\end{equation}
the Tauberian theorem for the survival probability~\cite{Widder2010} and the Prabhakar density asymptotics~\cite{MainardiGarrappa2015} give
\begin{subequations}
\begin{alignat}{3}
\Psi(t)&\;\sim\;&
\frac\gamma{\lambda\Gamma(1-\alpha)}t^{-\alpha},
\qquad &t\to\infty,
\label{eq:asymp_Psi}\\
w(t)&\;\sim\;&
\frac\gamma{\lambda|\Gamma(-\alpha)|}t^{-1-\alpha},
\qquad &t\to\infty.
\label{eq:asymp_wt}
\end{alignat}
\end{subequations}
For $0<\alpha<1$, $\alpha$ determines the decay orders in Eqs.~\eqref{eq:asymp_Psi} and~\eqref{eq:asymp_wt}, while
$\gamma/\lambda$ scales their leading amplitudes at fixed $\alpha$. The survival tail implies an infinite mean waiting time,
placing the law in the aging regime of Sec.~\ref{subsec:Physmot_fract}. Transmon dynamics depends on the reduced quantum resolvent
and the initial state (Sec.~\ref{sec:Prabhakar_asymptotics}); its implications for decoherence reshaping are discussed in
Sec.~\ref{sec:physical_interpretation}, which distinguishes directional storage from quantum-state preservation.

At fixed $\alpha$ and $\lambda$, Eq.~\eqref{eq:11e} identifies the waiting-time density for a positive integer
$\gamma=n$ as the $n$-fold convolution of the $\gamma=1$ density. This statistical compounding interpretation does not identify $\gamma$ with a microscopic TLS cluster size. Section~\ref{subsec:gamma_stable_subord} gives the
gamma-mixture representation for general $\gamma>0$~\cite{CahoyPolito2013}.

\subsection{\label{subsec:gamma_stable_subord}{Gamma-mixture representation of the Prabhakar waiting-time law
using a one-sided $\alpha$-stable L\'evy subordinator}}

Let $U$ be gamma distributed with shape $\gamma>0$ and rate $\lambda>0$, independent of the one-sided
$\alpha$-stable subordinator $D_\alpha$~\cite{Applebaum2013}. Its density is
\begin{equation}
h_{\lambda,\gamma}(u)
=
\frac{\lambda^\gamma}{\Gamma(\gamma)}
u^{\gamma-1}e^{-\lambda u},
\qquad u>0 .
\label{eq:gamma_directing_density}
\end{equation}
Here $u$ is the random subordinator argument, with $[u]=T^\alpha$ and $[\lambda]=T^{-\alpha}$ as required
by the one-sided stable Laplace identity and Prabhakar transform~\cite{PensonGorska2010,GarraGarrappa2018}.
It is therefore neither an effective Debye relaxation rate nor an individual TLS switching rate.

For $u>0$ and $0<\alpha<1$, let $\ell_\alpha(t\mid u)$ be the probability density of $D_\alpha(u)$ with respect
to physical waiting time $t$. At $\alpha=1$, the conditional law is the Dirac probability measure $\delta_u(dt)$ at $t=u$, written symbolically as
$\ell_1(t\mid u)=\delta(t-u)$. The transform and mixture integrals below use this measure interpretation at $\alpha=1$.
The conditional Laplace transform is
\begin{equation}
\int_0^\infty e^{-st}\ell_\alpha(t\mid u)\,dt
=
e^{-u s^\alpha},
\qquad 0<\alpha\le1.
\label{eq:stable_subordinator_density}
\end{equation}
For the physical renewal waiting time
\begin{equation}
T:=D_\alpha(U),\label{eq:physical_time}
\end{equation}
averaging the conditional law over $U$ gives
\begin{equation}
\begin{aligned}
w(t)
&=
\int_0^\infty
h_{\lambda,\gamma}(u)
\ell_\alpha(t\mid u)\,du\\
&=
\lambda^\gamma t^{\alpha\gamma-1}
E_{\alpha,\alpha\gamma}^{\gamma}(-\lambda t^\alpha).
\label{eq:gamma_stable_mixture}
\end{aligned}
\end{equation}
Consequently,
\begin{align}
\widetilde w(s)
&=
\int_0^\infty h_{\lambda,\gamma}(u)e^{-u s^\alpha}\,du
\nonumber\\
&=
\left(\frac{\lambda}{\lambda+s^\alpha}\right)^\gamma
=
\left(1+\frac{s^\alpha}{\lambda}\right)^{-\gamma},
\label{eq:prabhakar_gamma_stable_representation}
\end{align}
which recovers Eq.~\eqref{eq:11e}. Averaging normalized conditional laws against a normalized gamma density in Eq.~\eqref{eq:gamma_stable_mixture}
establishes nonnegativity and normalization for $0<\alpha\leq1$, $\gamma>0$, and $\lambda>0$. At $\alpha=1$, $T=U$, so $w(t)$ is the gamma density
in Eq.~\eqref{eq:gamma_directing_density} with $u=t$. The Poisson waiting-time law is recovered at $\alpha=\gamma=1$.

The self-similarity relation $D_\alpha(u)\overset{d}{=}u^{1/\alpha}D_\alpha(1)$ shows that $U$ randomizes the conditional waiting-time
scale~\cite{SamorodnitskyTaqqu1994}, providing a statistical representation of heterogeneous reconfiguration durations
in the effective TLS description.

The gamma mixing density constructs $w$ by averaging over the subordinator argument in
Eq.~\eqref{eq:gamma_stable_mixture}. Within the CM domain of Eq.~\eqref{eq:11ee}, the Bernstein
measure $\mu_{\mathrm{eff}}$ in Eq.~\eqref{eq:survival_Bernstein_measure} describes the exponential modes of the resulting no-renewal probability
and is determined after forming $\Psi$ from $w$ through Eq.~\eqref{eq:11psi}. Thus the gamma mixing variable $U$ and the effective
survival rate $R$ enter at different stages of the construction.

Section~\ref{sec:gen_master_qrenewal} uses the waiting-time density $w$ and survival probability $\Psi$ as the event and no-event factors, respectively, in chronologically ordered renewal histories to
construct the CPTP reduced evolution.

\section{Generalized master equation based on quantum renewal theory}\label{sec:gen_master_qrenewal}

Using the coarse-grained Prabhakar renewal model and state convention of Sec.~\ref{sec:PMRD}, we construct the reduced transmon dynamics.
\subsection{Drift and renewal jump maps}
\label{subsec:drift_jump}

The construction uses the longitudinal Markovian dephasing generator $\mathcal L_Z$, renewal waiting-time density $w(t)$, survival probability $\Psi(t)$, and
transverse CPTP map $\Phi_X$~\cite{Budini2004,Budini2005,Vacchini2020}. The drift semigroup is
\begin{equation}
\mathcal D_t
:=
e^{t\mathcal L_Z},
\label{eq:drift_semigroup}
\end{equation}
where $\mathcal L_Z$ is defined in
Eq.~\eqref{eq:LZ_generator}.

For a fixed renewal history
$\omega=\{t_1,t_2,\ldots\}$, the conditioned qubit state
$\rho_\omega(t)$ of
Eq.~\eqref{eq:conditioned_reduced_state}
evolves between events as
\begin{equation}
\frac{d}{dt}\rho_\omega(t)
=
\mathcal L_Z[\rho_\omega(t)],
\qquad
t\neq t_n,
\label{eq:conditioned_markovian_drift}
\end{equation}
and at each event obeys
\begin{equation}
\rho_\omega(t_n^+)
=
\Phi_X[\rho_\omega(t_n^-)].
\label{eq:conditioned_kick_qubit}
\end{equation}
Averaging these conditioned trajectories according to
Eq.~\eqref{eq:renewal_averaged_reduced_state} gives the
physical reduced state.

A renewal event represents a coarse-grained TLS ensemble
rearrangement that produces a statistically significant
change in the qubit's transverse environment, rather
than an individual microscopic TLS flip.

\subsection{Renewal map and renewal equation}\label{subsec:ren_map_equ}

The reduced state is
\begin{equation}
\rho(t)
=
\Lambda_{\mathrm{fwd}}(t)\,\rho(0),
\label{eq:renmap}
\end{equation}
where the forward map sums all chronological renewal
histories~\cite{Budini2004,Budini2005,Vacchini2020},
\begin{equation}
\begin{aligned}
\Lambda_{\mathrm{fwd}}(t)
&=
\Psi(t)\mathcal D_t
\\
&+\sum_{n=1}^{\infty}
\int_{0<t_1<\cdots<t_n<t}
\left[\prod_{k=1}^{n}w(t_k-t_{k-1})\right]
\\
&\times\Psi(t-t_n)
\mathcal D_{t-t_n}\Phi_X
\mathcal D_{t_n-t_{n-1}}\Phi_X
\cdots
\Phi_X\mathcal D_{t_1}\,
\\
&\times dt_1\cdots dt_n .
\label{eq:renmapdef}
\end{aligned}
\end{equation}
Here $\mathcal D_t$ is defined in
Eq.~\eqref{eq:drift_semigroup}.

Grouping these histories by the first waiting interval $\tau$ gives the map-level Volterra equation
\begin{equation}
\Lambda_{\mathrm{fwd}}(t)
=
\Psi(t)\mathcal D_t
+
\int_0^t
w(\tau)
\Lambda_{\mathrm{fwd}}(t-\tau)
\Phi_X\mathcal D_\tau\,d\tau .
\label{eq:23}
\end{equation}
The first term describes uninterrupted drift over $[0,t]$, weighted by the survival probability
$\Psi(t)$ of Eq.~\eqref{eq:11psi}. In the integral, drift $\mathcal D_\tau$ is followed
by the CPTP event map $\Phi_X$ of Eq.~\eqref{eq:PhiX_def} and an independent renewal
history of duration $t-\tau$.

For $[\mathcal L_Z,\Phi_X]\neq0$, Eq.~\eqref{eq:23} cannot in general be reduced to a
closed convolution involving only $\rho(t-\tau)$ without changing the chronological operator order.

Each chronological composition of drift and event maps is CPTP. The history weights are nonnegative
and sum, after integration over event times, to unity, so their average $\Lambda_{\mathrm{fwd}}(t)$ is CPTP.
This construction does not require pointwise positivity of the time-domain memory kernel in
Eq.~\eqref{eq:memKt}~\cite{Budini2004,Budini2005,Vacchini2020}. For the present model, Eq.~\eqref{eq:23} is our primary map-level renewal representation. Its equivalent homogeneous memory equation follows from the ordered Laplace-domain resolvent in Sec.~\ref{sec:kernel_resolvent_formulation} and
Appendix~\ref{app:renewal_structure}.

For a minimal transverse coupling, each coarse-grained renewal event is represented by a unitary rotation about
the $x$ axis,
\begin{equation}
\Phi_X[\rho]
=
U_x(\theta)\,\rho\,U_x^\dagger(\theta),
\qquad
U_x(\theta)
=
\exp\!\left(-\frac{i\theta}{2}\sigma_x\right).
\label{eq:PhiX_def}
\end{equation}
We assume the same fixed angle $\theta\in[-\pi,\pi]$, including its sign, and effective transverse orientation
at every event; randomness enters only through the renewal times. The microscopic Hamiltonian identifies
the coupling subspaces, while this common signed rotation is an additional mesoscopic assumption whose realization
by a native TLS ensemble remains to be justified. The unitary map is CPTP, introduces no independent
dissipative Lindblad term and models mixing between coherence and longitudinal polarization through the
angle and its sign.

Equations~\eqref{eq:LZ_generator} and~\eqref{eq:PhiX_def} give
\begin{equation}
[\mathcal{L}_Z,\Phi_X]\neq0,
\qquad
\Gamma_Z>0,
\quad
\sin\theta\neq0.
\label{eq:nonCommute}
\end{equation}
This noncommutativity underlies the nontrivial spectral structure analyzed below.
The Pauli-basis realization of $\Phi_X$ and the common invariant decomposition are given in
Secs.~\ref{subsec:Pauli_realization} and~\ref{sec:MCIS_qubit}, respectively.

\subsection{Exact noncommuting kernel--resolvent formulation}
\label{sec:kernel_resolvent_formulation}
Equation~\eqref{eq:23} is formulated in the reduced interaction picture associated with the reference
Hamiltonian in Eq.~\eqref{eq:H0}. The explicit coherent commutator is removed, leaving the longitudinal Markovian
generator $\mathcal L_Z$~\cite{BreuerPetruccione2007}. The map $\Phi_X$ is assigned after coarse-graining in
this picture within the quantum-renewal construction~\cite{Budini2004,Vacchini2013,Vacchini2020}
with clock statistics and preparation specified in Sec.~\ref{subsec:Physmot_fract}. It is not identified
with the literal microscopic interaction-picture transform of a fixed Schr\"odinger-picture transverse
unitary.

Define the shifted operator-valued Laplace transform
\begin{equation}
\widetilde{f}(s-\mathcal{L}_Z)
:=
\int_0^\infty e^{-st}e^{t\mathcal{L}_Z}f(t)\,dt,
\label{eq:operator_laplace_shift}
\end{equation}
where $s$ is understood as a scalar multiple of the
identity superoperator. In a right half-plane where
the Laplace transforms converge, set
\begin{equation}
\mathcal A(s):=s\mathbb I-\mathcal L_Z,
\quad
F(s):=\widetilde w(\mathcal A(s)),
\quad
G(s):=\widetilde\Psi(\mathcal A(s)).
\label{eq:operator_LT}
\end{equation}
The scalar functions act on $\mathcal A(s)$ through analytic 
functional calculus~\cite{Higham2008}; the operators
in Eq.~\eqref{eq:operator_LT} are assumed invertible
in the right half-plane considered.

Laplace transforming Eq.~\eqref{eq:23} gives
\begin{equation}
\widetilde\Lambda_{\mathrm{fwd}}
=
G(s)
+
\widetilde\Lambda_{\mathrm{fwd}}\Phi_XF(s).
\label{eq:LT_renewal}
\end{equation}
Factoring on the right yields
\begin{equation}
G(s)
=
\widetilde\Lambda_{\mathrm{fwd}}(s)
\left(
\mathbb I-\Phi_XF(s)
\right),
\end{equation}
and hence
\begin{equation}
\widetilde\Lambda_{\mathrm{fwd}}(s)
=
G(s)
\left(
\mathbb I-\Phi_XF(s)
\right)^{-1}.
\end{equation}

The survival identity~\eqref{eq:Psi_laplace_s} and
renewal-resolvent definition~\eqref{eq:30a} give
\begin{subequations}
\begin{align}
G(s)
&=
(\mathbb I-F(s))\mathcal A(s)^{-1},
\label{eq:Gs}
\\
K(s)
&:=
\widetilde K(\mathcal A(s))
=
\mathcal A(s)F(s)(\mathbb I-F(s))^{-1}.
\label{eq:Ks}
\end{align}
\end{subequations}
Since $\mathcal A(s)$ and $F(s)$ commute, whereas $\Phi_X$
need not commute with either,
\begin{equation}
\begin{aligned}
\widetilde\Lambda_{fwd}(s)^{-1}
&=
\left[
\mathbb I-\Phi_XF(s)
\right]G(s)^{-1}
\\
&=
\mathcal A(s)+K(s)-\Phi_XK(s)
\\
&=
\mathcal A(s)-(\Phi_X-\mathbb I)K(s).
\end{aligned}
\end{equation}
Consequently,
\begin{equation}
\widetilde\rho(s)
=
\left[
\mathcal A(s)
-
(\Phi_X-\mathbb I)
\widetilde K(\mathcal A(s))
\right]^{-1}
\rho(0).
\label{eq:exact_kernel_resolvent}
\end{equation}

The forward-ordered Laplace-domain inverse propagator is the noncommuting operator pencil
\begin{equation}
\mathcal M_{\mathrm{fwd}}(s)
:=
\mathcal A(s)-(\Phi_X-\mathbb I)\widetilde K(\mathcal A(s)).
\label{eq:M_def}
\end{equation}
The chronological histories in Eq.~\eqref{eq:renmapdef}
place the drift-dressed kernel to the right of
$\Phi_X-\mathbb I$.
The alternative ordering
$\widetilde K(s-\mathcal L_Z)(\Phi_X-\mathbb I)$
corresponds to a different quantum-renewal construction,
inequivalent when
$[\mathcal L_Z,\Phi_X]\neq0$~\cite{Vacchini2020}.

Equation~\eqref{eq:exact_kernel_resolvent} specializes the ordered quantum-renewal propagator of Ref.~\cite{Vacchini2020} to the present model and gives an exact shifted, operator-valued
analogue of the scalar Montroll--Weiss propagator for continuous-time random walks~\cite{MontrollWeiss1965,MetzlerKlafter2000}. Here the kernel $\widetilde K(s-\mathcal L_Z)$ dresses the semigroup generated by $\mathcal L_Z$ with renewal statistics, providing the reduced spectral signature of the broad distribution of configurational rearrangement times associated with the glassy TLS environment~\cite{Budini2004,Vacchini2020,MainardiGarrappa2015}.

\subsection{Operator-space matrix formulation}
\label{sec:exact_matrix_formulation}

For finite-dimensional $\mathcal H$, the operator space
$\mathcal B(\mathcal H)$ is also finite-dimensional and
complete under any norm. The Hilbert--Schmidt (HS)
inner product
\begin{equation}
(A,B)_{\mathrm{HS}}
=
\mathrm{Tr}(A^\dagger B)
\label{eq:HS_inner_product}
\end{equation}
makes it a Hilbert space~\cite{ManentiMotta2023}.

Assuming $\mathcal L_Z$ is diagonalizable on the
relevant operator space, choose right and left
eigenoperators $\{R_n\}$ and $\{L_m\}$ satisfying
\begin{equation}
\begin{aligned}
\mathcal L_Z[R_n]
&=
\lambda_n R_n,
\\
\mathcal L_Z^{\ddagger}[L_m]
&=
\lambda_m^{*}L_m,
\\
\operatorname{Tr}
\left(
L_m^\dagger R_n
\right)
&=
\delta_{mn}.
\end{aligned}
\label{eq:biorthogonality_basis}
\end{equation}
Completeness gives
$X=\sum_n R_n\operatorname{Tr}(L_n^\dagger X)$
for every $X\in\mathcal B(\mathcal H)$.
Thus we write
\begin{equation}
\begin{aligned}
\widetilde\rho(s)
&=
\sum_n c_n(s)R_n,
\\
r_m
&:=
\operatorname{Tr}
\left[
L_m^\dagger\rho(0)
\right].
\end{aligned}
\label{eq:state_expansion}
\end{equation}
Biorthogonality gives
$c_n(s)=\operatorname{Tr}
[L_n^\dagger\widetilde\rho(s)]$.
Projecting the forward resolvent equation of
Eqs.~\eqref{eq:exact_kernel_resolvent}--\eqref{eq:M_def}
with $\operatorname{Tr}[L_m^\dagger(\,\cdot\,)]$ yields
\begin{equation}
\sum_n
\mathcal M_{\mathrm{fwd},mn}(s)c_n(s)
=
r_m,
\label{eq:projected_system}
\end{equation}
where
\begin{equation}
\mathcal M_{\mathrm{fwd},mn}(s)
:=
\operatorname{Tr}
\left\{
L_m^\dagger
\mathcal M_{\mathrm{fwd}}(s)[R_n]
\right\}.
\label{eq:Mmn_def}
\end{equation}

Define the jump-map matrix elements
\begin{equation}
\phi_{mn}
:=
\operatorname{Tr}
\left\{
L_m^\dagger
\Phi_X[R_n]
\right\}.
\label{eq:phi_def}
\end{equation}
Completeness and biorthogonality give
\begin{equation}
\left(
\Phi_X-\mathbb I
\right)[R_n]
=
\sum_{\ell}
\left(
\phi_{\ell n}-\delta_{\ell n}
\right)
R_{\ell},
\label{eq:jump_expansion}
\end{equation}
where $\mathbb I$ is the identity superoperator on
$\mathcal B(\mathcal H)$.
Analytic functional calculus~\cite{Higham2008} gives
\begin{equation}
\widetilde K
\left(
s\mathbb I-\mathcal L_Z
\right)[R_n]
=
\widetilde K
\left(
s-\lambda_n
\right)R_n.
\label{eq:functional_calculus_action}
\end{equation}
Substituting
Eqs.~\eqref{eq:jump_expansion}--\eqref{eq:functional_calculus_action}
into Eq.~\eqref{eq:Mmn_def} yields the exact forward
matrix elements
\begin{equation}
\mathcal M_{\mathrm{fwd},mn}(s)
=
(s-\lambda_n)\delta_{mn}
-
\bigl(
\phi_{mn}-\delta_{mn}
\bigr)
\widetilde K(s-\lambda_n).
\label{eq:Mmn_explicit}
\end{equation}
The kernel acts first on the input (column) mode
$R_n$, whose waiting-interval drift eigenvalue
$\lambda_n$ fixes the shift
$\widetilde K(s-\lambda_n)$.
The jump map then transfers this input into the
output (row) mode $m$ through $\phi_{mn}$.

If $\mathcal L_Z$ and $\Phi_X$ commute, $\Phi_X$
preserves each drift eigenspace, so the jump matrix
is block diagonal with respect to these eigenspaces
and modes with distinct drift eigenvalues remain
uncoupled.
When they do not commute, off-diagonal $\phi_{mn}$
couple modes carrying different kernel shifts,
producing a matrix-valued spectral problem.

We therefore seek a common invariant decomposition
of the operator family
\begin{equation}
\bigl\{
\mathcal{L}_Z,\Phi_X,
\widetilde{K}(s-\mathcal{L}_Z)
\bigr\}.
\label{eq:operator_family}
\end{equation}
Under such a decomposition, the full pencil
$\mathcal M_{\mathrm{fwd}}(s)$ of Eq.~\eqref{eq:M_def}
reduces to the direct sum of its restrictions to the
invariant subspaces~\cite{Kato1995}.
Section~\ref{sec:dynamics_decoherence} gives the
explicit qubit realization.

\subsection{Drift-conjugated Caputo-type representation of the forward renewal memory}
\label{sec:caputo_prabhakar_series}

The kernel--resolvent formulation remains primary.
For comparison with fractional-calculus notation,
Eq.~\eqref{eq:exact_kernel_resolvent} admits an equivalent
time-domain, drift-conjugated Caputo-type form,
separating the memory into a convolution with the
drift-relative derivative and an explicit initial-value
term.
This auxiliary form preserves the forward order; CP
follows from the renewal-history construction in
Sec.~\ref{subsec:ren_map_equ}.

Let $m$ be the renewal density defined in
Eqs.~\eqref{eq:m_t} and~\eqref{eq:m_s}, related to the
memory kernel by Eq.~\eqref{eq:30a}.
For an operator-valued function $F$ that is absolutely
continuous on every finite interval $[0,T]$, define
a drift-conjugated counterpart of the scalar Caputo-type
operator in Eq.~(1.1) of Ref.~\cite{Kochubei2011}:
\begin{equation}
\begin{aligned}
{}^{\mathrm C}\mathcal D_{m,\mathcal L_Z}F(t)
&:=
\int_0^t
m(t-\tau)e^{(t-\tau)\mathcal L_Z}
\\
&\quad\times
\left[
\dot F(\tau)-\mathcal L_ZF(\tau)
\right]d\tau.
\label{eq:Capute_def}
\end{aligned}
\end{equation}
Since $m\in L^1(0,T)$ for every $T>0$, as established
in Appendix~\ref{app:prabhakar_forward_series}, the
assumed regularity of $F$ ensures that this convolution
belongs to $L^1(0,T)$ and is defined almost everywhere
in the finite-dimensional Liouville space, even if $m$
has an integrable singularity at the origin.

Assume that, for some $\sigma>0$, both
$e^{-\sigma t}F(t)$ and
$e^{-\sigma t}[\dot F(t)-\mathcal L_ZF(t)]$
belong to $L^1(0,\infty)$.
For $\operatorname{Re}s>\sigma$, the Laplace convolution
and differentiation rules, together with the semigroup
shift, give~\cite{ArendtEtAl2011}
\begin{equation}
\mathcal L
\left\{
{}^{\mathrm C}\mathcal D_{m,\mathcal L_Z}F
\right\}
=
\widetilde K(\mathcal A(s))\widetilde F(s)
-
\widetilde m(\mathcal A(s))F(0),
\label{eq:laplace_caputoD}
\end{equation}
where $\mathcal A(s)$ is defined in Eq.~\eqref{eq:operator_LT}.
Consequently,
\begin{equation}
\widetilde K(\mathcal A(s))\widetilde F(s)
=
\mathcal L
\left\{
{}^{\mathrm C}\mathcal D_{m,\mathcal L_Z}F
+
e^{t\mathcal L_Z}m(t)F(0)
\right\}.
\label{eq:Ks_Fs}
\end{equation}

Applying this identity to $F=\rho$ gives the exact
forward-order time-domain equation
\begin{equation}
\begin{aligned}
\dot\rho(t)
&=
\mathcal L_Z[\rho(t)]
\\
&\quad+
(\Phi_X-\mathbb I)
\left[
{}^{\mathrm C}\mathcal D_{m,\mathcal L_Z}\rho(t)
+
e^{t\mathcal L_Z}m(t)\rho(0)
\right].
\label{eq:rho_fractional}
\end{aligned}
\end{equation}
The jump-difference map remains outside both the
drift-conjugated operator and its boundary term.
For $[\mathcal L_Z,\Phi_X]\neq0$, the memory contribution
cannot in general be replaced by
$\mathcal D^{\mathrm C}_{m,\mathcal L_Z}
[(\Phi_X-\mathbb I)\rho]$~\cite{Vacchini2020}.

For the Prabhakar waiting-time law of
Sec.~\ref{sec:PMRD},
Appendix~\ref{app:prabhakar_forward_series} gives the
convergent renewal-series representation of the kernel
and its explicit Prabhakar-convolution expansion in
the forward memory term~\cite{CahoyPolito2013}.

\section{Coherence reshaping and storage--return dynamics from fractional renewal-driven transverse noise}
\label{sec:dynamics_decoherence}
We specialize the exact formulation of
Secs.~\ref{sec:kernel_resolvent_formulation}
and~\ref{sec:exact_matrix_formulation} to the transmon model,
obtaining a $1\oplus1\oplus2$ operator-space decomposition of
the noncommuting kernel--resolvent problem. We analyze the
poles, branch cuts, and long-time asymptotics of the
renewal-dependent $(\sigma_y,\sigma_z)$ sector to characterize
state-selective decoherence reshaping and storage--return
dynamics. We compare with Poisson renewal and discuss the
coarse-grained TLS interpretation, distinguishing directional
transfer from quantum-state fidelity.

\subsection{Pauli representation of the drift and renewal kick}
\label{subsec:Pauli_realization}
\label{sec:reduced_qubit_dynamics}

For $\mathcal H=\mathbb C^2$, we evaluate the forward operator
$\mathcal M_{\mathrm{fwd}}(s)$ in Eq.~\eqref{eq:M_def}
using Eq.~\eqref{eq:Mmn_explicit} in the Pauli operator basis
\begin{equation}
\{I,\sigma_x,\sigma_y,\sigma_z\},
\label{eq:Pauli_basis}
\end{equation}
which spans $\mathcal B(\mathbb C^2)$~\cite{ManentiMotta2023}.
In the biorthogonal expansion of
Eqs.~\eqref{eq:biorthogonality_basis}
and~\eqref{eq:state_expansion}, we choose
$R_I=I/2$, $L_I=I$, $R_j=\sigma_j/2$, and $L_j=\sigma_j$
for $j\in\{x,y,z\}$. The coefficients $c_j(t)$ are therefore
the Bloch components $r_j(t)$ plotted in the figures, with
$r_j=c_j(0)=r_j(0)$ as in Eq.~\eqref{eq:initial_bloch_state}.
The longitudinal dephasing generator acts diagonally:
\begin{subequations}
\begin{align}
\mathcal{L}_Z[I]&=0,\label{eq:LZ_I}\\
\mathcal{L}_Z[\sigma_x]&=-\Gamma_Z \sigma_x,
\label{eq:LZ_x}\\
\mathcal{L}_Z[\sigma_y]&=-\Gamma_Z \sigma_y,
\label{eq:LZ_y}\\
\mathcal{L}_Z[\sigma_z]&=0.\label{eq:LZ_z}
\end{align}
\end{subequations}
Thus, the transverse $x$ and $y$ components are damped,
while the $z$ component is immune to longitudinal pure
dephasing. The minimal transverse kick in
Eq.~\eqref{eq:PhiX_def} fixes the $x$ axis and rotates
the $yz$ plane, mixing $y$ and $z$ when $\sin\theta\ne0$:
\begin{subequations}
\begin{align}
\Phi_X[I] &= I, \label{eq:PhiX_I}\\
\Phi_X[\sigma_x] &= \sigma_x, \label{eq:PhiX_x}\\
\Phi_X[\sigma_y] &=
\cos\theta\,\sigma_y+\sin\theta\,\sigma_z,
\label{eq:PhiX_y}\\
\Phi_X[\sigma_z] &=
-\sin\theta\,\sigma_y+\cos\theta\,\sigma_z.
\label{eq:PhiX_z}
\end{align}
\end{subequations}
On the coefficient vector in the ordered basis
$(\sigma_y,\sigma_z)$, this action is
\begin{equation}
\Phi_X^{(yz)}=
\begin{pmatrix}
\cos\theta & -\sin\theta\\
\sin\theta & \cos\theta
\end{pmatrix}.
\label{eq:PhiX_yz_matrix}
\end{equation}
These drift and kick actions determine the exact
operator-space block structure derived below.

\subsection{Minimal common invariant decomposition}
\label{sec:MCIS_qubit}

A subspace $W\subset\mathcal B(\mathbb C^2)$ is common
invariant under $\mathcal L_Z$ and $\Phi_X$ if
$\mathcal L_Z(W)\subset W$ and
$\Phi_X(W)\subset W$~\cite{Kato1995}.
The actions in Eqs.~\eqref{eq:LZ_I}--\eqref{eq:LZ_z}
and~\eqref{eq:PhiX_I}--\eqref{eq:PhiX_z} give the
common invariant subspaces

\begin{subequations}
\begin{align}
W_0&:=\mathrm{span}\{I\},\label{eq:id_subspace}\\
W_x&:=\mathrm{span}\{\sigma_x\},\label{eq:x_subspace}\\
W_{yz}&:=\mathrm{span}\{\sigma_y,\sigma_z\},
\label{eq:yz_subspace}
\end{align}
\end{subequations}
and the decomposition
\begin{equation}
\begin{aligned}
\mathcal B(\mathbb C^2)
&=W_0\oplus W_x\oplus W_{yz}\\
&=\mathrm{span}\{I\}\oplus\mathrm{span}\{\sigma_x\}
  \oplus\mathrm{span}\{\sigma_y,\sigma_z\}.
\end{aligned}
\label{eq:MCIS_decomp}
\end{equation}
\paragraph{Minimality.}
For $\Gamma_Z>0$ and $\sin\theta\ne0$ given by
Eq.~\eqref{eq:nonCommute}, $\mathcal L_Z|_{W_{yz}}$
has distinct eigenvalues $-\Gamma_Z$ and $0$ as expressed by
Eqs.~\eqref{eq:LZ_y} and~\eqref{eq:LZ_z}.
Its only invariant lines are
$\mathrm{span}\{\sigma_y\}$ and
$\mathrm{span}\{\sigma_z\}$, neither preserved by
$\Phi_X$ as evident in Eqs.~\eqref{eq:PhiX_y}
and~\eqref{eq:PhiX_z}.
Thus, the two-dimensional $W_{yz}$ has no proper nonzero
common invariant subspace. At $\Gamma_Z=0$ or
$\sin\theta=0$, it admits further reduction over the
complex operator space.
\paragraph{Closure under resolvent.}
Invariance of $W_{yz}$ under $\mathcal L_Z$ extends to
operator functions defined through functional calculus
on the restricted spectrum, including
$\widetilde K(s-\mathcal L_Z)$~\cite{Higham2008,Kato1995}.
Together with invariance under $\Phi_X$, this gives
\begin{equation}
\mathcal M_{\text{fwd}}(s)(W_{yz})\subset W_{yz}.
\label{eq:closure_M}
\end{equation}
Under Eq.~\eqref{eq:nonCommute},
Eq.~\eqref{eq:MCIS_decomp} therefore yields a minimal
block-diagonal representation of the forward pencil
$\mathcal M_{\mathrm{fwd}}(s)$: $W_0$ is the
one-dimensional identity sector, $W_x$ is jump-invariant,
and $W_{yz}$ is the coupled sector where Markovian
dephasing and renewal-induced rotations compete.

\subsection{Exact block reduction and nonlinear spectral problem}
\label{sec:exact_block_reduction}

Define
\begin{subequations}
\begin{align}
K_0(s)&:=\widetilde{K}(s),\label{eq:K0_def}\\
K_\Gamma(s)&:=\widetilde{K}(s+\Gamma_Z),
\label{eq:KGamma_def}\\
\delta&:=1-\cos\theta.\label{eq:delta_def}
\end{align}
\end{subequations}
Appendix~\ref{subsec:shiftedK_Wyz} explains the kernel
arguments through analytic functional calculus and specifies
their physical-sheet branch structure.

In the ordered Pauli basis $(I,\sigma_x,\sigma_y,\sigma_z)$,
adapted to Eq.~\eqref{eq:MCIS_decomp}, the exact
Laplace-space inverse resolvent is
\begin{equation}
\mathcal M_{\mathrm{fwd}}(s)
=
\begin{pmatrix}
s & 0 & 0 & 0\\
0 & s+\Gamma_Z & 0 & 0\\
0 & 0 &
s+\Gamma_Z+\delta K_\Gamma(s)
&
\sin\theta\,K_0(s)\\
0 & 0 &
-\sin\theta\,K_\Gamma(s)
&
s+\delta K_0(s)
\end{pmatrix}.
\label{eq:M_full_block}
\end{equation}

Equivalently,
\begin{equation}
\mathcal{M}(s)
=
[s]\oplus[s+\Gamma_Z]\oplus\mathcal{M}_{yz}(s),
\label{eq:M_direct_sum}
\end{equation}
where the block on $W_{yz}$ defined by Eq.~\eqref{eq:yz_subspace} is
\begin{equation}
\mathcal M_{\mathrm{fwd},yz}(s)
=
\begin{pmatrix}
s+\Gamma_Z+\delta K_\Gamma(s)
&
\sin\theta\,K_0(s)\\
-\sin\theta\,K_\Gamma(s)
&
s+\delta K_0(s)
\end{pmatrix}.
\label{eq:Myz_exact}
\end{equation}
The corresponding resolvent is
\begin{equation}
\begin{aligned}
\mathcal R(s)
&:=
\mathcal M_{\mathrm{fwd}}(s)^{-1}
=
[s^{-1}]
\oplus
[(s+\Gamma_Z)^{-1}]\\
&\mathrel{\phantom{:=}}\oplus
\mathcal R_{\mathrm{fwd},yz}(s),\\
\mathcal R_{\mathrm{fwd},yz}(s)
&:=
\mathcal M_{\mathrm{fwd},yz}(s)^{-1}.
\end{aligned}
\label{eq:full_reduced_resolvents}
\end{equation}

The identity component obeys
\begin{equation}
\widetilde{c}_I(s)=\frac{r_I}{s}.
\label{eq:cI_exact}
\end{equation}
Since $(\Phi_X-I)[\sigma_x]=0$ as apparent in Eq.~\eqref{eq:PhiX_x}, the renewal kernel drops out of
$W_x$ defined in Eq.~\eqref{eq:x_subspace}, leaving the purely
Markovian dephasing evolution of Eq.~\eqref{eq:LZ_x}:
\begin{equation}
\widetilde{c}_x(s)=\frac{r_x}{s+\Gamma_Z}
\qquad\Longrightarrow\qquad
c_x(t)=r_x e^{-\Gamma_Z t}.
\label{eq:cx_exact}
\end{equation}

\begin{figure*}[t]
  \centering
  \includegraphics[width=\textwidth]{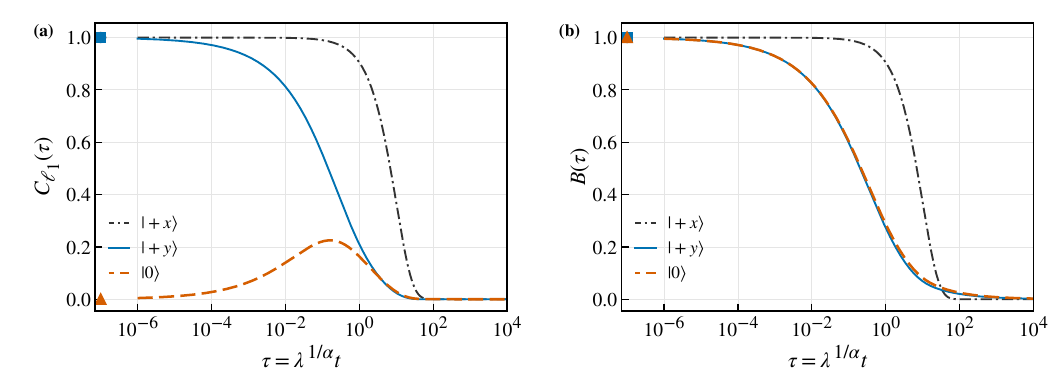}
  \caption{\label{fig:sim15c_initial_state_selectivity}Initial-state dependence under Prabhakar renewals: (a) $\ell_1$ coherence in the computational $\sigma_z$ basis,
$C_{\ell_1}=\sqrt{r_x^2+r_y^2}$, and (b) Bloch-vector length, $B=\sqrt{r_x^2+r_y^2+r_z^2}$, for initial states $|+x\rangle$, $|+y\rangle$, and $|0\rangle$.
Parameters are $(\alpha,\gamma)=(0.5,0.8)$, $\Gamma_Z/\lambda^{1/\alpha}=0.1$, and $\theta=\pi/2$, with $\tau=\lambda^{1/\alpha}t$.
Filled markers show exact initial values at $\tau=0$, displaced to the left edge for display on logarithmic axes. ChatGPT (OpenAI; GPT-5 Sol Ultra) generated the Python
simulation and plotting code from R.U.Erdogan's theoretical and numerical specifications; R.U.Erdogan reviewed it for consistency with the dynamical model and numerical methods.}
\end{figure*}

Thus, in the exact shifted-Montroll--Weiss formulation,
renewal dependence is confined to $W_{yz}$.
Figure~\ref{fig:sim15c_initial_state_selectivity} illustrates
the resulting initial-state selectivity: the traceless part
of $|+x\rangle\langle+x|$ lies in $W_x$ and follows the
jump-invariant control in Eq.~\eqref{eq:cx_exact}, whereas
those of $|+y\rangle\langle+y|$ and $|0\rangle\langle0|$
lie in $W_{yz}$ and activate $y\leftrightarrow z$
storage--return dynamics when $\sin\theta\ne0$.

Define
\begin{subequations}
\begin{align}
\widetilde{\mathbf c}_{yz}(s)&:=
\begin{pmatrix}
\widetilde c_y(s)\\
\widetilde c_z(s)
\end{pmatrix},\\
\mathbf r_{yz}&:=
\begin{pmatrix}
r_y\\
r_z
\end{pmatrix},
\label{eq:cyz_vector}
\end{align}
\end{subequations}
so that
\begin{equation}
\widetilde{\mathbf c}_{yz}(s)
=\mathcal{M}_{yz}(s)^{-1}\mathbf r_{yz}.
\label{eq:cyz_block_eq}
\end{equation}
The block determinant is
\begin{equation}
\begin{aligned}
\Delta(s)&:=\det\mathcal{M}_{yz}(s)\\
&=
\bigl[s+\Gamma_Z+\delta K_\Gamma(s)\bigr]
\bigl[s+\delta K_0(s)\bigr]\\
&+
\sin^2\theta\,K_\Gamma(s)K_0(s).
\label{eq:Delta_def}
\end{aligned}
\end{equation}
Using $\delta^2+\sin^2\theta=2\delta$ gives
\begin{equation}
\begin{aligned}
\Delta(s)
&=s(s+\Gamma_Z)\\
&+\delta\Bigl[(s+\Gamma_Z)K_0(s)+sK_\Gamma(s)\\
&+2K_\Gamma(s)K_0(s)\Bigr].
\end{aligned}
\label{eq:Delta_simplified}
\end{equation}
The component amplitudes are
\begin{subequations}
\begin{align}
\widetilde c_y(s)
&=
\frac{
\left[s+\delta K_0(s)\right]r_y
-\sin\theta\,K_0(s)r_z
}{
\Delta(s)
},
\label{eq:cy_exact}\\
\widetilde c_z(s)
&=
\frac{
\sin\theta\,K_\Gamma(s)r_y
+\left[s+\Gamma_Z+\delta K_\Gamma(s)\right]r_z
}{
\Delta(s)
}.
\label{eq:cz_exact}
\end{align}
\end{subequations}

Numerical inverse-Laplace reconstruction of the exact
kernel--resolvent expressions gives the time-domain curves.
Appendix~\ref{app:numerical_inverse_laplace} documents
Talbot-convergence tests and cross-checks against separately
implemented Hankel-contour and direct Garrappa OPC
evaluations~\cite{garrappa2015numerical}, with the validation
scope specified there.

\begin{figure*}[t]
  \centering
  \includegraphics[width=\textwidth]{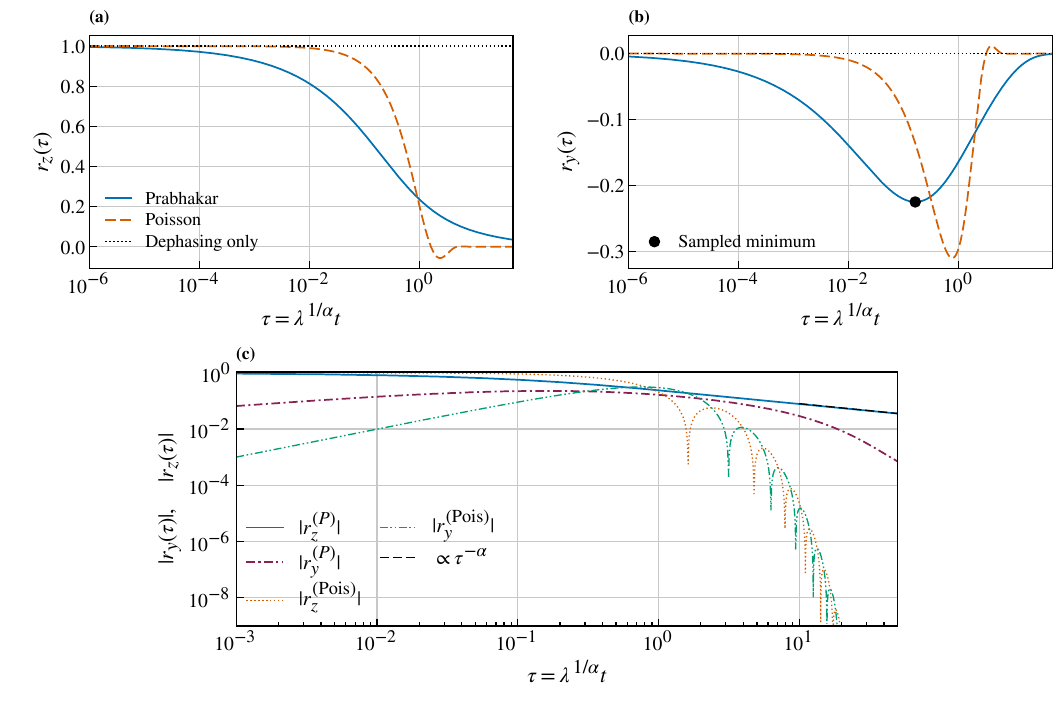}
  \caption{Longitudinal polarization retention and reverse $z\to y$ transfer from $|0\rangle$, with the Prabhakar parameters and time normalization of
Fig.~\ref{fig:sim15c_initial_state_selectivity}. The Poisson reference has $\nu/\lambda^{1/\alpha}=1$ and the same dephasing rate and kick angle. (a) Longitudinal polarization $r_z(\tau)$.
(b) Transverse Bloch component $r_y(\tau)$; the marker identifies its sampled Prabhakar minimum. Black dotted lines show the dephasing-only controls,
$r_z=1$ and $r_y=0$, respectively. (c) Component magnitudes on logarithmic axes; the dashed black $\tau^{-\alpha}$ guide indicates
the longitudinal Prabhakar asymptote. ChatGPT (OpenAI; GPT-5 Sol Ultra) generated the Python simulation and plotting code from R.U.Erdogan's theoretical and
numerical specifications; R.U.Erdogan reviewed it for consistency with the reduced dynamics and numerical methods.}
  \label{fig:sim3_bloch_storage_return}
\end{figure*}

Figure~\ref{fig:sim3_bloch_storage_return} shows the
$|0\rangle$ trajectories from
Eqs.~\eqref{eq:cy_exact}--\eqref{eq:cz_exact};
Sec.~\ref{sec:Prabhakar_asymptotics} derives their long-time
hierarchy. To quantify directional storage, write
\begin{equation}
\begin{pmatrix}
c_y(t)\\
c_z(t)
\end{pmatrix}
=
\begin{pmatrix}
T_{YY}^{(P)}(t) & T_{YZ}^{(P)}(t)\\
T_{ZY}^{(P)}(t) & T_{ZZ}^{(P)}(t)
\end{pmatrix}
\begin{pmatrix}
r_y\\
r_z
\end{pmatrix},
\label{eq:Tyz_transfer_matrix}
\end{equation}
where $T_{ZY}^{(P)}$ quantifies the contribution of the
initial transverse component $r_y$ to the longitudinal
polarization $r_z(t)=\langle\sigma_z\rangle_t
=\rho_{00}(t)-\rho_{11}(t)$ in the energy basis.
The storage-transfer amplitude is
\begin{equation}
S_{y\to z}^{(P)}(t;\theta)
=
\left|T_{ZY}^{(P)}(t;\theta)\right|.
\label{eq:Sy_to_z_definition}
\end{equation}
In Laplace space, Eq.~\eqref{eq:cz_exact} gives
\begin{equation}
\widetilde T_{ZY}^{(P)}(s;\theta)
=
\frac{\sin\theta\,K_\Gamma(s)}{\Delta(s)}.
\label{eq:Tzy_laplace_definition}
\end{equation}

\begin{figure*}[t]
  \centering
  \includegraphics[width=\textwidth]
    {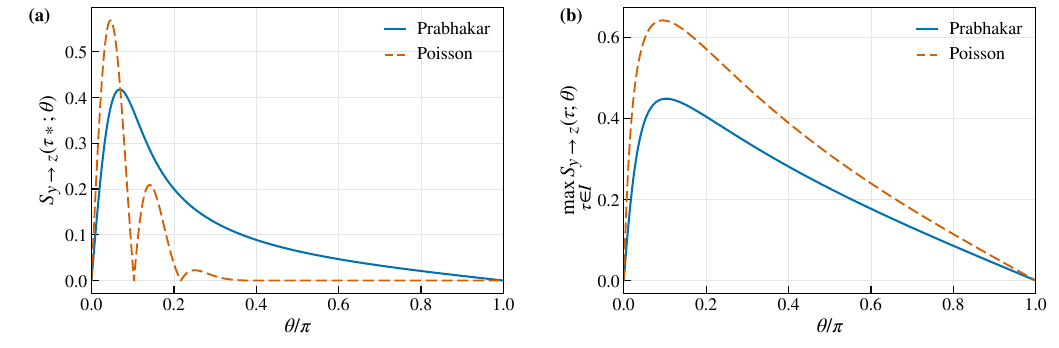}
  \caption{\label{fig:sim40_exact_tzy_prabhakar_poisson}
Exact $y\to z$ storage-transfer magnitude $S_{y\to z}(\tau;\theta)=|T_{ZY}(\tau;\theta)|$ from the initial state $|+y\rangle$, with $\tau=\lambda^{1/\alpha}t$.
Both renewal models use the same pure-dephasing drift and transverse kicks; the Poisson reference has $\nu=\lambda^{1/\alpha}$.
The parameters are $(\alpha,\gamma,\beta)=(0.5,0.8,0.4)$, $\Gamma_Z=0.1$, and $\lambda=1$. (a) Transfer at $\tau_\ast=10$.
(b) Maximum transfer over $10^{-3}\leq\tau\leq50$. The Prabhakar response is larger over part of the angle range in (a), whereas the Poisson finite-window maximum is larger at every
sampled $0<\theta<\pi$ in (b). Both models give zero transfer at $\theta=0,\pi$. ChatGPT (OpenAI; GPT-5 Sol Ultra) generated the Python evaluation and plotting code from R.U.Erdogan's theoretical and
numerical specifications; R.U.Erdogan reviewed it against the transfer expressions and numerical methods.}
\end{figure*}

Figure~\ref{fig:sim40_exact_tzy_prabhakar_poisson} compares the exact transfer with the scale-matched Poisson reference, $\nu=\lambda^{1/\alpha}$, using identical longitudinal drift and transverse kicks. The $\sin\theta$ prefactor gives $S_{y\to z}^{(P)}(t;0)=S_{y\to z}^{(P)}(t;\pi)=0$ with the same endpoint zeros for Poisson renewal. Figure~\ref{fig:sim40_exact_tzy_prabhakar_poisson}(a) compares transfer at the fixed normalized time $\tau_\ast=10$ and Fig.~\ref{fig:sim40_exact_tzy_prabhakar_poisson}(b) compares
maxima over $10^{-3}\leq\tau\leq50$.

For the displayed parameters, numerical interpolation of the forward-order curves in
Fig.~\ref{fig:sim40_exact_tzy_prabhakar_poisson}(a) gives one nontrivial interior crossing at $\theta_c/\pi\simeq0.06994$. At $\tau_\ast=10$, the Prabhakar transfer is larger for $\theta_c<\theta<\pi$, with the largest positive difference near $\theta/\pi\simeq0.103$: the Poisson response is near
a damped-oscillatory transient node, while the Prabhakar response remains finite. This advantage reflects persistence at the selected time; Poisson has the larger global lobe at $\tau_\ast$ and in
Fig.~\ref{fig:sim40_exact_tzy_prabhakar_poisson}(b) a finite-window maximum at least as large at every sampled kick angle.

The determinant in Eq.~\eqref{eq:Delta_def} defines the
exact nonlinear spectral problem: isolated poles of the
reduced propagator arise from loss of invertibility of
the $yz$ block,
\begin{equation}
\Delta(s)=0.
\label{eq:pole_equation}
\end{equation}
Branch points and cuts originate in the nonanalytic
structure of $K_0(s)$ and $K_\Gamma(s)$ and are not determined
by Eq.~\eqref{eq:pole_equation} alone.

\subsection{Prabhakar branch structure, long-time behavior, and Poisson-renewal limit in the \(W_{yz}\) block}
\label{sec:Prabhakar_asymptotics}

For the normalized Prabhakar law in
Eqs.~\eqref{eq:11e}--\eqref{eq:11ee}, the small-$s$ result
in Eq.~\eqref{eq:prabhakar_small_s} of
Sec.~\ref{subsec:ASLS} gives, through Eq.~\eqref{eq:30a},
the kernel asymptotic as $s\to0^+$:
\begin{equation}
\widetilde{K}(s)\sim\frac{\lambda}{\gamma}s^{1-\alpha},
\qquad 0<\alpha<1.
\label{eq:small_s_wK}
\end{equation}
The unshifted kernel $K_0(s)=\widetilde K(s)$ in
Eq.~\eqref{eq:K0_def} has a branch point at $s=0$.
For $\Gamma_Z>0$, $K_\Gamma(s)$ in
Eq.~\eqref{eq:KGamma_def} is analytic there, with nearest
branch point at $s=-\Gamma_Z$:
\begin{equation}
K_\Gamma(s)=\widetilde{K}(s+\Gamma_Z)
=K_\Gamma(0)+O(s),
\end{equation}
where
\begin{equation}
K_\Gamma(0)=
\frac{\Gamma_Z}
{\left(1+\Gamma_Z^\alpha/\lambda\right)^\gamma-1}.
\label{eq:KGamma_zero}
\end{equation}
The origin is the rightmost of these branch points.
Its cut controls the long-time behavior if all isolated
physical-sheet poles satisfy $\operatorname{Re}s_p<0$
and the state-dependent origin-branch coefficient is nonzero.

Substitution of Eqs.~\eqref{eq:small_s_wK}
and~\eqref{eq:KGamma_zero} into
Eq.~\eqref{eq:Delta_simplified} gives, for $\delta\neq0$,
\begin{equation}
\Delta(s)
\sim
\delta\Bigl[\Gamma_Z+2K_\Gamma(0)\Bigr]
\frac{\lambda}{\gamma}s^{1-\alpha},
\qquad s\to0.
\label{eq:Delta_small_s}
\end{equation}
The reduced forward-order inverse block is therefore
\begin{equation}
\begin{aligned}
\mathcal M_{\mathrm{fwd},yz}(s)^{-1}
=
&s^{\alpha-1}
\frac{\gamma}{
\lambda\delta
\left[\Gamma_Z+2K_\Gamma(0)\right]
}\\
&\times
\begin{pmatrix}
0&0\\
\sin\theta K_\Gamma(0)&
\Gamma_Z+\delta K_\Gamma(0)
\end{pmatrix}\\
&+o\left(s^{\alpha-1}\right).
\label{eq:Myz_inverse_small_s}
\end{aligned}
\end{equation}
Directional transfer through the off-diagonal entries
additionally requires $\sin\theta\neq0$.
For $\mathbf r_{yz}=(r_y,r_z)^{\mathsf T}$,
Eqs.~\eqref{eq:Delta_small_s}
and~\eqref{eq:Myz_inverse_small_s} give
\begin{equation}
\begin{aligned}
\widetilde{\mathbf c}_{yz}(s)
&=
s^{\alpha-1}\mathbf a_{\rm br}
+o(s^{\alpha-1}),\\
\mathbf a_{\mathrm{br}}
&=
\frac{\gamma}{
\lambda\delta
\left[\Gamma_Z+2K_\Gamma(0)\right]
}\\
&\times
\begin{pmatrix}
0\\
\sin\theta K_\Gamma(0)r_y
+\left(\Gamma_Z+\delta K_\Gamma(0)\right)r_z
\end{pmatrix}.
\label{eq:branch_amplitude_yz}
\end{aligned}
\end{equation}

Local branch-cut inversion under the boundary-value and
contour assumptions of Appendix~\ref{app:leading_tail} gives
\begin{equation}
\mathbf c_{yz}(t)
=
\frac{\mathbf a_{\rm br}}{\Gamma(1-\alpha)}t^{-\alpha}
+o(t^{-\alpha}),
\qquad t\to\infty.
\label{eq:longtime_vector_yz}
\end{equation}
The leading $s^{\alpha-1}$ singularity in
Eq.~\eqref{eq:branch_amplitude_yz} is confined to the
$z$-output row, with a coefficient depending on both
$r_y$ and $r_z$. For $r_y=1$ and $r_z=0$,
Eq.~\eqref{eq:longtime_vector_yz} yields the forward transfer
from coherence to longitudinal polarization,
\begin{equation}
T_{ZY}^{(P)}(t)
\sim
\frac{
\gamma\sin\theta K_\Gamma(0)
}{
\lambda\delta
\left[\Gamma_Z+2K_\Gamma(0)\right]
\Gamma(1-\alpha)
}
t^{-\alpha}.
\label{eq:T_ZY_P}
\end{equation}
The reverse $z\to y$ transfer lacks this leading term
and has the generic positive-time asymptotic
$T_{YZ}^{(P)}(t)=O(t^{-1-\alpha})$.
Exchanging the order of the drift and kick operators in
Eqs.~\eqref{eq:drift_semigroup} and~\eqref{eq:PhiX_def}
changes the directional amplitudes and residues but preserves
$\Delta(s)$ and its noncanceled pole locations.

The inverse transform combines residues at noncanceled physical-sheet zeros $s_p$ of Eq.~\eqref{eq:pole_equation},
away from branch cuts and zeros of the scalar-kernel denominators, with Hankel integrals along the cuts of
$K_0$ and $K_\Gamma$. The residues produce exponential or damped-oscillatory transients; the contribution associated with the shifted
branch point at $s=-\Gamma_Z$ is exponentially dressed. Appendix~\ref{app:Wyz_singularity_analysis} states the
exclusion conditions and contour reconstruction, and Appendix~\ref{app:leading_tail} derives both directional tails.

For the constant-rate Poisson-renewal limit~\cite{Cox1962,feller1991introduction},
\begin{equation}
w_{\rm exp}(t)=\nu e^{-\nu t},
\qquad
\widetilde w_{\rm exp}(s)=\frac{\nu}{s+\nu},
\label{eq:w_exp_summary}
\end{equation}
the renewal-resolvent kernel is constant:
\begin{equation}
\widetilde K_{\rm exp}(s)=\nu.
\label{eq:K_exp_summary}
\end{equation}
The resulting $W_{yz}$ resolvent is rational.
Appendix~\ref{app:Poisson_resolvent_pole} gives its
matrix, determinant, and poles, including the
repeated-pole case. Its pole-generated transients
decay exponentially, possibly with polynomial
prefactors, whereas the fractional Prabhakar kernel
also produces the algebraic branch-cut tails
derived above.

The comparisons in Figs.~\ref{fig:sim3_bloch_storage_return} and~\ref{fig:sim40_exact_tzy_prabhakar_poisson} set
$\nu=\lambda^{1/\alpha}$, matching the scalar renewal scale rather than the mean waiting time, which diverges
for $0<\alpha<1$. This calibration need not equalize expected event counts over a finite observation window.
The reported differences are therefore conditional on this calibration and can reflect both mean renewal
counts and higher renewal statistics. Section~\ref{sec:physical_interpretation} additionally compares processes with equal expected event counts
at every time.

\subsection{Physical interpretation: Liouville-mode-resolved Prabhakar memory and mode-selective polarization retention}
\label{sec:physical_interpretation}

During a waiting interval, longitudinal drift attenuates
an initial $y$ component. A transverse event can transfer
its surviving amplitude into the zero-drift $z$ sector,
where a subsequent long no-event interval retains
longitudinal polarization and contributes to the storage
tail in Eq.~\eqref{eq:T_ZY_P}. The reverse transfer enters
the dephasing-sensitive $y$ sector and lacks the same
leading origin-branch term, explaining the directional
hierarchy of Sec.~\ref{sec:Prabhakar_asymptotics}.

To assess energy relaxation, we add independent
zero-temperature amplitude damping toward
$r_z^{\mathrm{eq}}=1$ between events, retaining the same
transverse kicks and the Prabhakar process initialized
at zero renewal age. Define the normalized variables
\begin{equation}
\begin{aligned}
\omega_r&=\lambda^{1/\alpha},
&\qquad \tau&=\omega_r t,\\
g&=\frac{\Gamma_Z}{\omega_r},
&\qquad \eta&=\frac{\Gamma_1}{\omega_r},
\end{aligned}
\label{eq:finite_t1_scales}
\end{equation}
where $\Gamma_Z$ and $\Gamma_1=T_1^{-1}$ are the
interevent pure-dephasing and energy-relaxation rates.
The ratios $g$ and $\eta$ express them relative to
$\omega_r$; $\eta=0$ corresponds to $T_1=\infty$.

For opposite preparations $\mathbf r(0)=\pm\mathbf e_j$
subject to the same dynamical map, define
\begin{equation}
Q_{j\to i}(\tau)
=
\frac{r_i(\tau;+j)-r_i(\tau;-j)}{2},
\quad i,j\in\{y,z\},\ i\ne j.
\label{eq:finite_t1_pair_signal}
\end{equation}
Here $\mathbf e_j$ is the unit vector along axis $j$. The pair difference $Q_{j\to i}$ isolates the signed
response of the output $i$ component to the initial $j$ component. Subtraction cancels the common
relaxation-induced polarization offset while retaining the physical attenuation of this response.
Appendix~\ref{app:finite_t1_directional} derives the affine map and the factorization of its homogeneous
transfer block. For $0<\eta<2g$, this factorization exponentially cuts off both directional algebraic tails
under the stated asymptotic hypotheses. The uncut law in Eq.~\eqref{eq:T_ZY_P} therefore applies to the
pure-dephasing limit.

We compare Prabhakar ($\mathrm P$) and scale-matched
Poisson ($\mathrm{Pois}$) renewal with $\nu=\omega_r$
and identical interevent dynamics and kicks. The
differences at a prescribed readout time and between
separately evaluated maxima are
\begin{align}
\Delta S_\ast
&=
\left|Q_{y\to z}^{(\mathrm P)}(\tau_\ast)\right|
-
\left|Q_{y\to z}^{(\mathrm{Pois})}(\tau_\ast)\right|,
\label{eq:finite_t1_delta_star}\\
\Delta S_{\max}
&=
\max_{0\leq\tau\leq50}
\left|Q_{y\to z}^{(\mathrm P)}(\tau)\right|
\nonumber\\
&\quad-
\max_{0\leq\tau\leq50}
\left|Q_{y\to z}^{(\mathrm{Pois})}(\tau)\right|.
\label{eq:finite_t1_delta_max}
\end{align}
The constant-rate reference has expected event count $\tau$, whereas the Prabhakar mean is
$M(\tau/\omega_r)$. To test whether unequal mean event activity alone accounts for the readout difference,
we introduce an inhomogeneous Poisson reference that matches the expected count at every time, retaining
the same interevent generator $\mathcal L_{\mathrm d}$ in Eq.~\eqref{eq:finite_t1_generator} and the same
deterministic kick $\Phi_X$.

We use the label ``$\mathrm{match}$'' for the count-matched Poisson comparator and choose its deterministic
physical-time intensity as 
\begin{equation}
\begin{aligned}
\nu_{\mathrm{match}}(t)&=m(t),\\
\mathbb E[N_{\mathrm{match}}(t)]
&=\int_0^t m(u)\,du
=M(t)=\mathbb E[N_{\mathrm P}(t)].
\end{aligned}
\label{eq:count_matched_intensity}
\end{equation}
Here $m$ and $M$ are the renewal density and mean renewal
function defined in Sec.~\ref{subsec:Physmot_fract}; their
Prabhakar transforms follow from the renewal sums in
Ref.~\cite{CahoyPolito2013}. The Poisson count law and time-change construction
(see Definition~3.1 and Proposition~7.3 of Ref.~\cite{LastPenrose2018}) give $N_{\mathrm{match}}(t)=N_*(M(t))$, with $N_*$
a unit-rate Poisson process and
$\mathbb E[N_{\mathrm{match}}(t)]=M(t)$. Local integrability of $m$,
established in Appendix~\ref{app:prabhakar_forward_series},
makes this construction well defined on every finite interval.
The intensity depends on absolute time since preparation;
it is not reset after an event and is distinct from the
waiting-time hazard $w(t)/\Psi(t)$ and the memory kernel $K(t)$.

Averaging the infinitesimal event and no-event alternatives
gives the time-local master equation
\begin{equation}
\begin{aligned}
\frac{d\rho_{\mathrm{match}}}{dt}
&=\mathcal L_{\mathrm d}[\rho_{\mathrm{match}}]\\
&\quad+m(t)(\Phi_X-\mathcal I)[\rho_{\mathrm{match}}].
\end{aligned}
\label{eq:count_matched_master}
\end{equation}
The additional term is a Lindblad dissipator~\cite{BreuerPetruccione2007}
with jump operator $\sqrt{m(t)}\,U_x(\theta)$. Nonnegativity
and local integrability of $m$ therefore give a CP-divisible
propagator. The constant-rate Poisson pole classification in
Sec.~\ref{sec:Prabhakar_asymptotics} does not apply to this
time-dependent comparator.

For the count-matched reference, define
$\Delta S_\ast^{(\mathrm{match})}$ and
$\Delta S_{\max}^{(\mathrm{match})}$ by
Eqs.~\eqref{eq:finite_t1_delta_star}
and~\eqref{eq:finite_t1_delta_max}, replacing
$\mathrm{Pois}$ by $\mathrm{match}$.

\begin{figure*}[t]
\centering
\includegraphics[width=\textwidth]
{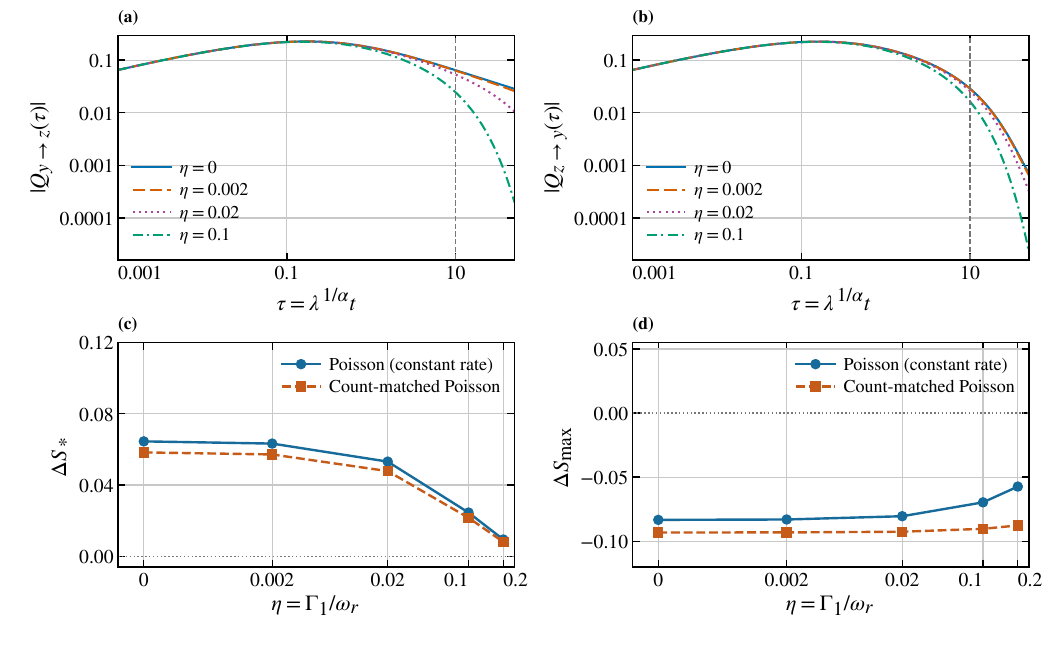}
\caption{
Directional transfer with independent zero-temperature
relaxation and zero-age Prabhakar renewal:
$(\alpha,\gamma,g)=(0.5,0.8,0.1)$, $\theta=\pi/2$,
and $\tau=\omega_r t$, $\omega_r=\lambda^{1/\alpha}$.
(a,b) Forward and reverse pair-difference magnitudes
for $\eta=0,0.002,0.02,0.1$, on logarithmic axes over
$10^{-3}\leq\tau\leq50$. Their signed signals are
respectively positive and negative on this interval
and vanish at $\tau=0$.
Gray dashed lines mark $\tau_\ast=10$. (c,d) Prabhakar-minus-reference differences in forward
magnitude at $\tau_\ast$ and between separately refined
numerical maxima on $0\leq\tau\leq50$, respectively
(Eqs.~\eqref{eq:finite_t1_delta_star}
and~\eqref{eq:finite_t1_delta_max}), using these rates
and the equal-damping control $\eta=0.2=2g$.
References are constant-rate, scale-matched Poisson
(blue circles; $\nu=\omega_r$, mean count $\tau$)
and count-matched Poisson (orange squares;
$\nu_{\mathrm{match}}(t)=m(t)$, mean count
$M(\tau/\omega_r)$ at every time).
Both use identical interevent dynamics, signed kicks,
and preparations. Equal mean counts do not imply
equal counting statistics. Panels~(c,d) have linear ordinates and symmetric-log
rate axes, linear for $|\eta|\leq0.002$ and logarithmic
beyond; connecting lines guide the eye.
Physical attenuation is retained without exponential
compensation. Both comparisons share one Prabhakar
calculation; peak refinement does not certify global
continuous-time optimality
(Appendix~\ref{app:count_matched_benchmark}).
ChatGPT (OpenAI; GPT-5 Sol Ultra) generated the original
Python simulation and plotting code from R.U.Erdogan's
theoretical and numerical specifications; he reviewed
it against the finite-$T_1$ model and numerical methods.
}
\label{fig:sim108_finite_t1}
\end{figure*}

For $(\alpha,\gamma,g)=(0.5,0.8,0.1)$ and
$\theta=\pi/2$, panels~(a,b) of
Fig.~\ref{fig:sim108_finite_t1} show the forward and
reverse Prabhakar pair-difference magnitudes at
$\eta=0,0.002,0.02,0.1$. Panels~(c,d) compare the
readout and peak differences against both Poisson
references at these rates and at the equal-damping
value $\eta=0.2=2g$. For every tested rate, both
references give positive readout differences at
$\tau_\ast=10$ and negative peak differences on
$0\leq\tau\leq50$. Prabhakar therefore retains the
larger forward-transfer magnitude at the prescribed
readout, while each Poisson reference attains the
larger evaluated peak within the observation window.
The readout difference persists after matching the
entire mean-count profile and thus cannot be
attributed solely to unequal mean event activity.
The signed readout responses are positive for
Prabhakar and negative for count-matched Poisson;
the metrics compare their magnitudes.

\begin{figure*}[t]
\centering
\includegraphics[width=\textwidth]
{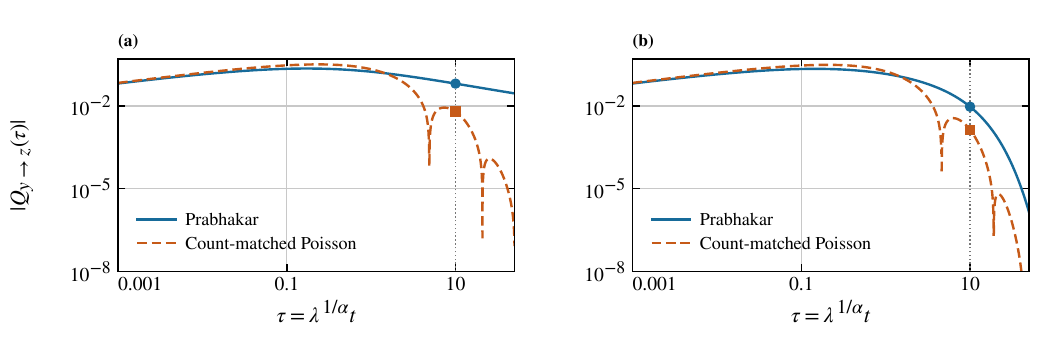}
\caption{
Forward-transfer magnitudes for Prabhakar renewal (solid blue) and count-matched Poisson dynamics
(dashed orange), with the parameters, zero-age initialization, and benchmark definition of
Fig.~\ref{fig:sim108_finite_t1}. (a) $\eta=0$; (b) $\eta=0.2=2g$, the equal-damping
control for the homogeneous difference generator. Curves evaluate Eqs.~\eqref{eq:count_matched_block}
and~\eqref{eq:count_matched_prabhakar_transfer}, retaining physical attenuation. Both axes are logarithmic, with
$10^{-3}\leq\tau\leq50$. Gray dotted lines mark $\tau_\ast=10$; filled circles and squares denote the respective readout magnitudes. The signed readouts have opposite signs. Signed data retain $\tau=0$ and zero crossings. ChatGPT (OpenAI; GPT-5 Sol Ultra) generated the Python simulation and plotting code from R.U.Erdogan's theoretical and numerical specifications.
}
\label{fig:count_matched_transfer}
\end{figure*}

Figure~\ref{fig:count_matched_transfer} shows the time-dependent forward-transfer comparison for
$\eta=0$ in panel~(a) and $\eta=0.2=2g$ in panel~(b). At $\tau_\ast=10$, the common Prabhakar and
count-matched Poisson expected count is $M(10/\omega_r)\simeq4.5787$, whereas the constant-rate
reference has expected count $10$. Appendix~\ref{app:finite_t1_directional} gives the
normalized comparator equations, the exact transfer expressions evaluated, and the numerical validation,
including the independent mean-count check and the equal-damping control.

At equal damping, $\eta=2g$, the homogeneous
interevent difference generator is $-\eta I_2$
and commutes with the kick rotation. The history
factorization in
Appendix~\ref{app:finite_t1_directional} therefore
gives, for $j\in\{\mathrm P,\mathrm{match}\}$,
\begin{equation}
Q_{y\to z}^{(j)}(\tau)
=e^{-\eta\tau}\,
\mathbb E\!\left[
\sin\!\left(\theta N_j(\tau/\omega_r)\right)
\right].
\label{eq:count_matched_equal_damping}
\end{equation}
The mean count does not determine this average of
a nonlinear function of the count. For $\theta=\pi/2$,
counts congruent to $1$ and $3$ modulo $4$ contribute
with opposite signs, whereas even counts give zero
forward transfer. The difference in
Fig.~\ref{fig:count_matched_transfer}(b) thus reveals
sensitivity to the count distribution beyond its
mean even when the homogeneous drift and rotation
commute. Away from equal damping, their noncommutation
for $\sin\theta\ne0$ additionally makes the response
sensitive to the event times within each ordered
history. These results, for the stated parameters
and observables, do not by themselves establish
CP-indivisibility of the Prabhakar dynamics.


The amplitude in Eq.~\eqref{eq:Sy_to_z_definition} and the pair differences in Eq.~\eqref{eq:finite_t1_pair_signal} quantify directional
transfer of a selected Bloch component. Persistence of the forward response describes retention of longitudinal polarization generated from initial coherence. The forward and reverse responses are
evaluated for different initial preparations; they do not by themselves establish a sequential write--store--retrieve protocol or improved
arbitrary-state storage or channel fidelity.

The initial-state dependence in Fig.~\ref{fig:sim15c_initial_state_selectivity} and transfer
from longitudinal polarization to coherence in Fig.~\ref{fig:sim3_bloch_storage_return}, together with
the renewal-insensitive $W_x$ control in Eq.~\eqref{eq:cx_exact}, identify mode- and
direction-dependent decoherence reshaping in $W_{yz}$. For this kick geometry, transfer requires a nonzero
initial $W_{yz}$ component and $\sin\theta\neq0$.

The material interpretation remains mesoscopic. A quantitative connection to a specific device requires
microscopic modeling or experimental characterization of the clock statistics and preparation specified in
Sec.~\ref{subsec:Physmot_fract}, together with the effective kick map defined in
Sec.~\ref{subsec:ren_map_equ}.

\section{\label{sec:CONCL}Conclusion}
We specialized established quantum-renewal theory to a transmon with longitudinal Markovian dephasing and transverse kicks governed
by Prabhakar renewal statistics. Forward chronological histories yield a CPTP map satisfying a first-event Volterra equation and
fix its noncommuting kernel--resolvent ordering. Under the stated regularity assumptions, the equivalent drift-conjugated Caputo-type
form retains the required initial-value term. The differentiated renewal-density series converges absolutely and uniformly on compact intervals in $(0,\infty)$, giving the ordinary positive-time
representative of the causal memory kernel. The drift-dressed Prabhakar-convolution expansion converges locally in $L^1$ and resums to the same forward resolvent. The exact $1\oplus1\oplus2$ operator-space
decomposition isolates the trace and renewal-insensitive $W_x$ subspaces from the coupled $W_{yz}$ subspace supporting memory-dependent
coherence--polarization transfer. Independent inverse-Laplace comparisons test selected scalar quantities and reduced components,
including an ordering-sensitive component, for the parameters and times in Appendix~\ref{app:numerical_inverse_laplace}.

For $0<\alpha<1$, $\Gamma_Z>0$, and $\sin\theta\ne0$, under the stated admissibility, branch, and pole conditions,
the origin branch point contributes a $t^{-\alpha}$ term only to longitudinal polarization, with a coefficient that
can vanish for particular initial states. In particular, $T_{ZY}^{(P)}(t)$ has a $t^{-\alpha}$ storage tail, whereas
$T_{YZ}^{(P)}(t)$ is generically $O(t^{-1-\alpha})$. Transverse events produce state-selective decoherence
reshaping by transferring an initial coherence component into longitudinal polarization unaffected by interevent
pure dephasing. A nonzero storage tail provides asymptotically more persistent retention of a selected Bloch component
than exponentially decaying constant-rate Poisson transients, without itself establishing larger peak transfer or improved
state-storage or channel fidelity.

The parameter dependence must be interpreted separately for the scalar kernel and the directional response. Within $0<\alpha<1$
and $0<\beta=\alpha\gamma<1$, $\beta$ governs the short-time kernel singularity, while $\alpha$ and $\gamma$ determine its
long-time power and prefactor (see Appendix~\ref{app:branch_renewal}). Increasing $\alpha$ slows the scalar kernel's algebraic decay
but steepens the directional tails stated above. The dimensionless kernel's $1/\gamma$ long-time prefactor at fixed $\alpha$ does not imply the same scaling for storage:
Eq.~\eqref{eq:T_ZY_P} contains both an explicit $\gamma$ factor and a $\gamma$-dependent shifted kernel.

In the independent zero-temperature relaxation extension of Appendix~\ref{app:finite_t1_directional}, opposite-preparation differences remove the affine
polarization offset while retaining physical finite-$T_1$ attenuation. For $0<\Gamma_1<2\Gamma_Z$, the exact
factorization of the homogeneous transfer block yields an exponential cutoff of the directional algebraic tails
under the stated asymptotic hypotheses. For the cases tested in Fig.~\ref{fig:sim108_finite_t1}, Prabhakar renewal gives a larger forward-transfer
magnitude at the prescribed readout than both the scale-matched constant-rate and count-matched inhomogeneous Poisson references, whereas both references
give larger evaluated finite-window peaks. The readout difference persists when expected event counts agree at every time and therefore cannot be
attributed solely to unequal mean event activity. Correlations between successive TLS reconfiguration intervals remain an open extension beyond the present
renewal model.

Future comparisons with the experiments discussed in Secs.~\ref{sec:DGTLSSQ} and~\ref{sec:GTSL} should relate
measured fluctuations to coarse-grained renewal events and test the waiting-time law and interval independence,
accounting for detection bandwidth and renewal age at state preparation. Quantitative fits to experimental data should yield a common parameter set whose predictions can be tested against
independent measurements of both transfer directions, with the renewal-insensitive $W_x$ response as a control. Together with
independently evaluated storage fidelities, these tests would assess whether tuning TLS environments can make the predicted
persistence useful for quantum storage.
\section*{Data availability}
The source code used to generate the numerical results and figures reported in this article has not been deposited in a public repository
and is available from the authors upon reasonable request.
\appendix
\section{Complete monotonicity, Bernstein admissibility and nonequilibrium interpretation}
\label{app:prabhakar_admissibility}
This appendix clarifies the relation among CM, scalar renewal admissibility, and CP of the reduced quantum dynamics. It also specifies the nonequilibrium interpretation assigned to the transverse TLS renewal environment.
\subsection{Scalar admissibility and its distinction from complete positivity}
\label{app:C6}
The renewal normalization conditions are derived in Eqs.~\eqref{eq:11c} and~\eqref{eq:11d}, and the resulting normalized
waiting-time transform and its admissible CM domain are given in Eqs.~\eqref{eq:11e} and~\eqref{eq:11ee}, respectively
~\cite{MainardiGarrappa2015,GorskaHorzelaBratekDattoliPenson2018}.

For completeness, a function $f:(0,\infty)\rightarrow\mathbb{R}$ is completely monotone if it is infinitely differentiable and satisfies
\begin{equation}
(-1)^n f^{(n)}(t)\geq 0,
\qquad
t>0,\quad n=0,1,2,\ldots .
\label{eq:C.6.1}
\end{equation}
The broader probability domain is distinguished from the CM subdomain in Sec.~\ref{subsec:ASLS}. In the CM subdomain specified by Eq.~\eqref{eq:11ee}, the
normalized waiting-time density $w(t)$ satisfies Eq.~\eqref{eq:C.6.1}. Moreover, the survival probability defined in Eq.~\eqref{eq:11psi} satisfies $\Psi'(t)=-w(t)$. Together with $\Psi(t)\geq0$, this implies that
$\Psi(t)$ is also CM.

Bernstein's theorem then gives the unique positive-measure representation already written in Eq.~\eqref{eq:survival_Bernstein_measure}
\cite{Widder2010,Schilling2010}. The measure $\mu_{\mathrm{eff}}$ appearing there describes effective exponential survival rates. It is neither the switching-rate distribution of individually resolved TLSs nor a microscopic bath spectral density.

These scalar properties should be distinguished from CP of the reduced map. CP follows from the renewal-history expansion in
Eq.~\eqref{eq:renmapdef}: each history is a composition of CPTP drift and event maps, and the admissible renewal law assigns nonnegative
normalized weights to those histories. Accordingly, the CM condition in Eq.~\eqref{eq:11ee} is not a pointwise-positivity requirement on
the renewal-resolvent kernel defined in Eqs.~\eqref{eq:30a} and~\eqref{eq:30b}.
\subsection{Nonequilibrium interpretation and the KMS condition}
\label{app:C.7}

Equation~\eqref{eq:renewal_averaged_reduced_state} distinguishes the two environmental descriptions used in the model: the longitudinal
reservoir is traced out as a quantum bath, whereas the transverse TLS contribution enters through an average over classical renewal
histories. The latter is therefore not modeled as a microscopic quantum reservoir in thermal equilibrium.

The Kubo--Martin--Schwinger (KMS) condition constrains equilibrium quantum correlation functions and produces the corresponding thermal
detailed-balance relation~\cite{BreuerPetruccione2007}. Because the transverse TLS contribution is represented by a classical,
nonstationary renewal clock, no equilibrium KMS relation follows from, or is imposed on, its renewal statistics. This statement applies only
to the transverse renewal component and does not modify the separate Born--Markov treatment of the longitudinal reservoir.

The Bernstein measure in Eq.~\eqref{eq:survival_Bernstein_measure} should consequently be interpreted as a classical effective-rate measure for the survival law, rather than as a thermal spectral density. For $0<\alpha<1$, the long-time behavior derived from Eqs.~\eqref{eq:prabhakar_small_s}--\eqref{eq:asymp_wt} gives the heavy-tailed, aging renewal regime used for the nonequilibrium coarse-grained TLS description.
\section{Forward quantum-renewal construction and Volterra structure}
\label{app:renewal_structure}
This appendix proves the first-event factorization underlying Eq.~\eqref{eq:23} and records the distributional form of its homogeneous memory closure. The drift semigroup, renewal statistics, event map, ordered-history expansion, and forward inverse-propagator pencil are defined in Sec.~\ref{subsec:Physmot_fract} and Eqs.~\eqref{eq:drift_semigroup},~\eqref{eq:renmapdef},~\eqref{eq:PhiX_def}, and~\eqref{eq:M_def}, respectively, and are not restated here. All map products are composed from right to left.

Under the normalization and local-integrability assumptions imposed in Sec.~\ref{subsec:Physmot_fract}, all finite-time operator integrals below are well defined.
Because the reduced qubit operator space is finite dimensional, the drift and event superoperators are bounded and the required Bochner-integrability conditions follow componentwise.
\subsection{First-event factorization}

Let $\Lambda_n(t)$ denote the $n$-event contribution appearing in Eq.~\eqref{eq:renmapdef}, with $\Lambda_0(t)$ denoting its no-event term. For
$n\geq 1$, we condition on the first renewal time $\tau:=t_1$. For $n\geq 2$, introduce the remaining event times relative to the
first event, $u_j:=t_{j+1}-\tau, \text{ for } j=1,\ldots,n-1$ and set $u_0:=0$. The ordered simplex then factorizes as $0<\tau<t,\text{ and } 0<u_1<\cdots<u_{n-1}<t-\tau$ with unit Jacobian. The waiting-time weight becomes
\begin{equation}
\prod_{k=1}^{n}w(t_k-t_{k-1})
=
w(\tau)
\prod_{j=1}^{n-1}w(u_j-u_{j-1}),
\end{equation}
the terminal survival factor becomes $\Psi(t-t_n)=\Psi[(t-\tau)-u_{n-1}]$, completing the weight of the residual $(n-1)$-event history. The chronological map is
\begin{equation}
\mathcal{D}_{(t-\tau)-u_{n-1}}
\Phi_X
\mathcal{D}_{u_{n-1}-u_{n-2}}
\cdots
\Phi_X
\mathcal{D}_{u_1}
\Phi_X
\mathcal{D}_{\tau}.
\end{equation}
The factors to the left of the final $\Phi_X\mathcal{D}_{\tau}$ form an $(n-1)$-event forward history over the remaining duration $t-\tau$. This is the map-valued counterpart of the classical first-renewal decomposition~\cite{Cox1962}. For $n=1$, that residual history is the no-event contribution. Hence, for every $n\geq 1$,
\begin{equation}
\Lambda_n(t)
=
\int_0^t
w(\tau)\,
\Lambda_{n-1}(t-\tau)\,
\Phi_X
\mathcal{D}_{\tau}
\,d\tau .
\label{eq:B_first_event_recursion}
\end{equation}

Summing Eq.~\eqref{eq:B_first_event_recursion} over $n\geq 1$ and including the no-event term yields the forward map-level Volterra
equation stated in Eq.~\eqref{eq:23}. Thus the position of $\Phi_X\mathcal{D}_{\tau}$ on the right of the residual map follows directly from the first-event decomposition.
\subsection{Laplace-domain homogeneous closure}
\label{app:laplace_homogeneous_closure}
Taking the Laplace transform of Eq.~\eqref{eq:23} gives Eq.~\eqref{eq:LT_renewal}; the survival and renewal-resolvent identities in Eqs.~\eqref{eq:Gs} and~\eqref{eq:Ks} then yield
the forward pencil in Eq.~\eqref{eq:M_def}. Acting with Eq.~\eqref{eq:exact_kernel_resolvent} on the initial state and rearranging gives the homogeneous Laplace-domain relation
\begin{equation}
\begin{aligned}
s\widetilde{\rho}(s)-\rho(0)
&=
\mathcal{L}_Z\widetilde{\rho}(s)
\\
&\quad+
(\Phi_X-\mathbb{I})
\widetilde{K}\!\left(\mathcal A(s)\right)
\widetilde{\rho}(s).
\end{aligned}
\label{eq:B_homogeneous_laplace}
\end{equation}

For later use, define the drift-dressed causal memory operation by
\begin{equation}
\mathcal{L}_{t\to s}
\left\{
\mathcal{K}_{\mathcal{L}_Z}[\rho](t)
\right\}
:=
\widetilde{K}\!\left(\mathcal A(s)\right)
\widetilde{\rho}(s),
\label{eq:B_dressed_memory_definition}
\end{equation}
where $\mathcal{L}_{t\to s}$ denotes the Laplace transform from $t$ to $s$. When the scalar kernel has a locally integrable representative,
the inverse transform of Eq.~\eqref{eq:B_dressed_memory_definition} is
\begin{equation}
\mathcal{K}_{\mathcal{L}_Z}[\rho](t)
=
\int_0^t
K(t-\tau)\,
\mathcal{D}_{t-\tau}[\rho(\tau)]
\,d\tau .
\label{eq:B_dressed_memory_time}
\end{equation}
For singular kernels, including the Poisson limit, the memory operation is understood causally in the distributional sense through
Eq.~\eqref{eq:B_dressed_memory_definition}. The equivalent drift-conjugated Caputo-type representation is given in Eq.~\eqref{eq:rho_fractional}.

A conditioned trajectory is piecewise differentiable but discontinuous at its renewal epochs. Its distributional derivative is
\begin{equation}
\frac{d}{dt}\rho_{\omega}(t)
=
\mathcal{L}_Z[\rho_{\omega}(t)]
+
\sum_n
\delta(t-t_n)
(\Phi_X-\mathbb{I})
[\rho_{\omega}(t_n^-)] .
\label{eq:B_pathwise_distributional}
\end{equation}
The jump contribution therefore cannot be discarded merely because the set of renewal times has Lebesgue measure zero. The boundary term in Eq.~\eqref{eq:rho_fractional} follows from Eqs.~\eqref{eq:Capute_def}--\eqref{eq:Ks_Fs}.

\section{Detailed singularity analysis of the Prabhakar-renewal $W_{yz}$ block}
\label{app:Wyz_singularity_analysis}
This appendix develops the complex-analytic structure of the exact $W_{yz}$ resolvent underlying the long-time results of
Sec.~\ref{sec:Prabhakar_asymptotics}.  The scalar Prabhakar renewal law, its normalization, and its scalar waiting-time and survival asymptotics
were derived in Sec.~\ref{subsec:ASLS} using the standard Prabhakar Laplace-transform structure and CM admissibility conditions~\cite{GarraGarrappa2018,MainardiGarrappa2015,GorskaHorzelaBratekDattoliPenson2018}. Hence they are not rederived here. Instead, we fix the physical-sheet branch conventions, determine the pole and branch-cut singularities of the reduced resolvent, and use the corresponding inverse-Laplace decomposition to obtain the storage--return asymptotics. The isolated poles describe exponential or damped-oscillatory transient contributions, whereas the branch cut at the origin determines the algebraic tail. The appendix closes by contrasting this fractional branch-cut structure with the Poisson-renewal limit. Throughout Secs.~\ref{app:branch_renewal}--\ref{app:leading_tail}, we assume $0<\alpha<1$, $\lambda>0$ and $\gamma>0$. For the memoryless case in Sec.~\ref{app:Poisson_resolvent_pole},  $\alpha=\gamma=1$ is assumed. The contour analysis below uses the CM subdomain specified by Eq.~\eqref{eq:11ee}, together with the stated branch and pole conditions. The broader domain of probabilistic admissibility is discussed in Sec.~\ref{subsec:ASLS}.
\subsection{Branch convention and scalar renewal-resolvent kernel}
\label{app:branch_renewal}
We use the normalized waiting-time transform $\widetilde w(s)$ in Eq.~\eqref{eq:11e} and the scalar renewal-resolvent kernel $\widetilde K(s)$ defined by Eq.~\eqref{eq:30a}.

The fractional power is defined on the principal Laplace sheet,
\begin{equation}
s^\alpha=\exp[\alpha\log s],
\qquad
\arg s\in(-\pi,\pi),
\label{eq:app_principal_branch}
\end{equation}
with principal cut
\begin{equation}
\mathcal C_0=(-\infty,0].
\end{equation}
We use the principal logarithm and its associated principal complex
power, with branch cut on the negative real axis~\cite{OlverEtAl2010}. The resulting negative-axis boundary structure of the Prabhakar
transform is discussed in Refs.~\cite{MainardiGarrappa2015,GarraGarrappa2018}, while the one-sided stable-law representation underlying the normalized renewal law is given in Ref.~\cite{PensonGorska2010}.
For \(s=-r\pm i0\), \(r>0\), define
\begin{equation}
\xi(r)=\frac{r^\alpha}{\lambda},
\qquad
Z_\pm(r)=1+\xi(r)e^{\pm i\pi\alpha}.
\end{equation}
Writing
\begin{equation}
Z_\pm(r)=R(r)e^{\pm i\varphi(r)},
\end{equation}
with
\begin{equation}
R(r)=\left[1+2\xi(r)\cos(\pi\alpha)+\xi(r)^2\right]^{1/2},
\end{equation}
and
\begin{equation}
\varphi(r)
=
\arg\!\left[1+\xi(r)e^{i\pi\alpha}\right],
\end{equation}
the boundary values of the waiting-time transform are
\begin{equation}
\widetilde w_\pm(-r)
=
R(r)^{-\gamma}e^{\mp i\gamma\varphi(r)}.
\label{eq:app_w_boundary}
\end{equation}

\begin{figure*}[t]
  \centering
  \includegraphics[width=\textwidth]{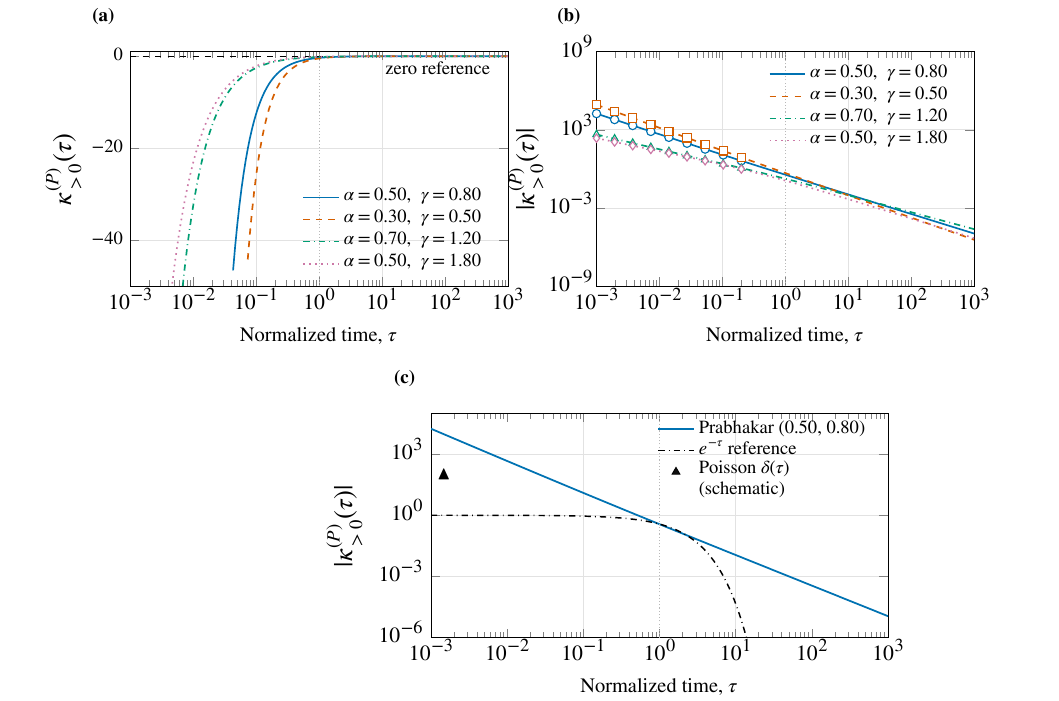}
  \caption{Dimensionless scalar Prabhakar renewal-resolvent kernel, with $\tau=\lambda^{1/\alpha}t$ and $\kappa_{>0}^{(P)}(\tau)=\lambda^{-2/\alpha}K_{>0}^{(P)}(\tau/\lambda^{1/\alpha})$.
(a) Signed positive-time representative for the indicated $(\alpha,\gamma)$ values, excluding contributions supported at $t=0$. (b) Its magnitude; open markers show an independent renewal-series
evaluation. (c) Magnitude for $(\alpha,\gamma)=(0.5,0.8)$ and $\lambda=1$, compared with the regular exponential-memory reference $e^{-\tau}$. The filled triangle schematically denotes the local Poisson term $\delta(\tau)$, which has no positive-time kernel curve. The vertical dotted line at $\tau=1$ marks the characteristic renewal scale. The signed kernel is not a probability density; complete positivity follows from the renewal-history construction. ChatGPT (OpenAI; GPT-5 Sol Ultra) generated the Python evaluation and plotting code from R.U.Erdogan's theoretical and numerical specifications; R.U.Erdogan reviewed it against the kernel expressions and numerical methods.}
\label{fig:sim2_kernel_diagnostics}
\end{figure*}

The time-domain diagnostics of the scalar kernel in Eq.~\eqref{eq:30a} are shown in Fig.~\ref{fig:sim2_kernel_diagnostics}. They compare the signed
Prabhakar contribution with a regular exponential-memory reference and the local Poisson-renewal limit, $\kappa_{\mathrm{Pois}}(\tau)=\delta(\tau)$.
Here $K_{>0}^{(P)}(t)$ denotes the positive-time representative in Eq.~\eqref{eq:memKt}; the subscript $>0$ excludes contributions supported at $t=0$, and the superscript $P$ identifies the
Prabhakar kernel. With $\tau=\lambda^{1/\alpha}t$, its dimensionless form is
\begin{equation}
\kappa_{>0}^{(P)}(\tau)
=
\lambda^{-2/\alpha}
K_{>0}^{(P)}(\tau/\lambda^{1/\alpha}).
\end{equation}
For the plotted cases with $0<\alpha<1$ and
$0<\beta=\alpha\gamma<1$, the short- and long-time
asymptotic forms are
\begin{equation}
\begin{aligned}
\kappa_{>0}^{(P)}(\tau)
&\propto
\frac{\tau^{\beta-2}}{\Gamma(\beta-1)},
&&\tau\to0^+,
\\
\kappa_{>0}^{(P)}(\tau)
&\propto
\frac{\tau^{\alpha-2}}
{\gamma\,\Gamma(\alpha-1)},
&&\tau\to\infty.
\end{aligned}
\end{equation}
Thus smaller $\beta$ gives a stronger negative short-time power-law singularity, whereas larger $\alpha$ gives slower
algebraic decay at long times. At fixed $\alpha$, the leading long-time magnitude scales as $1/\gamma$.
These asymptotic trends enter the coupled transmon dynamics through the unshifted and shifted kernels discussed in Sec.~\ref{subsec:shiftedK_Wyz}.

\subsection{Discontinuity of the scalar renewal-resolvent kernel}
With the principal-power convention fixed in Sec.~\ref{app:branch_renewal}, we follow the standard upper- and lower-boundary value construction for Prabhakar transforms on the negative real axis~\cite{MainardiGarrappa2015,Widder2010} and define the discontinuity across the principal cut by
\begin{equation}
\operatorname{Disc}\,f(-r)
:=
f(-r+i0)-f(-r-i0),
\qquad r>0.
\end{equation}
This is the Plemelj jump notation which is the standard upper--lower boundary value framework for functions analytic off an oriented cut~\cite{Muschelishvili2011}. From Eq.~\eqref{eq:app_w_boundary},
\begin{equation}
\operatorname{Disc}\,\widetilde w(-r)
=
-2i\,R(r)^{-\gamma}\sin[\gamma\varphi(r)].
\label{eq:app_disc_w}
\end{equation}
Let
\begin{equation}
A_\pm(r)=Z_\pm(r)^\gamma
=
R(r)^\gamma e^{\pm i\gamma\varphi(r)}.
\end{equation}
The boundary values of the kernel are
\begin{equation}
\widetilde K_\pm(-r)
=
\frac{-r}{A_\pm(r)-1}.
\end{equation}
Hence
\begin{equation}
\operatorname{Disc}\,\widetilde K(-r)
=
2i\,r\,
\frac{R(r)^\gamma\sin[\gamma\varphi(r)]}
{R(r)^{2\gamma}
-2R(r)^\gamma\cos[\gamma\varphi(r)]+1}.
\label{eq:app_disc_K}
\end{equation}

The discontinuity in Eq.~\eqref{eq:app_disc_K}, derived here, should be distinguished from the Bernstein spectral density of the Prabhakar response function derived in Ref.~\cite{MainardiGarrappa2015}. The latter is obtained from the boundary values of the Prabhakar transform itself, whereas  Eq.~\eqref{eq:app_disc_K} is the explicit negative-axis discontinuity of the renewal-resolvent kernel. It follows by applying the standard upper--lower
boundary-value construction to $\widetilde K(s)=s\widetilde m(s)$ in Eq.~\eqref{eq:30a}, with $\widetilde m(s)$ defined in Eq.~\eqref{eq:m_s}~\cite{MainardiGarrappa2015,
GarrappaMainardiMaione2016}. In the CM parameter domain of Eq.~\eqref{eq:11ee}, this renewal-density transform is a Stieltjes function~\cite{GorskaHorzela2021,GorskaHorzela2023}.

Equation~\eqref{eq:app_disc_K} is therefore the scalar boundary-value input used to construct the discontinuities of the
shifted kernels $K_0$ and $K_\Gamma$ and, through the reduced $W_{yz}$ resolvent defined in Sec.~\ref{subsec:shiftedK_Wyz}, the branch-cut contribution evaluated in Sec.~\ref{app:pole_branch_decomp}.

\subsection{Shifted kernels and analytic structure of the reduced $W_{yz}$ resolvent}
\label{subsec:shiftedK_Wyz}
The inverse-resolvent matrix $\mathcal M_{yz}(s)$ and its determinant $\Delta(s)$ are defined in Eqs.~\eqref{eq:Myz_exact} and~\eqref{eq:Delta_def}. The coupled $W_{yz}$ block derived in the main text depends on the two shifted scalar kernels $K_0(s)$ and $K_\Gamma(s)$ defined in Eqs.~\eqref{eq:K0_def} and~\eqref{eq:KGamma_def}, respectively. The two scalar kernel arguments appearing in that block follow from analytic functional calculus. Since $W_{yz}$ is invariant under $\mathcal L_Z=\text{diag}({-\Gamma_Z,0})$,
\begin{equation}
\begin{aligned}
\left.
\widetilde K(s-\mathcal L_Z)
\right|_{W_{yz}}
&=
\widetilde K
\left[
sI_{W_{yz}}-
\left.\mathcal L_Z\right|_{W_{yz}}
\right]
\\
&= \widetilde{K}\!\left(
\begin{pmatrix}
s+\Gamma_Z & 0 \\
0 & s
\end{pmatrix}
\right)
\\
&=
\begin{pmatrix}
\widetilde{K}(s+\Gamma_Z) & 0 \\
0 & \widetilde{K}(s)
\end{pmatrix}.
\end{aligned}
\end{equation}
in the ordered basis $(\sigma_y,\sigma_z)$, on a branch and spectral domain where the scalar kernel $z\mapsto\widetilde K(s-z)$ is analytic~\cite{Higham2008,Haase2006,Kato1995}.

Accordingly the unshifted kernel $K_0$ has a branch point at $s=0$, while $K_\Gamma$ has a shifted branch point at $s=-\Gamma_Z$. On the physical sheet, the corresponding cuts are
\begin{equation}
\mathcal C_0=(-\infty,0],
\qquad
\mathcal C_\Gamma=(-\infty,-\Gamma_Z].\label{eq:app_physical_cuts}
\end{equation}

For $\Gamma_Z>0$, the two cuts in Eq.~\eqref{eq:app_physical_cuts} are nested rather than disjoint: $\mathcal C_\Gamma\subset\mathcal C_0$. Hence the full reduced resolvent, expressed by Eq.~\eqref{eq:app_Myz_resolvent} in the following paragraph, is analyzed on cut-free components of the physical sheet, with a piecewise discontinuity along the negative real axis. This branch structure is used in Sec.~\ref{app:pole_equation} to restrict the isolated-pole condition to cut-free physical-sheet domains, in Sec.~\ref{app:pole_branch_decomp} to construct the piecewise branch-cut contribution to the inverse Laplace transform, and in Sec.~\ref{app:leading_tail} to isolate the contribution from the branch point at $s=0$. The latter produces the leading algebraic tail, whereas the contribution associated with the branch point at $s=-\Gamma_Z$ carries an additional factor $e^{-\Gamma_Z t}$ and is asymptotically subleading when the $s=0$ branch amplitude is nonzero.

The scalar trends in Fig.~\ref{fig:sim2_kernel_diagnostics} enter the reduced transmon dynamics through the unshifted and shifted kernels $K_0(s)$ and $K_\Gamma(s)$ of Eqs.~\eqref{eq:K0_def} and~\eqref{eq:KGamma_def}. The resulting coherence and storage--return amplitudes additionally depend on $\Gamma_Z$, the kick angle $\theta$, and the initial $W_{yz}$ component. 

For the contour analysis below, we expand the reduced resolvent defined in Eq.~\eqref{eq:full_reduced_resolvents} as
\begin{equation}
\begin{aligned}
\mathcal R_{yz}(s)
&:=\mathcal M_{yz}(s)^{-1}\\
&=\frac{1}{\Delta(s)}
\begin{pmatrix}
s+\delta K_0(s) & -\sin\theta\,K_0(s)\\
\sin\theta\,K_\Gamma(s) &
s+\Gamma_Z+\delta K_\Gamma(s)
\end{pmatrix},
\label{eq:app_Myz_resolvent}
\end{aligned}
\end{equation}
where \(\Delta(s)=\det\mathcal M_{yz}(s)\) is given in Eqs.~\eqref{eq:Delta_def} and~\eqref{eq:Delta_simplified}. Thus the $\sigma_y$ drift eigenmode is associated with the shifted kernel $K_\Gamma(s)$, whereas the zero-drift $\sigma_z$ eigenmode is associated with the unshifted kernel $K_0(s)$. Their physical storage--return interpretation is discussed in Sec.~\ref{sec:physical_interpretation}.

For the branch-cut analysis below, the upper and lower physical-sheet boundary values are obtained by substituting the corresponding boundary values of $K_0$ and $K_\Gamma$ into Eq.~\eqref{eq:app_Myz_resolvent}. Their difference determines the
matrix discontinuity $\operatorname{Disc}\left[\mathcal M_{yz}(s)^{-1}\mathbf r_{yz}\right]$ that enters the inverse-Laplace cut integral in Sec.~\ref{app:pole_branch_decomp}.

\subsection{Pole equation}\label{app:pole_equation}
To characterize the isolated poles summarized in Sec.~\ref{sec:Prabhakar_asymptotics}, consider the zeros of the exact
determinant
\begin{equation}
\Delta(s)=0
\label{eq:app_pole_equation}
\end{equation}
away from the branch cuts of \(K_0\) and \(K_\Gamma\). Introducing
\begin{equation}
A(s)=\left(1+\frac{s^\alpha}{\lambda}\right)^\gamma,
\qquad
A_\Gamma(s)=
\left(1+\frac{(s+\Gamma_Z)^\alpha}{\lambda}\right)^\gamma,
\end{equation}
one has
\begin{equation}
K_0(s)=\frac{s}{A(s)-1},
\qquad
K_\Gamma(s)=\frac{s+\Gamma_Z}{A_\Gamma(s)-1}.
\end{equation}
With 
\begin{equation}
D(s):=[A(s)-1][A_\Gamma(s)-1],\label{eq:Ds}
\end{equation}
direct substitution into the determinant gives
\begin{subequations}
\label{eq:app_Delta_factorized}
\begin{align}
\Delta(s)
&=
\frac{s(s+\Gamma_Z)}{D(s)}\,Q(s),
\label{eq:app_Delta_factorized_a}
\\
Q(s)
&=
\bigl[A_\Gamma(s)-\cos\theta\bigr]
\bigl[A(s)-\cos\theta\bigr]
+\sin^2\theta .
\label{eq:app_Delta_factorized_b}
\end{align}
\end{subequations}
Thus, at points where $D(s)\neq0$, and after the canceled factors $s=0$ and $s=-\Gamma_Z$ have been examined separately in the
original determinant, the nontrivial candidate poles satisfy
\begin{equation}
Q(s)=0.
\label{eq:app_transcendental_pole}
\end{equation}
Only noncanceled zeros of Eq.~\eqref{eq:app_transcendental_pole} that lie on the chosen physical sheet and away from the branch cuts are isolated poles of the reduced resolvent, in the sense of the analytic-operator framework of Ref.~\cite{GohbergSigal1971}. On each cut-free component of the chosen physical sheet, $\mathcal M_{yz}(s)$, which is obtained in Eq.~\eqref{eq:Myz_exact}, is an analytic matrix-valued function. Its isolated characteristic values are the points at which it loses invertibility; in finite dimension their multiplicities agree with the multiplicities of the corresponding zeros of Eq.~\eqref{eq:Delta_def}~\cite{GohbergSigal1971}. The local factorization at a normal characteristic value determines the order of the corresponding pole of $\mathcal R_{yz}(s)$ defined in Eq.~\eqref{eq:full_reduced_resolvents}~\cite{GohbergSigal1971}.

Standard analytic matrix-function and contour-resolvent treatments are reviewed in Refs.~\cite{GuttelTisseur2017,Beyn2012}. Zeros of Eq.~\eqref{eq:Ds}, canceled factors, branch points, and cut boundary values must be tested directly in the original determinant and are not automatically isolated poles. 

\subsection{Pole--branch-cut decomposition}
\label{app:pole_branch_decomp}
Near an isolated normal characteristic value, the principal part of the inverse matrix function admits a finite Laurent expansion determined by
the associated root-vector chains~\cite{GohbergSigal1971}. The Bromwich contour may be deformed into isolated pole residue contributions and Hankel contours around the physical cuts~\cite{Doetsch1974,MainardiGarrappa2015}. Let
\begin{equation}
\widetilde{\mathbf c}_{yz}(s)
=
\mathcal M_{yz}(s)^{-1}\mathbf r_{yz},
\qquad
\mathbf r_{yz}=(r_y,r_z)^{\mathsf T}.\label{eq:laplace_cyz_ryz}
\end{equation}
Then the inverse Laplace transform can be written as a sum of isolated pole residues and branch-cut contributions,
\begin{equation}
\mathbf c_{yz}(t)
=
\sum_{s_p}
\operatorname{Res}
\left[
e^{st}\widetilde{\mathbf c}_{yz}(s),s=s_p
\right]
+
\mathbf c_{\mathrm{bc}}(t),
\label{eq:app_pole_branch_decomp}
\end{equation}
where $s_p$ are poles satisfying Eq.~\eqref{eq:app_pole_equation}. The branch-cut term is obtained from the discontinuity of $\widetilde{\mathbf c}_{yz}(s)$ across the cuts, for example
\begin{equation}
\mathbf c_{\mathrm{bc}}(t)
=
\frac{1}{2\pi i}
\int_{\mathcal C}
e^{st}
\operatorname{Disc}
\left[
\mathcal M_{yz}(s)^{-1}\mathbf r_{yz}
\right]\,ds,
\label{eq:app_branch_integral}
\end{equation}
where $\mathcal C=(-\infty,0]$ is oriented from $0$ toward $-\infty$, consistently with the upper-minus-lower definition of $\operatorname{Disc}$. The shared segment
of the two cuts is integrated only once, using the discontinuity of the full reduced resolvent. On $(-\Gamma_Z,0]$, only $K_0$ is discontinuous, whereas on $(-\infty,-\Gamma_Z]$ both $K_0$ and $K_\Gamma$ contribute. The scalar discontinuity of the unshifted kernel is given by Eq.~\eqref{eq:app_disc_K}; for the shifted kernel, the same formula is evaluated at $s+\Gamma_Z$. Substitution of these upper and lower
kernel values into the reduced-resolvent formula Eq.~\eqref{eq:app_Myz_resolvent} yields $\operatorname{Disc}\left[\mathcal M_{yz}(s)^{-1}\mathbf r_{yz}\right]$ which is the matrix discontinuity entering Eq.~\eqref{eq:app_branch_integral}. The $\alpha$- and $\gamma$-dependent scalar behavior illustrated in Fig.~\ref{fig:sim2_kernel_diagnostics} enters the coupled dynamics through the upper and lower boundary values of $K_0$ and $K_\Gamma$ and therefore through the discontinuity of the reduced resolvent $M_{yz}^{-1}$.

Equations~\eqref{eq:app_pole_branch_decomp} and \eqref{eq:app_branch_integral} provide the exact contour decomposition invoked in Sec.~\ref{sec:Prabhakar_asymptotics}. The contribution of the origin cut is evaluated explicitly in the next subsection.

\subsection{Reconstruction of the directional algebraic tails from the $s=0$ branch cut}
\label{app:leading_tail}
We reconstruct the $y\to z$ and $z\to y$ transfer tails from the matrix elements $[\mathcal R_{yz}(s)]_{21}$ and $[\mathcal R_{yz}(s)]_{12}$, respectively, of the exact
reduced resolvent in Eq.~\eqref{eq:app_Myz_resolvent}. Within the parameter domain of Eq.~\eqref{eq:11ee}, fix $0<\alpha<1$, $\lambda>0$, $\Gamma_Z>0$, and $\sin\theta\neq0$.
With $\delta$ defined in Eq.~\eqref{eq:delta_def}, set
\begin{equation}
k_\Gamma:=K_\Gamma(0),\qquad
D_0:=\delta(\Gamma_Z+2k_\Gamma),
\end{equation}
where $K_\Gamma(0)$ is given in
Eq.~\eqref{eq:KGamma_zero}.
These assumptions imply $k_\Gamma>0$ and $D_0>0$.

The two transfer entries share the denominator $\Delta(s)$. The binomial expansion (see Eq.~(4.6.7) of~\cite{OlverEtAl2010}) gives
\begin{equation}
\frac{s}{K_0(s)}
=\left(1+\frac{s^\alpha}{\lambda}\right)^\gamma-1
=\frac{\gamma}{\lambda}s^\alpha+O(s^{2\alpha}),
\end{equation}
while analyticity of the shifted kernel near the origin gives $K_\Gamma(s)=k_\Gamma+O(s)$. Dividing Eq.~\eqref{eq:Delta_simplified} by $K_0(s)$ therefore yields
\begin{equation}
\begin{aligned}
\frac{\Delta(s)}{K_0(s)}
={}&\delta[s+\Gamma_Z+2K_\Gamma(s)]\\
&+[s+\Gamma_Z+\delta K_\Gamma(s)]
  \frac{s}{K_0(s)}\\
={}&D_0+\frac{\gamma}{\lambda}
  (\Gamma_Z+\delta k_\Gamma)s^\alpha
  +o(s^\alpha).
\end{aligned}
\label{eq:app_reverse_denominator}
\end{equation}
The discarded terms are $O(s)+O(s^{2\alpha})$. Both are $o(s^\alpha)$ for $0<\alpha<1$, since
$|s|^{1-\alpha}\to0$ and $|s|^\alpha\to0$ as $s\to0$ (see Eqs.~(2.1.2) and~(2.1.3) of~\cite{OlverEtAl2010}).
The convergent binomial expansion, analyticity of $K_\Gamma(s)$, and $D_0>0$ ensure that these expansions
extend uniformly to both sides of a sufficiently short origin cut.

For the $y\to z$ transfer, expanding $[\mathcal R_{yz}(s)]_{21}$ gives
\begin{equation}
\begin{aligned}
\widetilde T_{ZY}^{(P)}(s)
&=\frac{\sin\theta\,K_\Gamma(s)}{\Delta(s)}\\
&=b_{ZY}s^{\alpha-1}+o(s^{\alpha-1}).
\end{aligned}
\label{eq:app_storage_small_s}
\end{equation}
For the $z\to y$ transfer, expanding $[\mathcal R_{yz}(s)]_{12}$ gives
\begin{equation}
\begin{aligned}
\widetilde T_{YZ}^{(P)}(s)
&=-\frac{\sin\theta\,K_0(s)}{\Delta(s)}\\
&=-\frac{\sin\theta}{D_0}
  +b_{YZ}s^\alpha+o(s^\alpha),
\end{aligned}
\label{eq:app_reverse_small_s}
\end{equation}
where
\begin{equation}
\begin{aligned}
b_{ZY}
&:=\frac{\gamma\sin\theta\,k_\Gamma}{\lambda D_0},\\
b_{YZ}
&:=\frac{\gamma\sin\theta
  (\Gamma_Z+\delta k_\Gamma)}{\lambda D_0^2}.
\end{aligned}
\end{equation}
Thus the first nonanalytic powers are $s^{\alpha-1}$ and $s^\alpha$, respectively. The constant in Eq.~\eqref{eq:app_reverse_small_s} has no local cut discontinuity. As a coefficient of a local small-$s$ expansion, it does not establish a separate Dirac delta contribution to the physical transfer map.

Choose a sufficiently small $0<\varepsilon<\Gamma_Z$. For $s=-r\pm i0$, $0<r<\varepsilon$, the principal
logarithmic boundary values and power convention (see Eqs.~(4.2.7) and~(4.2.28) of~\cite{OlverEtAl2010}) give
\begin{equation}
(-r\pm i0)^\eta=r^\eta e^{\pm i\pi\eta},
\qquad
\eta\in\{\alpha-1,\alpha\}.
\end{equation} 
Taking the upper-minus-lower difference in
Eqs.~\eqref{eq:app_storage_small_s}
and~\eqref{eq:app_reverse_small_s} therefore gives
\begin{equation}
\begin{aligned}
\operatorname{Disc}\widetilde T_{ZY}^{(P)}(-r)
={}&-2i\,b_{ZY}\sin(\pi\alpha)r^{\alpha-1}\\
&+o(r^{\alpha-1}),\\
\operatorname{Disc}\widetilde T_{YZ}^{(P)}(-r)
={}& 2i\,b_{YZ}\sin(\pi\alpha)r^\alpha\\
&+o(r^\alpha).
\end{aligned}
\label{eq:app_directional_discontinuities}
\end{equation}
The remainders are uniform over the two boundary
values as $r\to0^+$.

For either $(i,j)=(Z,Y)$ or $(Y,Z)$, let
$T_{ij,\mathrm{bc},0}^{(P)}(t)$ denote the contribution
of the cut segment $-\varepsilon<s<0$.
With the orientation in
Eq.~\eqref{eq:app_branch_integral}, this contribution is
\begin{equation}
\begin{aligned}
T_{ij,\mathrm{bc},0}^{(P)}(t)
={}&-\frac{1}{2\pi i}
\int_0^\varepsilon e^{-rt}
\operatorname{Disc}
\widetilde T_{ij}^{(P)}(-r)\,dr .
\end{aligned}
\label{eq:app_origin_cut_integral}
\end{equation}
Watson's lemma (see Sec.~2.3(ii), Eqs.~(2.3.7) and~(2.3.8) of~\cite{OlverEtAl2010}) applies to both local integrals because $\alpha-1>-1$ and $\alpha>-1$. Euler's gamma integral and reflection formula (see Eqs.~(5.2.1) and~(5.5.3) of~\cite{OlverEtAl2010}) give, for the two exponents specified above,
\begin{equation}
\begin{aligned}
\int_0^\infty e^{-rt}r^\eta\,dr
&=\Gamma(1+\eta)t^{-1-\eta},\\
-\frac{\sin(\pi\eta)}{\pi}\Gamma(1+\eta)
&=\frac{1}{\Gamma(-\eta)}.
\end{aligned}
\end{equation}
These identities convert the two discontinuities in Eq.~\eqref{eq:app_directional_discontinuities} into
the corresponding algebraic tails. For fixed $\varepsilon>0$, extending the upper limits of the leading power integrals to infinity introduces only exponentially small errors.

Provided the cut integral in Eq.~\eqref{eq:app_branch_integral} converges absolutely for some positive time, its contribution from $r\geq\varepsilon$, including the shifted threshold at $r=\Gamma_Z$, is exponentially small as $t\to\infty$. Under the stated contour assumptions, with exponentially decaying pole contributions, the full
transfer coefficients therefore satisfy
\begin{align}
T_{ZY}^{(P)}(t)
&=\frac{b_{ZY}}{\Gamma(1-\alpha)}t^{-\alpha}
  +o(t^{-\alpha}),
\label{eq:app_storage_tail}\\
T_{YZ}^{(P)}(t)
&=\frac{b_{YZ}}{\Gamma(-\alpha)}t^{-1-\alpha}
  +o(t^{-1-\alpha}),
\label{eq:app_reverse_tail}
\end{align}
as $t\to\infty$.
Equation~\eqref{eq:app_storage_tail} recovers Eq.~\eqref{eq:T_ZY_P}, while Eq.~\eqref{eq:app_reverse_tail} establishes the next algebraic order of the reverse transfer. Both coefficients follow from the exact reduced
resolvent in Eq.~\eqref{eq:app_Myz_resolvent}.

For a general initial vector $\mathbf r_{yz}=(r_y,r_z)^{\mathsf T}$, the same componentwise inversion applied to
Eq.~\eqref{eq:branch_amplitude_yz} gives 
\begin{equation}
\begin{aligned}
\mathbf c_{\mathrm{bc},0}(t)
={}&\frac{\mathbf a_{\mathrm{br}}}
{\Gamma(1-\alpha)}t^{-\alpha}\\
&+o(t^{-\alpha}).
\end{aligned}
\label{eq:app_origin_cut_tail}
\end{equation}
Under the same contour and pole assumptions, this also yields the full-response result in
Eq.~\eqref{eq:longtime_vector_yz}.

For $0<\theta<\pi$, both $b_{ZY}$ and $b_{YZ}$ are positive. Since $\Gamma(1-\alpha)>0$ and
$\Gamma(-\alpha)<0$, the leading storage coefficient is positive and the leading reverse-transfer
coefficient is negative. When $\sin\theta=0$, both off-diagonal transfers vanish identically. The cases $\alpha=1$ and $\Gamma_Z=0$ require separate analysis.

\subsection{Poisson-renewal resolvent and pole structure}
\label{app:Poisson_resolvent_pole}
For the memoryless renewal law introduced in Eqs.~\eqref{eq:w_exp_summary} and \eqref{eq:K_exp_summary}, one has
\begin{equation}
K_0^{(\mathrm{exp})}(s)
=
K_\Gamma^{(\mathrm{exp})}(s)
=
\nu .
\end{equation}
The reduced inverse-resolvent matrix is therefore
\begin{equation}
\mathcal M_{yz}^{(\mathrm{exp})}(s)
=
\begin{pmatrix}
s+\Gamma_Z+\delta\nu & \nu\sin\theta\\
-\nu\sin\theta       & s+\delta\nu
\end{pmatrix},
\label{eq:app_poisson_block}
\end{equation}
with determinant
\begin{equation}
\Delta_{\mathrm{exp}}(s)
=
(s+\Gamma_Z+\delta\nu)(s+\delta\nu)
+
\nu^2\sin^2\theta .
\label{eq:app_poisson_determinant}
\end{equation}
The corresponding reduced resolvent is
\begin{equation}
\begin{aligned}
\mathcal R_{yz}^{(\mathrm{exp})}(s)
&=
\left[
\mathcal M_{yz}^{(\mathrm{exp})}(s)
\right]^{-1}\\
&=
\frac{1}{\Delta_{\mathrm{exp}}(s)}
\begin{pmatrix}
s+\delta\nu & -\nu\sin\theta\\
\nu\sin\theta & s+\Gamma_Z+\delta\nu
\end{pmatrix}.
\label{eq:app_poisson_resolvent}
\end{aligned}
\end{equation}
Its poles are
\begin{equation}
s_\pm
=
-\frac{\Gamma_Z+2\delta\nu}{2}
\pm
\frac12
\sqrt{
\Gamma_Z^2-4\nu^2\sin^2\theta
}.
\label{eq:app_poisson_poles}
\end{equation}
For distinct real poles, the inverse Laplace transform is a finite sum of exponential terms; a complex-conjugate pair gives damped
oscillations. At the repeated-pole condition
\begin{equation}
\Gamma_Z^2=4\nu^2\sin^2\theta,
\end{equation}
a polynomial prefactor proportional to $t$ may multiply the common exponential. Since $\mathcal R_{yz}^{(\mathrm{exp})}(s)$ is rational, the Poisson case contains no branch-cut contribution~\cite{Doetsch1974}. This is the technical basis of the memoryless-renewal comparison summarized in Sec.~\ref{sec:Prabhakar_asymptotics}. 

\section{Local convergence and Prabhakar-convolution expansion of the
forward renewal memory}
\label{app:prabhakar_forward_series}

This appendix supplies the two technical results used, but not proved, in the main text: local convergence of the renewal-order expansion introduced in Sec.~\ref{sec:PMRD} and the explicit drift-dressed Prabhakar representation of the memory term appearing in Eq.~\eqref{eq:rho_fractional}. The one-interval law and its material motivation are discussed in Sec.~\ref{sec:PMRD}, the chronological construction and proof of CP are given in Sec.~\ref{subsec:ren_map_equ}, Appendix~\ref{app:renewal_structure}, the singularity analysis and storage--return physics are given in Appendix~\ref{app:Wyz_singularity_analysis} and Sec.~\ref{sec:dynamics_decoherence}, respectively.

With $\mathcal A(s)=s\mathcal I-\mathcal L_Z$ from Sec.~\ref{sec:kernel_resolvent_formulation} and
$\mathcal J_X:=\Phi_X-\mathcal I$, we expand the forward product $\mathcal J_X\widetilde K(\mathcal A(s))\widetilde\rho(s)$ occurring in Eq.~\eqref{eq:exact_kernel_resolvent} whose time-domain counterpart is~\eqref{eq:rho_fractional}.

\subsection{Renewal-order expansion and local convergence}
This subsection establishes local-$L^1$ convergence of the series defining the renewal density $m(t)$ in Eq.~\eqref{eq:m_t} and justifies its termwise differentiation
on every compact interval $[\varepsilon,T]\subset(0,\infty)$. The first result ensures a finite expected renewal count on finite intervals and supports the drift-dressed convolution expansion below;
the second establishes the ordinary positive-time kernel series in Eq.~\eqref{eq:memKt}.

The positivity and transform identity used below follow from the generalized Mittag--Leffler renewal law. Specifically, with $(\nu,\delta,m)=(\alpha,\gamma,n)$, Eq.~(2.5) of Ref.~\cite{CahoyPolito2013} gives
$\widetilde q_n(s)=[\widetilde w(s)]^n$, while Eq.~(2.6) identifies $q_n$ as the nonnegative density of the $n$th renewal epoch. Moreover, Eq.~(2.4) of the same reference implies $0<\widetilde w(s)<1$ for every $s>0$. Define
\begin{align}
\widetilde q_n(s)
&:=[\widetilde w(s)]^n
=\frac{\lambda^{n\gamma}}
{(s^\alpha+\lambda)^{n\gamma}},
\nonumber\\
q_n(t)
&:=\lambda^{n\gamma}t^{\alpha n\gamma-1}
E_{\alpha,\alpha n\gamma}^{n\gamma}
(-\lambda t^\alpha),
\qquad n\geq1,\quad t>0,
\label{eq:D_qn}
\end{align}
and
\begin{equation}
m_{N_0}(t):=\sum_{n=1}^{N_0}q_n(t).
\end{equation}

To bound the integral in Eq.~\eqref{app:m_N_laplace_bound} over $[0,T]$ by a Laplace transform, for $T,s>0$ positivity and the elementary estimate $\mathbf{1}_{[0,T]}(t)\leq e^{s T}e^{-s t}$ give
\begin{align}
\mathbb{E}[\min\left\{N(T),N_0\right\}]&=\int_0^T m_{N_0}(t)\,dt
\leq
e^{s T}
\sum_{n=1}^{N_0}\widetilde q_n(s)
\nonumber\\
&=
e^{s T}
\sum_{n=1}^{N_0}[\widetilde w(s)]^n
\nonumber\\
&\leq
e^{s T}
\frac{\widetilde w(s)}
{1-\widetilde w(s)}.\label{app:m_N_laplace_bound}
\end{align}
The first equality follows from $\text{Pr}[t_n\leq T]=\text{Pr}[N(T)\geq n]$ and the finite tail-sum identity, with $t_n$ denoting the $n$th renewal epoch. The first inequality follows directly from the indicator estimate and the definition of the Laplace transform. The second equality uses the $n$th-renewal transform in Eq.~(2.5) of Ref.~\cite{CahoyPolito2013}, and the final inequality is the elementary bound by the corresponding infinite geometric series. Thus, this finite-sum estimate is derived here from the cited renewal transform.

Since $\lim_{N_0\to\infty}m_{N_0}=m:=\sum_{n\geq1}q_n$ and $\lim_{N_0\to\infty}\min\left\{N(T),N_0\right\}=N(T)$, the monotone convergence theorem (see Theorem~2.14 of Ref.~\cite{Folland1999}), yields
\begin{align}
\mathbb{E}[N(T)]&=\lim_{N_0\to\infty}\mathbb{E}[\min\left\{N(T),N_0\right\}]\nonumber\\&=\lim_{N_0\to\infty}\int_0^T m_{N_0}(t)\,dt=\int_0^T m(t)\,dt\nonumber\\
&\leq
e^{s T}
\frac{\widetilde w(s)}
{1-\widetilde w(s)}
<\infty.
\end{align}
Hence $m\in L^1(0,T)$. Furthermore, $m-m_{N_0}=\sum_{n=N_0+1}^{\infty}q_n\geq0$. Applying Tonelli's theorem (see Theorem~2.37(a) of Ref.~\cite{Folland1999}) gives
\begin{align}
\|m-m_{N_0}\|_{L^1(0,T)}
&=
\int_0^T\sum_{n=N_0+1}^{\infty}q_n(t)\,dt
\nonumber\\
&\leq
e^{s T}
\sum_{n=N_0+1}^{\infty}\widetilde q_n(s)
\nonumber\\
&=
e^{s T}
\frac{[\widetilde w(s)]^{N_0+1}}
{1-\widetilde w(s)}
\xrightarrow[N_0\to\infty]{}0 .
\label{eq:D_L1_tail}
\end{align}
Here the inequality again follows from the indicator estimate, whereas the last equality combines Eq.~(2.5) of Ref.~\cite{CahoyPolito2013} with the scalar geometric-tail identity. Because $0<\widetilde w(s)<1$, the
final expression tends to zero. Thus $m_{N_0}\to m$ in $L^1_{\mathrm{loc}}(\mathbb R_+)$. Consequently,
\begin{equation}
m(t)=\sum_{n=1}^{\infty}
\lambda^{n\gamma}t^{\alpha n\gamma-1}
E^{n\gamma}_{\alpha,\alpha n\gamma}(-\lambda t^\alpha)
\quad\text{in }L^1_{\mathrm{loc}}(\mathbb R_+).
\label{eq:D_m_series}
\end{equation}
Thus Eq.~\eqref{eq:D_m_series} is the local-$L^1$ form of the renewal-density definition in Eq.~\eqref{eq:m_t}, obtained here from Eqs.~\eqref{eq:D_qn} and \eqref{eq:D_L1_tail}. The associated integrated renewal sums are given in Eqs.~(2.33), (2.35), and (2.36) of Ref.~\cite{CahoyPolito2013}. Equations~\eqref{app:m_N_laplace_bound}--\eqref{eq:D_L1_tail} provide the bridge from Eq.~\eqref{eq:m_t} to the well-defined drift dressed Prabhakar convolution series used in the master equation. The estimate does not imply pointwise or locally uniform convergence at $t=0$; for $\alpha\gamma<1$, its leading term is unbounded there but remains locally integrable.

Equation~\eqref{eq:D_L1_tail} controls the integrated renewal-series tail but does not by itself justify ordinary
termwise differentiation. We therefore establish uniform convergence of both the series and its derivative series on
$[\varepsilon,T]\subset(0,\infty)$ for the normalized law of Sec.~\ref{subsec:ASLS}. Each $q_n$ in
Eq.~\eqref{eq:D_qn} is smooth for $t>0$. Proposition~1, Eq.~(5), of Ref.~\cite{GarraGarrappa2018},
written with dummy parameters $b,c$, states the identity
\begin{equation}
\begin{aligned}
\alpha z\frac{d}{dz}E_{\alpha,b}^{c}(z)
={}&E_{\alpha,b-1}^{c}(z)
\\
&+(1-b)E_{\alpha,b}^{c}(z),\qquad z\neq0.
\end{aligned}
\label{eq:D_Prabhakar_derivative_identity}
\end{equation}
Substituting $(b,c,z)=(n\beta,n\gamma,-\lambda t^\alpha)$ and
applying the product and chain rules identifies $q_n'(t)$ with
the $n$th summand in Eq.~\eqref{eq:memKt}.

Choose $\sigma>0$ and an integer $n_\star$ with $n_\star\beta>2$.
For $s=\sigma+i\omega$, positivity of $w$ and
$\operatorname{Re}(s^\alpha)>0$ on the principal branch give
\begin{align}
a_\sigma&:=\widetilde w(\sigma)\in(0,1),\nonumber\\
|\widetilde w(s)|&\leq a_\sigma,
\qquad
|\widetilde w(s)|\leq\lambda^\gamma|s|^{-\beta}.
\label{eq:D_vertical_transform_bounds}
\end{align}
The bound by $a_\sigma<1$ controls the sum over renewal orders; the second bound, following from
$|s^\alpha+\lambda|\geq|s|^\alpha$, controls the large-$|\omega|$ behavior. One time derivative contributes
a factor $s$ to the inverse-transform integrand, so $n_\star\beta>2$ ensures
\begin{equation}
\begin{aligned}
C_{\sigma,n_\star}&:=
\frac{1}{2\pi}\int_{\mathbb R}|\sigma+i\omega|\,
\\
&\qquad\times
|\widetilde w(\sigma+i\omega)|^{n_\star}\,d\omega
<\infty,
\end{aligned}
\label{eq:D_derivative_contour_constant}
\end{equation}
since its integrand is bounded by
$\lambda^{n_\star\gamma}|\sigma+i\omega|^{1-n_\star\beta}$.

For $n\geq n_\star$, the complex inversion formula in Theorem~24.4, Sec.~24, Eq.~(15) of Ref.~\cite{Doetsch1974},
applied to the transform of the smooth density $q_n$ in Eq.~\eqref{eq:D_qn}, gives the case $j=0$ below.
On $[\varepsilon,T]$, the differentiated integrand has the $t$-independent bound $e^{\sigma T}|s|\,|\widetilde w(s)|^{n_\star}
a_\sigma^{n-n_\star}$, which is integrable in $\omega$ by Eq.~\eqref{eq:D_derivative_contour_constant}. Thus Theorem~2.27(b) of Ref.~\cite{Folland1999}
justifies differentiation under the integral and yields the case $j=1$:
\begin{equation}
\begin{aligned}
q_n^{(j)}(t)&=\frac{1}{2\pi}\int_{\mathbb R}
e^{(\sigma+i\omega)t}(\sigma+i\omega)^j
\\
&\qquad\times
[\widetilde w(\sigma+i\omega)]^n\,d\omega,
\end{aligned}
\label{eq:D_qn_differentiated_inversion}
\end{equation}
where $q_n^{(0)}=q_n$. Factoring
$\widetilde w^{\,n}=\widetilde w^{\,n_\star}
\widetilde w^{\,n-n_\star}$ and using
Eq.~\eqref{eq:D_vertical_transform_bounds} together with
$|\sigma+i\omega|\geq\sigma$, for $N_0\geq n_\star$ we obtain
\begin{equation}
\begin{aligned}
&\sum_{n=N_0+1}^{\infty}
\sup_{t\in[\varepsilon,T]}|q_n^{(j)}(t)|
\\
&\quad\leq
e^{\sigma T}\sigma^{j-1}C_{\sigma,n_\star}
\frac{a_\sigma^{N_0+1-n_\star}}{1-a_\sigma}
\\[-2pt]
&\quad\xrightarrow[N_0\to\infty]{}0,
\qquad j=0,1.
\end{aligned}
\label{eq:D_differentiated_series_tail}
\end{equation}
The finitely many terms with $n<n_\star$ are smooth on $[\varepsilon,T]$. Equation~\eqref{eq:D_differentiated_series_tail}
and the Weierstrass $M$-test therefore give absolute and uniform convergence of both $\sum_n q_n$ and $\sum_n q_n'$ there.
Passing to the limit in the finite-sum identity $m_{N_0}(t)-m_{N_0}(\varepsilon)=\int_\varepsilon^t m_{N_0}'(u)\,du$
then establishes
\begin{equation}
m\in C^1((0,\infty)),\qquad
\dot m(t)=\sum_{n=1}^{\infty}q_n'(t),\quad t>0.
\label{eq:D_ordinary_differentiated_series}
\end{equation}

Together with Eq.~\eqref{eq:D_Prabhakar_derivative_identity}, Eq.~\eqref{eq:D_ordinary_differentiated_series}
justifies Eq.~\eqref{eq:memKt} as a series converging absolutely and uniformly on every compact positive-time
interval. This sum equals the restriction to $t>0$ of the causal kernel defined in Sec.~\ref{subsec:Physmot_fract}.

\subsection{Drift-dressed Prabhakar-convolution series}
The purpose of this subsection is to promote the locally convergent scalar renewal expansion to the finite-dimensional Liouville space by incorporating the longitudinal drift, thereby obtaining the explicit drift-dressed Prabhakar-convolution representation of the generalized-Caputo memory operator used in the main text.
For $\beta,\delta>0$ and locally integrable $H$, define
\begin{align}
&\mathcal E^{\delta}_{\alpha,\beta,-\lambda}
[\mathcal L_Z]H(t)
\nonumber\\
&\quad:=\int_0^t
(t-\tau)^{\beta-1}
E^{\delta}_{\alpha,\beta}
[-\lambda(t-\tau)^\alpha]
e^{(t-\tau)\mathcal L_Z}H(\tau)\,d\tau .
\label{eq:D_dressed_prabhakar}
\end{align}
The scalar kernel in Eq.~\eqref{eq:D_dressed_prabhakar} uses the three-parameter Mittag--Leffler function of Eq.~(1.3) in Ref.~\cite{prabhakar1971singular} and the Prabhakar integral of Definition~5, Eq.~(11), in Ref.~\cite{PolitoTomovski2016}. The semigroup factor is the drift dressing defined here for the forward renewal problem.

Let \(H:[0,\infty)\to\mathcal{B}(\mathcal{H})\) be strongly
measurable and suppose that, for some \(\sigma>0\),
\begin{equation}
  \int_{0}^{\infty}
  e^{-\sigma t}\,
  \lVert H(t)\rVert_{\mathrm{HS}}\,dt
  <\infty .
\end{equation}
In particular, $H$ is locally Bochner integrable; this hypothesis permits locally integrable singularities at the
origin. The scalar transform in Eq.~(3) of Ref.~\cite{GarraGarrappa2018}, Fubini's theorem applied componentwise in the
finite-dimensional Liouville space (see Theorem~2.37(a) and (b) of Ref.~\cite{Folland1999}) and the semigroup shift then yield, for $\operatorname{Re}s>\sigma$ sufficiently large,
\begin{align}
&\mathcal L\!\left\{
\mathcal E^{\delta}_{\alpha,\beta,-\lambda}
[\mathcal L_Z]H(t)
\right\}
\nonumber\\
&\quad=
\mathcal A(s)^{\alpha\delta-\beta}
\left[\mathcal A(s)^\alpha+\lambda\mathcal I\right]^{-\delta}
\widetilde H(s).
\label{eq:D_dressed_transform_general}
\end{align}
Equation~\eqref{eq:D_dressed_transform_general} is derived here by dressing the scalar Prabhakar convolution with the drift semigroup $e^{t\mathcal L_Z}$ and evaluating the resulting scalar transform at
$\mathcal A(s)=s\mathcal I-\mathcal L_Z$ based on Definition~1.11 and Theorems~1.12, 1.15, and 1.17 in Ref.~\cite{Higham2008}. Principal powers use the principal logarithm of Theorem~1.31 therein. We choose $\operatorname{Re}s$ sufficiently large that $\widetilde H(s)$ exists and
\begin{equation}
\begin{aligned}
\operatorname{spec}\!\left[\mathcal A(s)\right]
\cap(-\infty,0]
=
\varnothing,\\
\operatorname{spec}\!\left[
\mathcal A(s)^\alpha+\lambda\mathcal I
\right]
\cap(-\infty,0]
=
\varnothing.
\end{aligned}
\label{app:spect_ops}
\end{equation}
These conditions in Eq.~\eqref{app:spect_ops}, where $\operatorname{spec}[.]$ stands for the spectrum of the associated Liouville-space operator, ensure that the required principal matrix powers are
well defined. The fractional powers are defined on their principal branches by the functional calculus on the finite-dimensional Liouville space~\cite{Higham2008}; therefore this construction does not rely on an
eigenbasis of $\mathcal L_Z$ and diagonalization.

For $\beta=\alpha n\gamma$ and $\delta=n\gamma$, the prefactor in Eq.~\eqref{eq:D_dressed_transform_general} is the identity, and hence
\begin{equation}
\mathcal L\!\left\{
\mathcal E^{n\gamma}_{\alpha,\alpha n\gamma,-\lambda}
[\mathcal L_Z]H(t)
\right\}
=
\left[\mathcal A(s)^\alpha+\lambda\mathcal I\right]^{-n\gamma}
\widetilde H(s).
\label{eq:D_dressed_transform_matched}
\end{equation}
Equation~\eqref{eq:D_dressed_transform_matched} is the matched-parameter specialization of Eq.~\eqref{eq:D_dressed_transform_general}; it is not a separate operator-valued transform asserted in the scalar references.

Let $F$ be a Liouville-space-valued function that is absolutely continuous on every finite time interval. Introduce the drift-relative derivative
\begin{equation}
H_F(t):=\dot F(t)-\mathcal L_Z[F(t)],\quad F\in AC_{\text{loc}},
\label{eq:D_HF_definition}
\end{equation}
where $AC_{\text{loc}}$ is the space of locally absolutely continuous operator-valued functions. This definition is chosen so that
$d[e^{-t\mathcal L_Z}F(t)]/dt=e^{-t\mathcal L_Z}H_F(t)$, using the matrix-exponential derivative in Sec.~10.1 of Ref.~\cite{Higham2008}. Substituting the finite sum $m_{N_0}$ into the operator defined in
Eq.~\eqref{eq:Capute_def} and then taking the limit gives
\begin{align}
({}^{\mathrm C}\mathcal D_{m,\mathcal L_Z}F)(t)
&=
\sum_{n=1}^{\infty}\lambda^{n\gamma}
\Bigl(
\mathcal E^{n\gamma}_{\alpha,\alpha n\gamma,-\lambda}
[\mathcal L_Z]H_F
\Bigr)(t).
\label{eq:D_D_series}
\end{align}
Equation~\eqref{eq:D_D_series} is derived here from
Eqs.~\eqref{eq:D_m_series},~\eqref{eq:D_dressed_prabhakar} and~\eqref{eq:Capute_def}. The scalar initial-value subtraction in that definition is the generalized-Caputo form of Eq.~(1.1) in Ref.~\cite{Kochubei2011}; the drift conjugation and the series in Eq.~\eqref{eq:D_D_series} are specific to the present model.

Let $\mathcal P_n[F]$ denote the $n$th summand in Eq.~\eqref{eq:D_D_series}, including $\lambda^{n\gamma}$, and set $M_T:=\sup_{0\leq u\leq T}\|e^{u\mathcal L_Z}\|_{\text{HS}\to\text{HS}}$. Throughout the paper, $||.||_{\text{HS}}$ and $||.||_{\text{HS}\to\text{HS}}$ represent Hilbert-Schmidt (HS) norm of an operator and induced HS norm of a superoperator respectively. In the Pauli operator basis normalized with respect to the HS inner product, Eqs.~\eqref{eq:LZ_I}--\eqref{eq:LZ_z} give 
\begin{equation}
[e^{u\mathcal L_Z}]
=
\operatorname{diag}(1,e^{-\Gamma_Zu},e^{-\Gamma_Zu},1),\label{eq:diag_uLz}
\end{equation}
and hence
\begin{equation}
M_T=1,\quad\forall T>0.\label{eq:M_T}
\end{equation}

Young's convolution inequality gives
\begin{equation}
\sum_{n=1}^{\infty}
\|\mathcal P_n[F]\|_{L^1(0,T)}
\leq
M_T\|m\|_{L^1(0,T)}\|H_F\|_{L^1(0,T)}.
\label{eq:D_dressed_L1_bound}
\end{equation}
The corresponding initial-value series is bounded by $M_T\|F(0)\|\|m\|_{L^1(0,T)}$. Thus
Eq.~\eqref{eq:D_dressed_L1_bound}, derived here from Young's inequality (see Proposition~8.7 of Ref.~\cite{Folland1999}), proves absolute convergence in $L^1(0,T)$ and justifies the limiting step in
Eq.~\eqref{eq:D_D_series}. The equality therefore holds in $L^1_{\mathrm{loc}}$, and hence almost everywhere.

The terms in Eq.~\eqref{eq:D_D_series} are drift-dressed Prabhakar convolutions; only terms satisfying
$0<n\alpha\gamma<1$ fall within the standard regularized Hilfer--Prabhakar derivative definition of
Refs.~\cite{PolitoTomovski2016} and~\cite{GarraGorenfloPolitoTomovski2014}.

\subsection{Equivalence of the Prabhakar series and the forward kernel–resolvent}
The purpose of this subsection is to prove that the locally convergent drift-dressed Prabhakar renewal series obtained above is exactly equivalent to the forward-ordered kernel–resolvent master equation of Secs.~\ref{sec:kernel_resolvent_formulation} and~\ref{sec:caputo_prabhakar_series}, including the generalized-Caputo initial-value contribution, and to verify that this equivalence preserves the noncommuting operator order and reduces to the standard Poisson generator in the memoryless limit.

Let $\rho\in AC_{\text{loc}}([0,\infty);\mathcal{B}(\mathcal{H}))$, a locally absolutely continuous, operator valued function of time, be the solution of Eq.~\eqref{eq:rho_fractional}. For the density operator $\rho$, define
\begin{equation}
H_\rho(t):=\dot\rho(t)-\mathcal L_Z[\rho(t)]
=H_F(t)\big|_{F=\rho}.
\label{eq:D_Hrho_definition}
\end{equation}
This is the specialization of Eq.~\eqref{eq:D_HF_definition}, not an additional dynamical assumption. Substituting Eqs.~\eqref{eq:D_m_series} and \eqref{eq:D_D_series} into the exact forward equation derived in Eq.~\eqref{eq:rho_fractional} gives
\begin{align}
H_\rho(t)
=\mathcal J_X\sum_{n=1}^{\infty}
\Biggl\{
&\lambda^{n\gamma}
\Bigl(
\mathcal E^{n\gamma}_{\alpha,\alpha n\gamma,-\lambda}
[\mathcal L_Z]H_\rho(t)
\Bigr)
\nonumber\\
&+e^{t\mathcal L_Z}q_n(t)\rho(0)
\Biggr\}.
\label{eq:D_forward_series}
\end{align}
Equation~\eqref{eq:D_forward_series}, which is derived here, holds in $L^1_{\mathrm{loc}}$, and by Eq.~\eqref{eq:D_dressed_L1_bound} almost everywhere. Its factor order is the forward order in Eq.~(27) of Ref.~\cite{Vacchini2020}: $\mathcal J_X$ acts after each completed drift-dressed renewal contribution and its initial-value term. In general, $\mathcal J_X$ cannot be interchanged with functions of $\mathcal L_Z$ when $[\mathcal L_Z,\Phi_X]\neq0$.

For the $n$th convolution contribution in Eq.~\eqref{eq:D_forward_series}, define
\begin{equation}
\begin{aligned}
\mathcal P_n[\rho](t)
&:=
\lambda^{n\gamma}\,
\mathcal E_{\alpha,\alpha n\gamma,-\lambda}^{n\gamma}
[\mathcal L_Z]H_\rho(t)\\
&=
\int_0^t
q_n(t-\tau)e^{(t-\tau)\mathcal L_Z}
H_\rho(\tau)\,d\tau .\label{eq:P_n_rho_t}
\end{aligned}
\end{equation}
The integral representation follows by setting $\beta=\alpha n\gamma$ and $\delta=n\gamma$ in Eq.~\eqref{eq:D_dressed_prabhakar} and using \(q_n\) from Eq.~\eqref{eq:D_qn}.

By Eq.~\eqref{eq:D_m_series}, the initial-value boundary contribution in Eq.~\eqref{eq:rho_fractional} has the renewal-order expansion
\begin{equation}
e^{t\mathcal L_Z}[m(t)\rho(0)]
=
\sum_{n=1}^{\infty}
e^{t\mathcal L_Z}[q_n(t)\rho(0)].\label{eq:IV_renewal_exp}
\end{equation}
Accordingly, denote its $n$th summand by
\begin{equation}
\mathcal I_n(t)
:=
e^{t\mathcal L_Z}[q_n(t)\rho(0)]
=
q_n(t)e^{t\mathcal L_Z}[\rho(0)].\label{eq:I_n_t}
\end{equation}
The final equality follows because $q_n(t)$ is scalar and \(e^{t\mathcal L_Z}\) is linear. This boundary term
restores the generalized-Caputo initial-value subtraction in Eq.~\eqref{eq:laplace_caputoD}, as expressed by Eq.~\eqref{eq:Ks_Fs}; the cancellation is verified at each renewal order in the termwise identity below.

To justify termwise Laplace transformation of the two component series in Eq.~\eqref{eq:D_forward_series}, we use the Banach-space-valued Laplace integral framework (see Chap.~1 of Ref.~\cite{ArendtEtAl2011}). For $\sigma>0$, define the exponentially weighted Bochner $L^1$ norm
\begin{equation}
\|G\|_{1,\sigma}
:=
\int_0^\infty
e^{-\sigma t}\|G(t)\|_{\mathrm{HS}}\,dt,
\end{equation}
and for $T>0$, introduce the truncated norm
\begin{equation}
\begin{aligned}
 \lVert G\rVert_{1,\sigma;(0,T)}
 &:=
 \int_{0}^{T}e^{-\sigma t}
 \lVert G(t)\rVert_{\mathrm{HS}}\,dt,
 \\
 Y_T(\sigma)
 &:=
 \lVert H_\rho\rVert_{1,\sigma;(0,T)}.\label{eq:truncated_norm}
 \end{aligned}
\end{equation}

Since $\rho\in AC_{\mathrm{loc}}$, Eq.~\eqref{eq:D_Hrho_definition}, together with the boundedness of $\mathcal L_Z$ in the finite-dimensional Liouville space, implies that $H_\rho\in L^1(0,T), \forall T>0$. Hence, for
$\sigma\geq0$,
\begin{equation}
Y_T(\sigma)\leq \int_0^T
  \lVert H_\rho(t)\rVert_{\mathrm{HS}}\,dt<\infty .\label{eq:finite_Y_T}
\end{equation}
Thus the finite-$T$ quantity appearing below is well defined; no global exponentially weighted integrability is being assumed. Now define,
\begin{equation}
C_X :=\lVert\mathcal{J}_X\rVert_{\mathrm{HS}\to\mathrm{HS}}.\label{eq:C_X}
\end{equation}
Since the transverse kick map in Eq.~\eqref{eq:PhiX_def} acts by unitary conjugation, unitary invariance of the HS (Frobenius) norm (see Sec.~5.6 of~\cite{HornJohnson2017}) implies
\begin{equation}
\begin{aligned}
&\lVert\Phi_X[A]\rVert_{\mathrm{HS}}
=
\lVert U_x(\theta)AU_x^\dagger(\theta)\rVert_{\mathrm{HS}}
=
\lVert A\rVert_{\mathrm{HS}},
\\
&\lVert\Phi_X\rVert_{\mathrm{HS}\to\mathrm{HS}}=1 .
\end{aligned}
\end{equation}
The triangle inequality for the induced norm, and using $C_X$ as defined in Eq.~\eqref{eq:C_X}, the preceding estimates give
\begin{equation}
\begin{aligned}
C_X
=
\lVert\Phi_X-\mathcal{I}\rVert_{\mathrm{HS}\to\mathrm{HS}}
&\leq
\lVert\Phi_X\rVert_{\mathrm{HS}\to\mathrm{HS}}
+
\lVert\mathcal{I}\rVert_{\mathrm{HS}\to\mathrm{HS}}\\
&=2 .
\end{aligned}
\end{equation}
As established in Eq.~\eqref{eq:D_qn}, using Eqs.~(2.5) and (2.6) of Ref.~\cite{CahoyPolito2013}, we have $q_n(t)\geq0,\text{ and } \widetilde q_n(\sigma)
=\bigl[\widetilde w(\sigma)\bigr]^n$. Moreover, applying Tonelli's theorem to Eq.~\eqref{eq:D_m_series} gives
\begin{equation}
\begin{aligned}
\widetilde m(\sigma)
&=\int_0^\infty e^{-\sigma t}m(t)\,dt\\
&=\sum_{n=1}^{\infty}\widetilde q_n(\sigma)
=\frac{\widetilde w(\sigma)}
       {1-\widetilde w(\sigma)},
\qquad \sigma>0.
\end{aligned}
\end{equation}
This agrees with the scalar renewal identity in Eq.~\eqref{eq:m_s}.

For almost every $t\in(0,T)$, Eq.~\eqref{eq:M_T} and $e^{-\sigma t}=e^{-\sigma(t-\tau)}e^{-\sigma\tau}$ give the pointwise estimate
\begin{equation}
\begin{aligned}
e^{-\sigma t}\left\|\mathcal{P}_n[\rho](t)\right\|_{\mathrm{HS}}
\le
\int_0^t
&\left[e^{-\sigma(t-\tau)}q_n(t-\tau)\right]\\
&\times\left[e^{-\sigma\tau}\left\|H_\rho(\tau)\right\|_{\mathrm{HS}}\right]
\,d\tau\\
&=\left(a_{n,T}\ast b_T\right)(t),
\end{aligned}
\end{equation}
where $a_{n,T}(u):=e^{-\sigma u}q_n(u)\mathbf 1_{(0,T)}(u)$ and $b_T(u):=e^{-\sigma u}\lVert H_\rho(u)\rVert_{\mathrm{HS}}\mathbf 1_{(0,T)}(u)$ both of which are extended to $\mathbb R$ by inclusion of zero. Integrating over $0<t<T$,
\begin{equation}
\begin{aligned}
\left\|\mathcal{P}_n[\rho]\right\|_{1,\sigma;(0,T)}
:=
&\int_0^T
e^{-\sigma t}
\left\|\mathcal{P}_n[\rho](t)\right\|_{\mathrm{HS}}
\,dt\\
&\le
\lVert a_{n,T}\ast b_T\rVert_{L^1},
\end{aligned}
\end{equation}
and using Young's convolution inequality $\|a_{n,T}\ast b_T\|_{L^1}\le \|a_{n,T}\|_{L^1}\|b_T\|_{L^1}$ (see Proposition 8.7 of~\cite{Folland1999}), we get
\begin{equation}
\begin{aligned}
\left\|\mathcal{P}_n[\rho]\right\|_{1,\sigma;(0,T)}
\le
&\left(
\int_0^T e^{-\sigma u} q_n(u)\,du
\right)\\
&\times\left(
\int_0^T e^{-\sigma \tau}
\left\|H_\rho(\tau)\right\|_{\mathrm{HS}}
\,d\tau
\right).
\end{aligned}
\end{equation}
Considering Eq.~\eqref{eq:truncated_norm},
\begin{equation}
\left\|\mathcal{P}_n[\rho]\right\|_{1,\sigma;(0,T)}
\le
\left(
\int_0^T e^{-\sigma u} q_n(u)\,du
\right)
Y_T(\sigma),
\end{equation}
$q_n(u)\ge 0$ and the monotonicity of the Lebesgue integral,
\begin{equation}
\int_0^T e^{-\sigma u} q_n(u)\,du
\le
\int_0^\infty e^{-\sigma u} q_n(u)\,du
=
\widetilde{q}_n(\sigma)
\end{equation}
we get
\begin{equation}
\|\mathcal{P}_n\left[\rho\right]\|_{1,\sigma;\left(0,T\right)}\le\tilde{q}_n\left(\sigma\right)Y_T\left(\sigma\right),\label{eq:P_n_rho}
\end{equation}
which is a finite value by Eq.~\eqref{eq:finite_Y_T}. For the initial-value term defined in Eq.~\eqref{eq:I_n_t}, since $\|e^{t\mathcal{L}_Z}\|\le 1$ by Eq.~\eqref{eq:M_T}, we have,
\begin{equation}
\left\|\mathcal{I}_n(t)\right\|_{\mathrm{HS}}
\le
q_n(t)\left\|\rho(0)\right\|_{\mathrm{HS}}
\end{equation}
resulting in
\begin{equation}
\begin{aligned}
\left\|\mathcal{I}_n\right\|_{1,\sigma;(0,T)}
&\le
\int_0^T e^{-\sigma t} q_n(t)\,dt\,
\left\|\rho(0)\right\|_{\mathrm{HS}}
\\
&\le
\widetilde{q}_n(\sigma)
\left\|\rho(0)\right\|_{\mathrm{HS}}.\label{eq:I_n_rho}
\end{aligned}
\end{equation}

Taking the truncated weighted norm of Eq.~\eqref{eq:D_forward_series}, with the series understood in the $L^1_{\mathrm{loc}}$ sense
established above, using the definition in Eq.~\eqref{eq:truncated_norm} and the induced-norm inequality associated with the definition in Eq.~\eqref{eq:C_X}, gives
\begin{equation}
\begin{aligned}
Y_T(\sigma)
&=
\left\lVert\mathcal J_X\sum_{n=1}^{\infty}
\left[
\mathcal P_n[\rho]+\mathcal I_n
\right]
\right\rVert_{1,\sigma;(0,T)}
\\
&\leq
C_X
\sum_{n=1}^{\infty}
\left[
\lVert\mathcal P_n[\rho]\rVert_{1,\sigma;(0,T)}
+\lVert\mathcal I_n\rVert_{1,\sigma;(0,T)}\right]\\
&\leq
C_X
\sum_{n=1}^{\infty}\widetilde q_n(\sigma)\left[Y_T(\sigma)+
\lVert\rho(0)\rVert_{\mathrm{HS}}\right]\\
&=
C_X\widetilde m(\sigma)\left[Y_T(\sigma)+
\lVert\rho(0)\rVert_{\mathrm{HS}}
\right].\label{eq:Y_T_est}
\end{aligned}
\end{equation}
Here the first inequality follows from the triangle inequality and Eq.~\eqref{eq:C_X}, the second uses the two weighted
estimates Eqs.~\eqref{eq:P_n_rho} and~\eqref{eq:I_n_rho}, and the last equality uses the Tonelli identity (see Theorem~2.37(a) of Ref.~\cite{Folland1999}) established in the preceding paragraph.

For the present renewal law,
\begin{equation}
\lim_{\sigma\to\infty}\widetilde w(\sigma)  =
\lim_{\sigma\to\infty}\frac{\lambda^\gamma}{(\sigma^\alpha+\lambda)^\gamma}=0.\label{eq:wsigma_0}
\end{equation}
Consequently, $\widetilde m(\sigma)\to0$ as $\sigma\to\infty$. We may therefore choose $\sigma>0$ such that $C_X\widetilde m(\sigma)<1$. Rearranging the estimate in Eq.~\eqref{eq:Y_T_est}, based on the finiteness of $Y_T$ shown in Eq.~\eqref{eq:finite_Y_T}, yields
\begin{equation}
Y_T(\sigma)\leq\frac{C_X\widetilde m(\sigma)}{1-C_X\widetilde m(\sigma)}
\lVert\rho(0)\rVert_{\mathrm{HS}},
\qquad T>0 .\label{eq:Y_T_est2}
\end{equation}

To obtain the exponentially weighted $L^1$ norm from the finite time estimate in Eq.~\eqref{eq:Y_T_est2}, choose an increasing
sequence of cutoff times \(T_N\uparrow\infty\) and define
\begin{equation}
f_N(t):=e^{-\sigma t}\lVert H_\rho(t)\rVert_{\mathrm{HS}}\mathbf 1_{[0,T_N]}(t).
\end{equation}
Then $0\leq f_N(t)\leq f_{N+1}(t)$ and
\begin{equation}
\lim_{N\to\infty}f_N(t)
=e^{-\sigma t}\lVert H_\rho(t)\rVert_{\mathrm{HS}},
\quad \text{for almost every }t>0.
\end{equation}
Moreover, based on the definition in Eq.~\eqref{eq:truncated_norm},
\begin{equation}
\int_0^\infty f_N(t)\,dt=Y_{T_N}(\sigma),
\end{equation}
leads to
\begin{equation}
\begin{aligned}
\lVert H_\rho\rVert_{1,\sigma}
&=
\int_0^\infty
e^{-\sigma t}\lVert H_\rho(t)\rVert_{\mathrm{HS}}\,dt \\
&=
\lim_{N\to\infty}Y_{T_N}(\sigma).\label{eq:H_rho_1_sigma}
\end{aligned}
\end{equation}

Then the monotone convergence theorem~\cite{Folland1999}, which states that $\int f=\lim_{N\to\infty}\int f_N$ whenever
$0\leq f_N\uparrow f$, therefore Eqs.~\eqref{eq:Y_T_est2} and~\eqref{eq:H_rho_1_sigma} yield,
\begin{equation}
\lVert H_\rho\rVert_{1,\sigma}
\leq
\frac{C_X\widetilde m(\sigma)}{1-C_X\widetilde m(\sigma)}
\lVert\rho(0)\rVert_{\mathrm{HS}}<\infty .
\end{equation}
Thus the weighted integrability required for the termwise Laplace transformation follows from Eq.~\eqref{eq:D_forward_series}. Letting \(T\to\infty\) in Eqs.~\eqref{eq:P_n_rho} and~\eqref{eq:I_n_rho} and applying the monotone convergence theorem to the nonnegative norm integrands gives
\begin{equation}
\begin{aligned}
\lVert\mathcal P_n[\rho]\rVert_{1,\sigma}
&\leq
\widetilde q_n(\sigma)
\lVert H_\rho\rVert_{1,\sigma},\\
\lVert\mathcal I_n\rVert_{1,\sigma}
&\leq
\widetilde q_n(\sigma)
\lVert\rho(0)\rVert_{\mathrm{HS}}.\label{eq:Pn_In}
\end{aligned}
\end{equation}
Summing estimates in Eq.~\eqref{eq:Pn_In} over $n$ and using the Tonelli identity established above yields
\begin{equation}
\begin{aligned}
\sum_{n=1}^{\infty}
\bigl[
 \|\mathcal P_n[\rho]\|_{1,\sigma}
 &+
 \|e^{(\cdot)\mathcal L_Z}
   q_n(\cdot)\rho(0)\|_{1,\sigma}
\bigr]
\\
&\leq
\left(
 \|H_\rho\|_{1,\sigma}
 +
 \|\rho(0)\|_{\mathrm{HS}}
\right)
\frac{\widetilde w(\sigma)}
     {1-\widetilde w(\sigma)}
<\infty .\label{eq:sum_Pn_In}
\end{aligned}
\end{equation}
Hence the two component series in Eq.~\eqref{eq:D_forward_series} are absolutely summable in the weighted $L^1$ norm.

Let $\sigma_0>0$ be a weight satisfying $C_X\widetilde m(\sigma_0)<1$, as selected following Eq.~\eqref{eq:wsigma_0}. Equation~\eqref{eq:sum_Pn_In} permits termwise Laplace transformation of the two component series for $\operatorname{Re}s>\sigma_0$~\cite{ArendtEtAl2011}. We now transform the series renewal order by renewal order and resum it to recover the forward kernel–resolvent equation. Let $\mathcal B_n[\rho]$ denote the $n$th expression in braces in Eq.~\eqref{eq:D_forward_series}. Equations~\eqref{eq:D_qn} and \eqref{eq:D_dressed_transform_matched}, together with the Laplace and
boundary identities in Eqs.~\eqref{eq:laplace_caputoD} and~\eqref{eq:Ks_Fs}, give
\begin{align}
\mathcal L\!\left\{\mathcal B_n[\rho]\right\}(s)
&=\widetilde q_n(\mathcal A(s))
\left[\mathcal A(s)\widetilde\rho(s)-\rho(0)\right]
\nonumber\\
&\quad+\widetilde q_n(\mathcal A(s))\rho(0)
\nonumber\\
&=\widetilde q_n(\mathcal A(s))
\mathcal A(s)\widetilde\rho(s).
\label{app:D_single_term_transform}
\end{align}
Thus Eq.~\eqref{app:D_single_term_transform} is a termwise consequence of the scalar transforms in Refs.~\cite{GarraGarrappa2018},~\cite{CahoyPolito2013} and the manuscript's generalized-Caputo boundary construction in Eqs.~\eqref{eq:laplace_caputoD}--~\eqref{eq:rho_fractional}.

To resum the transformed renewal-order series, set $B(s):=\widetilde w(\mathcal A(s))$. For the present dephasing Liouvillian, $\operatorname{spec}[\mathcal A(s)]=\{s,s+\Gamma_Z\}$. The spectral-mapping property (see Theorem~1.13(d) of Ref.~\cite{Higham2008}) therefore gives
\begin{equation}
\operatorname{spec}[B(s)]
=
\left\{
\widetilde w(s),
\widetilde w(s+\Gamma_Z)
\right\}.
\label{eq:D_spectral_mapping}
\end{equation}

Moreover, for $\operatorname{Re}s>0$, the principal power satisfies $\operatorname{Re}s^\alpha>0$ for
$0<\alpha\leq1$. Hence, using Eqs.~\eqref{eq:11e} and~\eqref{eq:11ee},
$
|\widetilde w(s)|
=
\left|
1+\frac{s^\alpha}{\lambda}
\right|^{-\gamma}
<1.
$
It follows that
\begin{equation}
r_{\mathrm{sp}}[B(s)]
=
\max
\left\{
|\widetilde w(s)|,
|\widetilde w(s+\Gamma_Z)|
\right\}
<1,
\qquad
\operatorname{Re}s>0,
\label{app:spect_radius}
\end{equation}
where $r_{\mathrm{sp}}$ denotes the spectral radius. By Theorem~5.6.12 of Ref.~\cite{HornJohnson2017},
Eq.~\eqref{app:spect_radius} implies $B(s)^n\to0$ as $n\to\infty$. The finite-dimensional Neumann-series identity (see Problem~5.6.P26 in Ref.~\cite{HornJohnson2017}) then yields
\begin{figure*}[!t]
  \centering
  \includegraphics[width=\textwidth]{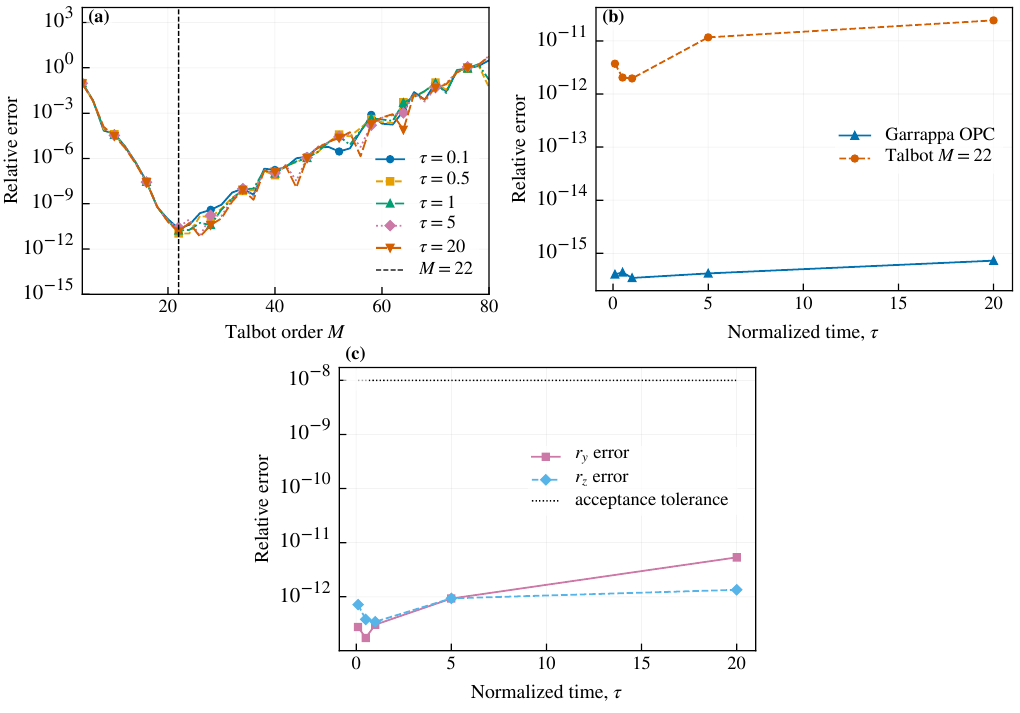}
  \caption{Relative errors of inverse-Laplace reconstructions against independent Hankel-contour references. Parameters and the error definition are given in
Appendix~\ref{app:numerical_inverse_laplace}. (a) Talbot reconstruction of the positive-time kernel $\kappa_{>0}^{(P)}(\tau)$ versus order $M$.
The vertical line marks the production choice $M=22$; the high-order error increase reflects finite-precision cancellation. (b) Waiting-time density evaluated by Garrappa OPC
(solid triangles) and Talbot inversion at $M=22$ (dashed circles). (c) Componentwise errors of the forward-ordered evolution from $|0\rangle$, reconstructed by Talbot inversion at $M=22$.
Squares and diamonds denote $r_y$ and $r_z$, respectively; the dotted line marks the acceptance tolerance $10^{-8}$. Here $\tau=\lambda^{1/\alpha}t$. ChatGPT (OpenAI; GPT-5 Sol Ultra) generated the Python
benchmarking and plotting code from R.U.Erdogan's theoretical and numerical specifications; R.U.Erdogan reviewed it against the inverse-Laplace formulas and numerical methods.}
\label{fig:sim7_inverse_laplace_benchmarks}
\end{figure*}
\begin{equation}
\sum_{n=1}^{\infty}B(s)^n
=
B(s)[\mathcal I-B(s)]^{-1},\qquad\operatorname{Re}s>0.
\label{app:D_Neumann}
\end{equation}

Since $\widetilde q_n(\mathcal A(s))=B(s)^n$, the left-hand side of Eq.~\eqref{app:D_Neumann} is the operator-valued
renewal-density series whereas its right-hand side is $\widetilde m(\mathcal A(s))$. This completes the resummation in Eq.~\eqref{eq:D_forward_transform_check}. By the spectral radius and Neumann series argument established in Eqs.~\eqref{app:spect_radius} and~\eqref{app:D_Neumann}, summing Eq.~\eqref{app:D_single_term_transform} yields
\begin{align}
\widetilde H_\rho(s)
&=\mathcal J_X
\left[\sum_{n=1}^{\infty}
\widetilde q_n(\mathcal A(s))\right]
\mathcal A(s)\widetilde\rho(s)
\nonumber\\
&=\mathcal J_X\widetilde m(\mathcal A(s))
\mathcal A(s)\widetilde\rho(s)
\nonumber\\
&=\mathcal J_X\widetilde K(\mathcal A(s))
\widetilde\rho(s).
\label{eq:D_forward_transform_check}
\end{align}
The last line uses $\widetilde K(s)=s\widetilde m(s)$ from Eq.~\eqref{eq:30a} and the matrix-function product property (see Theorem~1.15 of Ref.~\cite{Higham2008}). It recovers the forward order in Eq.~\eqref{eq:exact_kernel_resolvent}; the same generic order is Eq.~(27) of Ref.~\cite{Vacchini2020}.

Equation~\eqref{eq:D_forward_transform_check} has so far been established for $\operatorname{Re}s>\sigma_0$. By the CPTP property established in Sec.~\ref{subsec:ren_map_equ} and Appendix~\ref{app:renewal_structure}, $\rho(t)$ remains a density operator. Writing its spectral decomposition as
\begin{equation}
  \rho(t)
  =
  \sum_j p_j(t)\lvert j,t\rangle\langle j,t\rvert,
  \qquad
  p_j(t)\geq0,
  \quad
  \sum_jp_j(t)=1,
\end{equation}
the definition of the Schatten \(2\)-norm (Eq.~(1.165) of Ref.~\cite{Watrous2018}) gives
\begin{equation}
  \lVert\rho(t)\rVert_{\mathrm{HS}}^2
  =
  \operatorname{Tr}\rho(t)^2
  =
  \sum_jp_j(t)^2
  \leq
  \left[\sum_jp_j(t)\right]^2
  =
  1.
\end{equation}
Consequently, for \(\operatorname{Re}s>0\),
\begin{equation}
  \int_0^\infty
  e^{-\operatorname{Re}(s)t}
  \lVert\rho(t)\rVert_{\mathrm{HS}}\,dt
  \leq
  \int_0^\infty e^{-\operatorname{Re}(s)t}\,dt
  =
  \frac{1}{\operatorname{Re}s}.
\end{equation}
The results for vector-valued Laplace transforms in Secs.~1.4 and 1.5 of Ref.~\cite{ArendtEtAl2011} therefore imply that
$\widetilde\rho(s)$ is analytic in the half-plane $\operatorname{Re}s>0$. Hence
\begin{equation}
\widehat H_\rho(s)
:=
\mathcal A(s)\widetilde\rho(s)-\rho(0)\label{eq:widehat_H},\qquad\text{Re}(s)>0,
\end{equation}
defines an analytic drift-relative transform in $\operatorname{Re}s>0$.

For $\operatorname{Re}s>\sigma_0$, the weighted integrability established above permits integration by
parts. Since $e^{-sT}\rho(T)\to0$ as $T\to\infty$, Eq.~\eqref{eq:D_Hrho_definition} and the operational derivative formula of Sec.~1.6 of Ref.~\cite{ArendtEtAl2011} give
\begin{equation}
\begin{aligned}
\mathcal L\{H_\rho\}(s)
&=
\mathcal L\{\dot\rho-\mathcal L_Z\rho\}(s)\\
&=
s\widetilde\rho(s)-\rho(0)
-
\mathcal L_Z\widetilde\rho(s)\\
&=
\mathcal A(s)\widetilde\rho(s)-\rho(0)
=
\widehat H_\rho(s).\label{eq:laplace_Hrho}
\end{aligned}
\end{equation}
Equation~\eqref{eq:laplace_Hrho} shows that
$\widehat H_\rho(s)=\widetilde H_\rho(s)$ on $\operatorname{Re}s>\sigma_0$; hence $\widehat H_\rho$ is the analytic continuation of the
Laplace transform of $H_\rho$ to the right half-plane.

The matrix functional calculus used in Eqs.~\eqref{eq:D_dressed_transform_general}--\eqref{eq:D_dressed_transform_matched} shows that $B(s)=\widetilde w(\mathcal A(s))$ is analytic for
$\operatorname{Re}s>0$. Together with Eqs.~\eqref{app:spect_radius} and~\eqref{app:D_Neumann}, the analytic-inversion result of Chap.~VII, Sec.~1.1 of Ref.~\cite{Kato1995} implies that $[\mathcal I-B(s)]^{-1}$ is analytic throughout this half-plane. It follows that
\begin{equation}
\widetilde K(\mathcal A(s))=B(s)[\mathcal I-B(s)]^{-1}\mathcal A(s)\label{eq:K_A_s}
\end{equation}
is analytic for $\operatorname{Re}s>0$.

Combining Eqs.~\eqref{eq:D_forward_transform_check} and~\eqref{eq:laplace_Hrho} gives, for $\operatorname{Re}s>\sigma_0$,
\begin{equation}
\widehat H_\rho(s)
=
\mathcal J_X
\widetilde K(\mathcal A(s))
\widetilde\rho(s).
\end{equation}
Both sides of this identity are analytic in the connected half-plane \(\operatorname{Re}s>0\): the left-hand side is
analytic by the definition in Eq.~\eqref{eq:widehat_H} whereas the right-hand side is analytic by the analyticity of $\widetilde\rho(s)$ and Eq.~\eqref{eq:K_A_s}. Since the two sides agree on the nonempty open subset $\operatorname{Re}s>\sigma_0$, the identity theorem, applied componentwise in the finite-dimensional Liouville space (Appendix~A of Ref.~\cite{ArendtEtAl2011}), extends this equality to the entire half-plane $\operatorname{Re}s>0$. Using Eq.~\eqref{eq:widehat_H} and rearranging gives
\begin{equation}
\left[
\mathcal A(s)-\mathcal J_X\widetilde K(\mathcal A(s))
\right]\widetilde\rho(s)
=
\rho(0),
\end{equation}
which coincides with the kernel--resolvent equation derived in Eq.~\eqref{eq:exact_kernel_resolvent}.

Finally, the Poisson specialization $\alpha=\gamma=1$ gives $w(t)=\lambda e^{-\lambda t}$ and $m(t)=\lambda$, consistently with Eqs.~\eqref{eq:w_exp_summary} and~\eqref{eq:K_exp_summary} where $\lambda=\nu$. Applying definition given by Eq.~\eqref{eq:Capute_def} then yields
\begin{equation}
{}^{\mathrm C}\mathcal D_{m,\mathcal L_Z}\rho(t)
=\lambda[\rho(t)-e^{t\mathcal L_Z}\rho(0)].
\label{eq:D_poisson_caputo_check}
\end{equation}
Equation~\eqref{eq:D_poisson_caputo_check} is a direct consistency check derived here; it uses the matrix-exponential identity in Sec.~10.1 of Ref.~\cite{Higham2008}. The boundary term in Eq.~\eqref{eq:rho_fractional} restores $\lambda\rho(t)$, so
\begin{equation}
\dot\rho(t)=\mathcal L_Z[\rho(t)]
+\lambda(\Phi_X-\mathcal I)[\rho(t)].
\label{eq:D_poisson_master_check}
\end{equation}
Equation~\eqref{eq:D_poisson_master_check} is also the Poisson quantum-renewal generator obtained by differentiating Eq.~(16) of Ref.~\cite{Vacchini2020} with $\mathcal E=\Phi_X$. Hence all three representations---the ordered kernel resolvent, the generalized-Caputo equation, and the convergent Prabhakar series---have the same Markovian limit.

\section{Numerical inverse-Laplace validation}
\label{app:numerical_inverse_laplace}
R.U.Erdogan developed the theoretical formulation and numerical design of the simulations and designed Fig.~\ref{fig:conceptual_framework}.
Under his direction, ChatGPT (OpenAI; GPT-5 Sol Ultra) generated the Python simulation and plotting code, as well as the MATLAB
and \LaTeX{}/TikZ code used to produce Fig.~\ref{fig:conceptual_framework}. R.U.Erdogan reviewed the generated code for consistency with the underlying mathematical and physical
theory and the numerical methods described in this work. The authors take responsibility for the implementations, figures, and reported results.

The three panels in Fig.~\ref{fig:sim7_inverse_laplace_benchmarks} provide numerical cross-checks of the inverse-Laplace implementation.  They use
\(\alpha=0.5\), \(\gamma=0.8\), and
\(\beta=\alpha\gamma=0.4\), with normalized time
\(\tau=\lambda^{1/\alpha}t\).  The dimensionless
implementation sets \(\lambda=1\); panel~(c) additionally
uses
\(\Gamma_Z/\lambda^{1/\alpha}=0.1\) and
\(\theta=\pi/2\).  All panels are evaluated at $\tau\in\{0.1,0.5,1,5,20\}$.

For a scalar reference value $f_{\mathrm H}(\tau)$, the implemented reference-relative error is
\begin{equation}
\varepsilon_f^{(m)}(\tau)
:=\frac{
\left|f_m(\tau)-f_{\mathrm H}(\tau)\right|
}{\max\!\left\{\left|f_{\mathrm H}(\tau)\right|,10^{-20}\right\}},\label{eq:E_reference_relative_error}
\end{equation}
where $m$ denotes the Talbot or optimal-parabolic-contour (OPC) evaluation and ${\mathrm H}$ denotes the Hankel
branch-cut reference.  In panel~(c), Eq.~\eqref{eq:E_reference_relative_error} is applied separately to $r_y$ and $r_z$; no vector-norm error is
used.

The fixed-Talbot reconstructions use IEEE-754 binary64 arithmetic.  For a Laplace-domain quantity $F$, the
implemented Abate--Valk\'o rule is 
\begin{equation}
f_{\mathrm T}(\tau;M)
=\frac{2}{5\tau}
\sum_{k=0}^{M-1}
\operatorname{Re}\!\left[
\gamma_k F\!\left(\frac{\delta_k}{\tau}\right)
\right],
\label{eq:E_Talbot_rule}
\end{equation}
where, following Eqs.~(42)--(43) of Ref.~\cite{AbateValko2004},
\begin{equation}
\begin{aligned}
\delta_0&=\frac{2M}{5},\quad
\gamma_0=\frac{1}{2}e^{\delta_0},\\
\delta_k&=\frac{2k\pi}{5}
\left(\cot\frac{k\pi}{M}+i\right),\\
\gamma_k&=\left[
1+i\frac{k\pi}{M}
\left(1+\cot^2\frac{k\pi}{M}\right)
-i\cot\frac{k\pi}{M}
\right]e^{\delta_k},
\end{aligned}
\label{eq:E_Talbot_weights}
\end{equation}
and the second line applies for $k=1,\ldots,M-1$. The exponential factor $e^{\delta_k}$, including the half-weighted $k=0$ contribution, is retained in the executed implementation.

The high-precision Hankel reference uses the principal branch, $-\pi<\arg s<\pi$, and the upper lip $s=-r+i0^+=re^{i\pi}$ of the negative-real-axis cut. Combining Eqs.~(6) and~(7) of Ref.~\cite{MainardiGarrappa2015} fixes the corresponding upper-lip sign.  Thus the cut contribution is evaluated as
\begin{equation}
f_{\mathrm{cut}}(\tau)
=-\frac{1}{\pi}
\int_0^\infty
e^{-r\tau}
\operatorname{Im}F(-r+i0^+)\,dr.
\label{eq:E_Hankel_reference}
\end{equation}
The boundary values required for the present waiting-time density and renewal kernel are given explicitly in
Eqs.~\eqref{eq:app_disc_w} and~\eqref{eq:app_disc_K}, respectively; isolated-pole contributions, when present, are added separately.
The production references use 60-decimal-digit arithmetic, radial cutoff $R=800$, endpoint-removing substitutions,
and adaptive subintervals.  Convergence is checked at 90 decimal digits and by comparing $R=400$ with
$R=1200$.  A separate de Hoog reconstruction uses 75-decimal-digit arithmetic and degree 110.

Panel~(a) directly evaluates, for $\tau>0$ and the dimensionless choice $\lambda=1$, the inverse of the scalar renewal-resolvent kernel $\widetilde K(s)$ already defined in Eq.~\eqref{eq:30a}.  Under the dimensionless scaling used in Fig.~\ref{fig:sim2_kernel_diagnostics}, this positive-time
inverse is precisely the signed time-domain representative $\kappa_{>0}^{(P)}(\tau)$ plotted there; the tilde is reserved for the Laplace-domain kernel.  The inversion is performed without explicitly subtracting a delta, constant, pole, or asymptotic contribution.  The displayed scan uses the even orders
$M=4,6,\ldots,80$.  For $\tau=\{0.1,0.5,1,5,20\}$, the orders minimizing the  implemented error are, respectively, $M=\{22,24,22,26,26\}$.  Thus $M=22$ is a practical common working order, but it is not the individual optimum for all five times.  The stored and independently replayed
orders $M=74,76,78,80$ reproduce the high-order loss of accuracy caused by finite-precision cancellation in the Talbot sum.

Panel~(b) checks the normalized dimensionless waiting-time density $w_\tau(\tau)$ shown in Fig.~\ref{fig:sim9-prabhakar-renewal-clock}(a).  This
density is obtained from the Prabhakar law in Eq.~\eqref{eq:11a} after imposing $\beta=\alpha\gamma$ and $C=\lambda^\gamma$ as in
Eqs.~\eqref{eq:11c} and~\eqref{eq:11d}; its normalized Laplace transform is given in Eq.~\eqref{eq:11e}. The plotted Talbot values are the $M=22$ fixed-order estimates; the higher-order checks did not satisfy the $10^{-10}$ consecutive-order criterion.  The direct Prabhakar values are computed with the embedded three-parameter Garrappa OPC implementation~\cite{garrappa2015numerical} using $\log\epsilon=-34.54$.  At each of the five evaluation points, the OPC algorithm selects $N_{\mathrm{opt}}=27$, $\mu=1.5436533891171536$, and $h=0.17896814752774612$, giving the $55$ nodes $s_k=\mu(1+ihk)^2,\text{for} -27\leq k\leq27$.

For panel~(c), we evaluate the chronological forward-order $W_{yz}$ resolvent of Eqs.~\eqref{eq:exact_kernel_resolvent} and~\eqref{eq:Myz_exact}, using the shifted kernels defined in
Eqs.~\eqref{eq:K0_def}--\eqref{eq:delta_def}.  For the initial vector $\mathbf r_{yz}=(0,1)^{\mathsf T}$, the transformed components are the corresponding specialization of
Eqs.~\eqref{eq:cy_exact} and~\eqref{eq:cz_exact}, consistently with the corrected inverse resolvent in Eq.~\eqref{eq:app_Myz_resolvent}.  The panel labels ``$r_y$ error'' and ``$r_z$ error'' denote the errors in
the reconstructed components $c_y(t)$ and $c_z(t)$, respectively.

Both components use the $M=22$ fixed-Talbot procedure in Eqs.~\eqref{eq:E_Talbot_rule}--\eqref{eq:E_Talbot_weights} and the high-precision Hankel convention in Eq.~\eqref{eq:E_Hankel_reference}.  The required kernel
boundary values and nested physical cuts are given in Eqs.~\eqref{eq:app_disc_K} and~\eqref{eq:app_physical_cuts}; the coupled reference follows the pole--branch-cut decomposition of
Eqs.~\eqref{eq:app_pole_branch_decomp}--\eqref{eq:app_branch_integral}. Keyhole tests at outer radii $100$, $4000$, and $10^5$, with inner radius $10^{-12}$ and angular offset $0.01$, gave zero winding number after sample doubling, so no off-cut pole contribution was included.  A separately implemented de Hoog inversion~\cite{deHoogKnightStokes1982} also reconstructed the forward operator pencil of Eq.~\eqref{eq:exact_kernel_resolvent}.

Over the five normalized times specified above, the largest componentwise Talbot error was $5.35\times10^{-12}$, below the $10^{-8}$ tolerance. For this initial vector, interchanging the off-diagonal
kernels leaves $\Delta(s)$ and $\widetilde c_z(s)$ unchanged but changes $\widetilde c_y(s)$; hence the $y$ component is the ordering-sensitive validation.

These three numerical cross-checks test the implemented inverse-Laplace reconstructions over the displayed parameter set.

\section{Finite energy relaxation and directional transfer benchmarks}
\label{app:finite_t1_directional}

Finite energy relaxation damps the longitudinal mode supporting
the algebraic storage law in Eq.~\eqref{eq:T_ZY_P} and adds a
preparation-independent polarization. We separate this contribution
from the directional response and relate that response to the
pure-dephasing transfer block, establishing the basis for
Fig.~\ref{fig:sim108_finite_t1}. We then formulate the Poisson
benchmark with equal expected event counts introduced in
Sec.~\ref{sec:physical_interpretation} and give its numerical
controls.
\subsection{Relaxation and directional pair differences}
The Prabhakar process starts at zero renewal age at $t=0$, with all waiting times independently drawn from $w(t)$ and unchanged
deterministic transverse kicks. Adding independent zero-temperature amplitude damping gives the physical-time interevent Lindblad generator~\cite{GoriniKossakowskiSudarshan1976,Lindblad1976}
\begin{align}
\mathcal L_{\mathrm d}[\rho]
&=
\mathcal L_Z[\rho]
+\Gamma_1
\left(
L\rho L^\dagger
-\frac12\{L^\dagger L,\rho\}
\right),
\label{eq:finite_t1_generator}
\\
L&=|0\rangle\langle1|,
\qquad
g_2=g+\frac{\eta}{2}.
\label{eq:finite_t1_g2}
\end{align}

Under Eq.~\eqref{eq:finite_t1_scales}, $g_2$ is the normalized transverse interevent decay rate, combining pure dephasing and amplitude damping, while $\eta$ is the normalized longitudinal
relaxation rate~\cite{KrantzKjaergaardYanOrlandoGustavssonOliver2019}. Their contrast,
\begin{equation}
q:=g_2-\eta=g-\frac{\eta}{2},
\end{equation}
is not an additional dissipative rate. The corresponding interevent generator for Bloch-vector differences in the $yz$ subspace, with respect to $\tau$, is
\begin{equation}
B
=
\operatorname{diag}(-g_2,-\eta)
=
-\eta I_2+\operatorname{diag}(-q,0).
\end{equation}
The scalar term commutes with the remaining matrix; hence, for normalized duration $u$, Theorem~10.2 of Ref.~\cite{Higham2008} gives
\begin{equation}
e^{Bu}=e^{-\eta u}D_q(u),
\qquad
D_q(u):=\operatorname{diag}(e^{-qu},1).
\end{equation}
The kick rotation in this subspace is $\Phi_X^{(yz)}$ of Eq.~\eqref{eq:PhiX_yz_matrix}. A history $\omega=\{t_1,\ldots,t_n\}$ up to time $t$ has completed intervals
$u_0,\ldots,u_{n-1}$ and residual interval $u_n$. With $t_0=0$ and the endpoint convention $t_{n+1}=t$, these normalized durations are $u_j=\omega_r(t_{j+1}-t_j)$, $j=0,\ldots,n$,
where $\omega_r$ and $\tau$ are defined in Eq.~\eqref{eq:finite_t1_scales}, satisfying
\begin{equation}
\sum_{j=0}^{n}u_j=\tau.
\end{equation}
For $n\geq1$, the homogeneous propagator, ordered with the earliest operation on the right, is
\begin{equation}
\begin{aligned}
M_{\omega}^{(\eta,g)}(\tau)
&=e^{Bu_n}\Phi_X^{(yz)}e^{Bu_{n-1}}
  \cdots \Phi_X^{(yz)}e^{Bu_0}\\
&=e^{-\eta\tau}
  D_q(u_n)\Phi_X^{(yz)}D_q(u_{n-1})
  \cdots \Phi_X^{(yz)}D_q(u_0)\\
&=e^{-\eta\tau}M_{\omega}^{(0,q)}(\tau).
\end{aligned}
\end{equation}
This identity also holds for zero events, with $u_0=\tau$; extracting scalar factors preserves chronological ordering. For the renewal-history average~\cite{Vacchini2020}, use the
normalized waiting-time density
\begin{equation}
w_\tau(u)=\frac{1}{\omega_r}w(u/\omega_r).
\end{equation}
For $n\geq1$, the completed intervals belong to
\begin{equation}
\mathcal S_n(\tau)
:=
\left\{
(u_0,\ldots,u_{n-1})>0:
\sum_{j=0}^{n-1}u_j\leq\tau
\right\},
\end{equation}
with residual $u_n$ fixed by the interval sum. The scalar renewal weight in Eq.~\eqref{eq:renmapdef}, expressed in the normalized interval variables, defines the
history measure
\begin{equation}
dP_n(\omega)
=
\Psi(u_n/\omega_r)
\prod_{j=0}^{n-1}w_\tau(u_j)\,du_j.
\end{equation}
The survival factor $\Psi(u_n/\omega_r)$ is the probability of no further event during the residual interval. For a matrix-valued
history observable,
\begin{equation}
\begin{aligned}
\mathbb E_\omega[X_\omega(\tau)]
&=
\Psi(\tau/\omega_r)X_{\varnothing}(\tau)
\\
&\quad+
\sum_{n=1}^{\infty}
\int_{\mathcal S_n(\tau)}
X_\omega(\tau)\,dP_n(\omega),
\end{aligned}
\end{equation}
where $X_{\varnothing}(\tau)$ is the zero-event value. Independent relaxation leaves these normalized weights unchanged, so the
historywise factorization yields
\begin{equation}
\begin{aligned}
A_{yz}^{(\eta,g)}(\tau)
&=\mathbb E_\omega
  \!\left[M_{\omega}^{(\eta,g)}(\tau)\right]\\
&=e^{-\eta\tau}
  \mathbb E_\omega
  \!\left[M_{\omega}^{(0,q)}(\tau)\right]\\
&=e^{-\eta\tau}
  \mathsf T_{yz}^{(0,q)}(\tau).
\end{aligned}
\label{eq:finite_t1_factorization}
\end{equation}

Here $\mathsf T_{yz}^{(0,q)}(\tau)$ denotes the transfer matrix
with entries $T_{ij}^{(P)}$ in Eq.~\eqref{eq:Tyz_transfer_matrix},
evaluated at $t=\tau/\omega_r$ and $\Gamma_Z=\omega_r q$.
Only for $q\geq0$ does $\mathsf T_{yz}^{(0,q)}$ represent a physical
pure-dephasing reference, with normalized rate $q$ and the same
renewal law and kick angle. The finite-$T_1$ model remains physical
for all $g,\eta\geq0$.

Equation~\eqref{eq:finite_t1_factorization} concerns the homogeneous
block of the affine Bloch map~\cite{BethRuskaiSzarekWerner2002}
\begin{equation}
\mathbf r(\tau)
=
A(\tau)\mathbf r(0)+\mathbf b(\tau).
\end{equation}
Here $A_{yz}^{(\eta,g)}(\tau)
=[A_{ij}(\tau)]_{i,j\in\{y,z\}}$.
For opposite preparations $\mathbf r(0)=\pm\mathbf e_j$ subject
to the same channel,
\begin{equation}
\frac{
\mathbf r(\tau;+j)
-
\mathbf r(\tau;-j)
}{2}
=
A(\tau)\mathbf e_j.
\end{equation}
The pair observables in Eq.~\eqref{eq:finite_t1_pair_signal}
therefore select off-diagonal homogeneous entries: the displacement
cancels, but the exponential attenuation remains.

At $q=0$, scalar homogeneous interevent evolution gives
\begin{equation}
A_{yz}^{(\eta,g)}(\tau)
=
e^{-\eta\tau}
\mathbb E_\omega
\!\left[
\bigl(\Phi_X^{(yz)}\bigr)^{N(\tau/\omega_r)}
\right],
\end{equation}
where $N(\tau/\omega_r)$ counts renewals up to physical time
$t=\tau/\omega_r$. Thus $Q_{y\to z}=-Q_{z\to y}$, without implying
commutation of the full affine interevent and kick maps.

For $q>0$, $0<\alpha<1$, and $\sin\theta\neq0$, the tails in
Eqs.~\eqref{eq:app_storage_tail} and~\eqref{eq:app_reverse_tail}
apply to $\mathsf T_{yz}^{(0,q)}$ under the pole and cut hypotheses
of Appendix~\ref{app:Wyz_singularity_analysis}. Absorbing $\lambda$
into the normalization, Eq.~\eqref{eq:30a} becomes
\begin{equation}
\widetilde{\kappa}(z)
:=
\frac{K_0(\omega_r z)}{\omega_r}
=
\frac{z}{(1+z^\alpha)^\gamma-1},
\qquad \Re z>0,
\label{eq:finite_t1_normalized_kernel}
\end{equation}
where $K_0$ is defined in Eq.~\eqref{eq:K0_def}, $z$ is conjugate
to $\tau$, and complex powers are taken on the principal branch.
With $\delta$ defined in Eq.~\eqref{eq:delta_def}, set
\begin{equation}
k_q=\frac{K_0(\omega_r q)}{\omega_r},
\qquad
H_q=\delta(q+2k_q).
\end{equation}
The directional coefficients are
\begin{align}
a_{ZY}
&=
\frac{\gamma\sin\theta\,k_q}
{H_q\Gamma(1-\alpha)},
\label{eq:finite_t1_aZY}
\\
a_{YZ}
&=
\frac{
\gamma\sin\theta
\left[q+\delta k_q\right]
}{
H_q^2\Gamma(-\alpha)
}.
\label{eq:finite_t1_aYZ}
\end{align}
Equation~\eqref{eq:finite_t1_factorization} then gives, as
$\tau\to\infty$,
\begin{align}
Q_{y\to z}(\tau)
&\sim
e^{-\eta\tau}a_{ZY}\tau^{-\alpha},
\label{eq:finite_t1_forward_tail}
\\
Q_{z\to y}(\tau)
&\sim
e^{-\eta\tau}a_{YZ}\tau^{-1-\alpha}.
\label{eq:finite_t1_reverse_tail}
\end{align}
These asymptotics exclude the $q=0$ control and $\alpha=1$ limit,
which require separate treatment.

Numerical methods and validation are summarized at the end of Appendix~\ref{app:count_matched_benchmark}.
\subsection{Poisson benchmark with equal expected event counts}
\label{app:count_matched_benchmark}

For the comparator defined by
Eq.~\eqref{eq:count_matched_intensity}, introduce the
normalized intensity and mean count
\begin{equation}
\begin{aligned}
\mu(\tau)&=\frac{m(\tau/\omega_r)}{\omega_r},\\
\widetilde\mu(s)
&=\frac{1}{(1+s^\alpha)^\gamma-1},\\
\mathcal M(\tau)&=M(\tau/\omega_r)
=\int_0^\tau\mu(v)\,dv,\\
\widetilde{\mathcal M}(s)&=\frac{\widetilde\mu(s)}{s}.
\end{aligned}
\label{eq:count_matched_normalized_rate}
\end{equation}
These transforms follow from Eq.~\eqref{eq:m_s} and the
waiting-time transform in Eq.~\eqref{eq:11e}; $s$ is conjugate
to $\tau$, with $\Re s>0$ and principal-branch complex powers.
Using $g_2$ from Eq.~\eqref{eq:finite_t1_g2} and $\delta$
from Eq.~\eqref{eq:delta_def}, the homogeneous $yz$ block of
Eq.~\eqref{eq:count_matched_master} satisfies
\begin{equation}
\begin{aligned}
\frac{d A_{yz}^{(\mathrm{match})}}{d\tau}
&=
\begin{pmatrix}
-g_2-\delta\mu&-\mu\sin\theta\\
\mu\sin\theta&-\eta-\delta\mu
\end{pmatrix}
A_{yz}^{(\mathrm{match})},\\
A_{yz}^{(\mathrm{match})}(0)&=I_2.
\end{aligned}
\label{eq:count_matched_block}
\end{equation}
Here $\mu=\mu(\tau)$, and the $zy$ entry gives
$Q_{y\to z}^{(\mathrm{match})}$ through
Eq.~\eqref{eq:finite_t1_pair_signal}.

For the Prabhakar process, the chronological quantum-renewal
resolvent~\cite{Vacchini2020} in Eq.~\eqref{eq:Myz_exact},
together with Eq.~\eqref{eq:finite_t1_factorization}, gives
the exact expression evaluated for the additional curves:
\begin{equation}
Q_{y\to z}^{(\mathrm P)}(\tau)
=e^{-\eta\tau}\,
\mathcal L^{-1}_{s\to\tau}
\left\{
\frac{\sin\theta\,\widetilde\kappa(s+q)}{\Delta_q(s)}
\right\},
\label{eq:count_matched_prabhakar_transfer}
\end{equation}
where $q$ is the damping contrast defined above,
$\widetilde\kappa$ is given by
Eq.~\eqref{eq:finite_t1_normalized_kernel}, and
\begin{equation}
\begin{aligned}
\Delta_q(s)
&=\bigl[s+q+\delta\widetilde\kappa(s+q)\bigr]
  \bigl[s+\delta\widetilde\kappa(s)\bigr]\\
&\quad+\sin^2\theta\,
\widetilde\kappa(s+q)\widetilde\kappa(s).
\end{aligned}
\label{eq:count_matched_determinant}
\end{equation}
The shifted kernel in the numerator fixes the forward
$y\to z$ ordering. Equations~\eqref{eq:count_matched_block}
and~\eqref{eq:count_matched_prabhakar_transfer} retain the
same drift and kick geometry and the physical relaxation
attenuation; only the stochastic event law changes.

The Poisson generating function (see Eq.~(1.8) of Ref.~\cite{LastPenrose2018}) gives
$\mathbb E[u^{N_{\mathrm{match}}(\tau/\omega_r)}] =\exp[\mathcal M(\tau)(u-1)]$.
Its defining series converges absolutely for every $u\in\mathbb C$. At equal damping, $q=0$, setting
$u=e^{i\theta}$, taking the imaginary part, and using Eq.~\eqref{eq:count_matched_equal_damping}
gives the independent exact control
\begin{equation}
Q_{y\to z}^{(\mathrm{match})}(\tau)
=e^{-\eta\tau-\delta\mathcal M(\tau)}\sin\!\left[\mathcal M(\tau)\sin\theta\right].
\label{eq:count_matched_equal_damping_control}
\end{equation}

The Prabhakar curves, readouts, and peak values in
Figs.~\ref{fig:sim108_finite_t1} and
\ref{fig:count_matched_transfer} use a shared
22-term fixed-Talbot calculation in binary64
arithmetic~\cite{AbateValko2004}.
Signed arrays retain $\tau=0$ and positive times
from $10^{-6}$ to $50$.

For the count-matched comparator, Hankel inversion
of $\widetilde\mu$ in
Eq.~\eqref{eq:count_matched_normalized_rate},
supplemented by the renewal series near zero,
supplied the intensity. Its integrable initial
behavior
$\mu(\tau)\sim
\tau^{\alpha\gamma-1}/\Gamma(\alpha\gamma)$
was regularized by integrating in
$\xi=\tau^{\alpha\gamma}$.
The comparator equations were integrated with
DOP853 and checked by tolerance refinement,
Radau integration, and the exact control in
Eq.~\eqref{eq:count_matched_equal_damping_control}.

For the independent mean-count check,
22-term fixed-Talbot inversion of
$\widetilde{\mathcal M}$ in
Eq.~\eqref{eq:count_matched_normalized_rate}
was compared with high-precision de Hoog inversion
and independent integration of $\mu$.
On the respective validation samples, the largest
observed absolute discrepancies were approximately
$3.94\times10^{-13}$ and $1.21\times10^{-12}$.

For every simulated relaxation rate, both
Prabhakar directional-transfer entries were checked
against de Hoog inversion~\cite{deHoogKnightStokes1982}
at $\tau=10^{-3},10^{-2},10^{-1},1,10,50$
and at the separately refined forward-transfer
peak for that rate.
The checks used inversion degree 80 with a
requested precision of 60 decimal digits,
refined to degree 120 and 90 digits at the
readout and peaks.
The largest observed absolute discrepancy was
approximately $6.48\times10^{-14}$.
The earlier de Hoog validation also included
affine-map and chronological-history controls.

For Fig.~\ref{fig:sim108_finite_t1}(d),
peak values were estimated separately for each
model and relaxation rate by refining each
detected local maximum and checking grid doubling.
The reported discrepancies apply to the tested
samples; neither uniform error bounds nor
global continuous-time optimality are certified.
\bibliography{./Fractional_Glassy}
\end{document}